\documentclass[a4paper,fleqn]{cas-dc}
\usepackage[numbers]{natbib}
\usepackage{amsfonts}
\usepackage{algorithmic}
\usepackage{amsmath} 
\usepackage{amssymb} 
\usepackage{caption}
\usepackage{subcaption}
\usepackage{etoolbox}
\usepackage{forest}
\usepackage{lingmacros}
\usepackage{textcomp}
\usepackage{tree-dvips}
\usepackage{CJKutf8}
\usepackage{tikz}
\usepackage{tikz-cd}
\usepackage[arrowdel]{physics}
\usepackage{graphicx}
\usepackage{wrapfig}
\usepackage{listings}
\usepackage{pgfplots, pgfplotstable}
\usepackage{diagbox} 
\usepackage[usestackEOL]{stackengine}
\usepackage{makecell}
\usepackage{mathrsfs}
\usepackage{moresize}
\usepackage{multirow}
\usepackage{multicol}
\usepackage[numbers]{natbib}
\usepackage[T1]{fontenc}
\usepackage{xcolor}
\allowdisplaybreaks[1]
\definecolor{orchid}{rgb}{0.7, 0.4, 1.1}
\definecolor{comment_color}{rgb}{0, 0.5, 0}
\definecolor{keyword_color}{rgb}{0.3, 0, 0.6}
\definecolor{string_color}{rgb}{0.5, 0, 0.1}

\begin{document}

\shorttitle{Propagation electrodynamics and differential conduction of action potentials in geometrically branched squid giant axons}
\shortauthors{Xi Liu \begin{CJK*}{UTF8}{bsmi}劉錫\end{CJK*}$^a$, Wenxi Fang \begin{CJK*}{UTF8}{bsmi}方文希\end{CJK*}$^b$, Ken Perlin$^c$}
\title{{\Large Propagation electrodynamics and differential conduction of action potentials in geometrically branched squid giant axons and neurons}}
\author[1]{\color{black}Xi Liu \begin{CJK*}{UTF8}{bsmi}劉錫\end{CJK*}}
\author[2]{\color{black}Wenxi Fang \begin{CJK*}{UTF8}{bsmi}方文希\end{CJK*}}
\author[3]{\color{black}Ken Perlin}
\address{$^a$xl3467@columbia.edu, Columbia University; $^b$u3013972@connect.hku.hk, Inner Mongolia University of Science and Technology; $^c$perlin@nyu.edu, New York University}
\begin{abstract}
Classical neuronal cable theory relies on quasi-static electric field approximations and neglects magnetic induction, Lorentz force coupling, and transient electromagnetic currents, limiting its ability to fully characterize action potential propagation within geometrically branched axons and dendrites. This work develops a coupled Maxwell-electromagnetic cable framework by integrating finite-difference time-domain (FDTD) solutions of Maxwell's equations with extended Hodgkin-Huxley and Fitzhugh-Nagumo membrane dynamics, incorporating magnetic gating perturbations, electromagnetic trans-membrane currents $I_{\text{EM}}$, and nanoscale quantum corrections for thin neural segments. Controlled propagation experiments are designed to quantify deviations from standard cable predictions across asymmetric and symmetric axonal bifurcation geometries. Numerical results demonstrate that inductive magnetic effects lower the critical branch radius for junction conduction failure and break symmetric action potential invasion in geometrically identical child branches under external transverse magnetic fields. An electromagnetic corrected geometric ratio $GR_{\text{EM}}$ is proposed to revise impedance-matching conditions at branch points, accounting for size-dependent axial current imbalance induced by magnetic and displacement currents. Parent axon conduction velocity deviates substantially from the canonical $\sqrt{d}$ scaling law when electromagnetic feedback and quantum charge distributions are included, triggering early signal blockage at large cable diameters. Collectively, this study establishes that quasi-static cable models underestimate electromagnetic corrections to propagation speed, waveform shape, and bifurcation transmission fidelity; the coupled Maxwell-cable framework provides a comprehensive multi-physics tool for modeling electrodynamic signal behavior in complex neuronal architectures.
\end{abstract}
\maketitle
\pagestyle{plain}
\thispagestyle{empty}
\section{Introduction}
\begin{figure}
\centering
\begin{subfigure}{0.45\textwidth}
\centering
\includegraphics[width=\textwidth,height=0.3\textwidth]{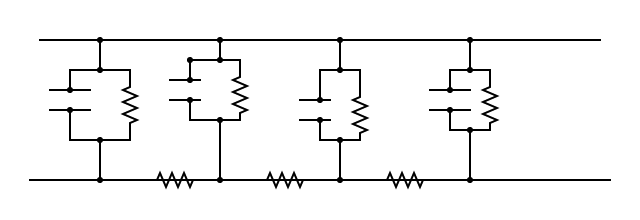}
\caption{Cable circuit of dendrites and axons.}
\label{fig:cable_circuit}
\end{subfigure}
\begin{subfigure}{0.45\textwidth}
\centering
\includegraphics[width=0.4\textwidth,height=0.2\textwidth]{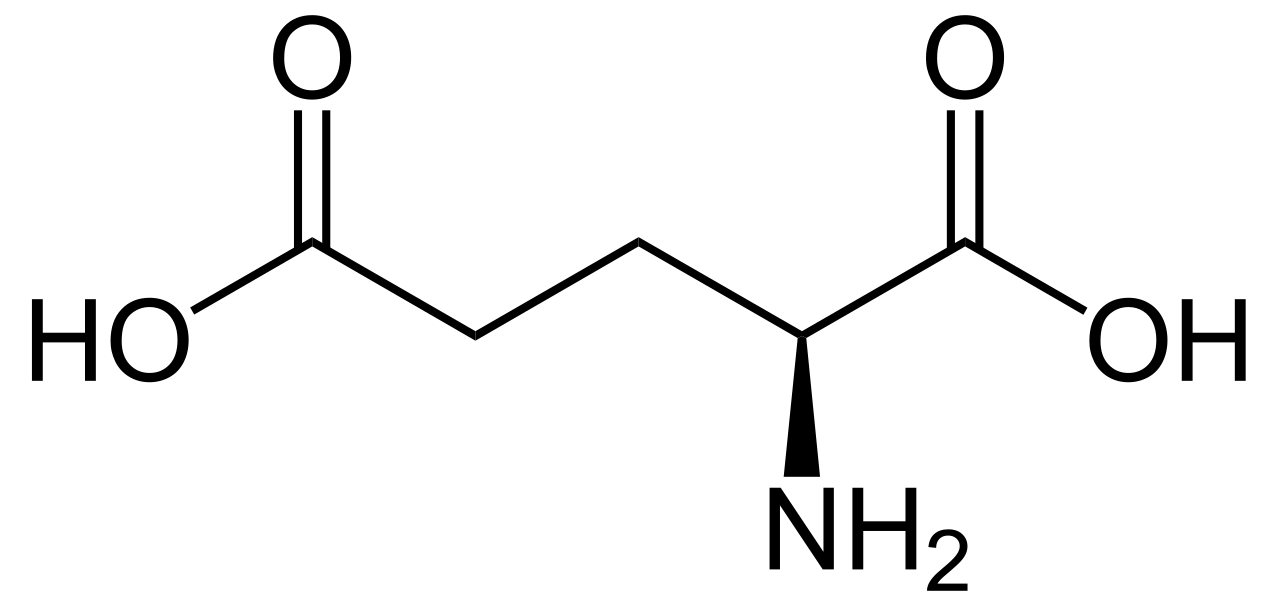}
\includegraphics[width=0.4\textwidth,height=0.2\textwidth]{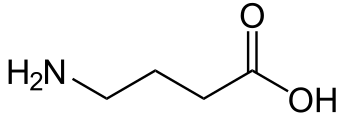}
\includegraphics[width=0.4\textwidth,height=0.2\textwidth]{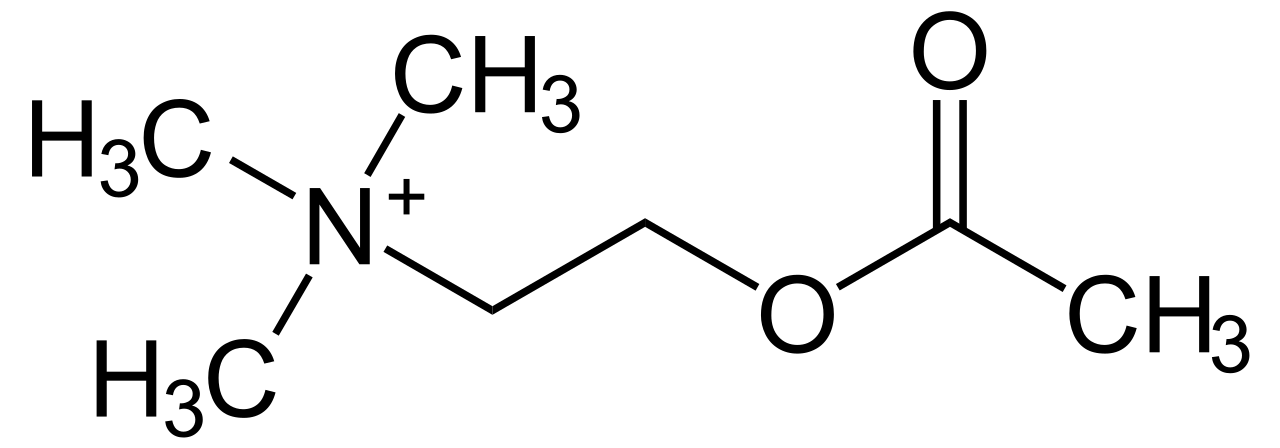}
\includegraphics[width=0.4\textwidth,height=0.2\textwidth]{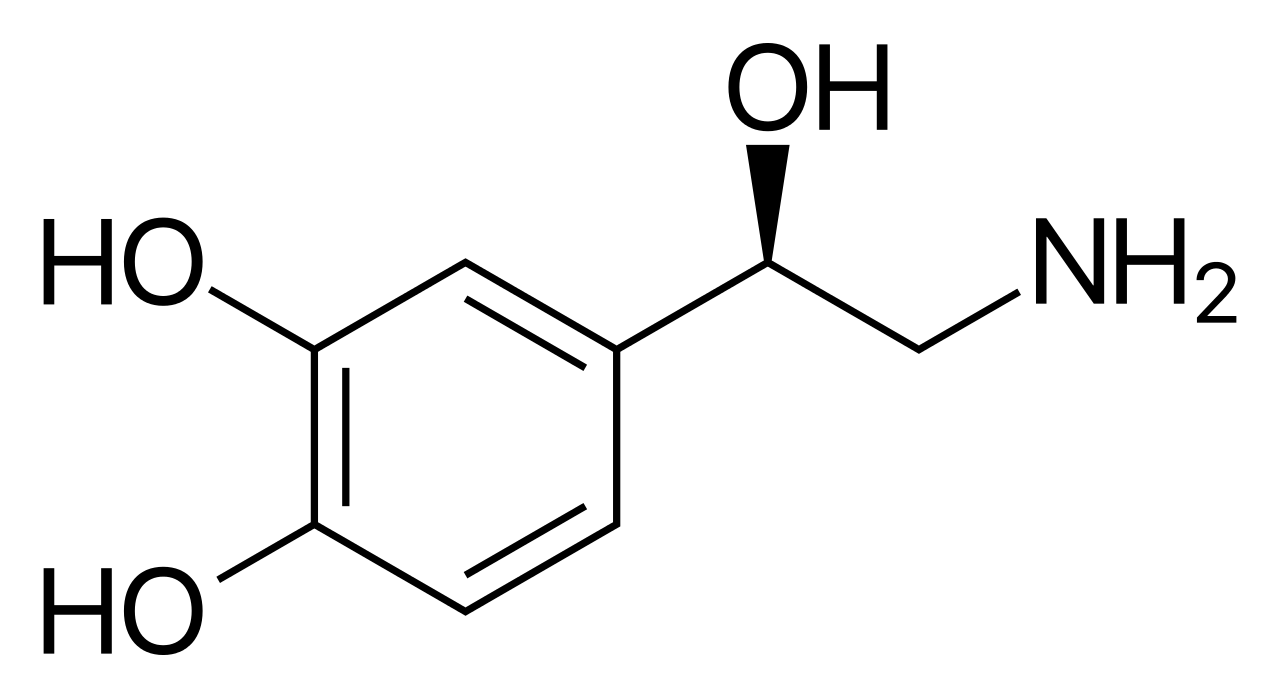}
\caption{Glutamate, $\gamma$-aminobutyric acid, acetylcholine, and norepinephrine are neurotransmitters passing through the cable in interneuron communication.}
\label{fig:neurotransmitters}
\end{subfigure}
\caption{Neural cable circuit and associated neurotransmitters.}
\label{fig:neural_cable}
\end{figure}

The interneuron communication and data processing in the brain depends on the signal propagation among cells that involves varying geometry, which can be modeled by cable theory \cite{rall_2011}. Interneuron communication includes electrical and chemical synapses. Electrical synapses involve direct connections between the presynaptic and postsynaptic cell membranes via gap junctions that allow the flow of electric current between cells, enabling rapid signal transmission and action potential propagation \cite{waxman_1980}. In a chemical synapse, the electrical activity of the presynaptic neuron triggers the release of neurotransmitters such as glutamate, $\gamma$-aminobutyric acid, acetylcholine, or norepinephrine, which bind to receptors on the postsynaptic cell (figure \ref{fig:neurotransmitters}). Computational investigations in action potentials often involves simplifying assumptions on space clamping conditions that halt the action potential propagation, uniform properties of the cable, and constant velocity propagation. These assumption can simplify the partial differential equation of the spread of membrane potential, but they cannot be used here since changing the geometry of the cable conductor affects the action potential propagation, its shape and velocity \cite{goldstein_1974}, \cite{lindsay_2004}. The cable equation can be modified taking into account of parameters such as charge inhomogeneities \cite{lazarevich_2013}. The cable equation can be augmented with external driving forces from applied electromagnetic fields, synaptic excitation, and active membrane properties. The Fitzbugh-Nagumo model contains an external driving force that is dependent on potential or the spatiotemporal varying external driving force \cite{luscher_1990}. To solve the nonhomogeneous cable equation with a source or forcing term, the method of eigenfunction expansions can be used. We investigated how altering cable conductor geometry impacts action potential propagation, observing that changing the diameter of a branch can either accelerate or hinder propagation in adjacent branches, contingent on matching current flow at branch points. A geometric ratio quantifying branch diameter relationships aids in classifying equivalent cylinders, guiding analysis of propagation behavior.


\section{Cable theory}
In cable theory, dendrites and axons are modeled by cylinders with RC circuits (containing resistors and capacitors) connected in parallel. See figure \ref{fig:cable_circuit} for the cable circuit. The top half of the entire long cable is interfacing with the extracellular fluid, the bottom half of the entire long cable is inside cytosol, cytoplasmic matrix, or the intracellular fluid. Let $r_n$ be the resistance of the $n$th RC circuit in the figure, $c_n$ be the capacitance of the $n$th RC circuit, $R_n$ ($\Omega\cdot cm ^ 2$) and $C_n$ ($F/cm ^ 2$) be the specific resistance and capacitance of an unit area of membrane, $a$ be the radius of the axon
$r_n = \frac{R_n}{2\pi a},\quad c_n = C_n 2\pi a$.
As the radius $a$ of the axon increases, a greater area for current to pass through the membrane, so the resistance $r_n$ becomes lower. As the circumference $2\pi a$ of the axon increases, more membrane can be used to store charge, so the $c_n$ becomes higher. Let $\rho_l$ be the specific electrical resistance of the axoplasm (cytoplasm inside the axon), then the intracellular resistance $r_l$ per unit length ($\Omega\cdot cm ^ {-1}$) in the longitudinal direction is
$r_l = \frac{\rho_l}{\pi a ^ 2}$.
As the axon cross sectional area $\pi a ^ 2$ increases, there are more paths for the current flow the axoplasm, so the axoplasmic resistance decreases \cite{traub_1977}. By Ohm's law for voltage $V$, current $I$, and resistance $R$, $V = IR$
\begin{align*}
\Delta V = -i_l r_l\Delta x,\quad
\frac{\partial V}{\partial x} = -i_l r_l,\quad
\frac{1}{r_l}\frac{\partial V}{\partial x} = -i_l
\end{align*}
Let $i_n$ be the current passing through the membrane per unit length $n$, then the total current passing through $x$ units is $x\cdot i_n$. so the change of current in axon cytoplasm $\Delta i_l$ at distance $\Delta x$ is
\begin{align*}
\Delta i_l = -i_n\Delta x,\quad
\frac{\partial i_l}{\partial x} = -i_n
\end{align*}
on the side of the cytoplasm, the capacitance causes a current towards the membrane, this current is displacement current
$i_c = c_n\frac{\partial V}{\partial t}$.
$i_r = \frac{V}{r_n}$ is current through the membrane.
since $i_n = i_r + i_c$, $\frac{\partial i_l}{\partial x}$ is the change of axoplasm current per unit length
\begin{align*}
\frac{\partial i_l}{\partial x} = -i_n = \frac{V}{r_n} + c_n\frac{\partial V}{\partial t}
\end{align*}
substituting $\frac{1}{r_l}\frac{\partial V}{\partial x} = -i_l$, $\frac{1}{r_l}\frac{\partial ^ 2 V}{\partial x ^ 2} = -\frac{\partial i_l}{\partial x}$
\begin{align*}
\frac{1}{r_l}\frac{\partial ^ 2 V}{\partial x ^ 2} = c_n\frac{\partial V}{\partial t} + \frac{V}{r_n}
\end{align*}
using a length constant $\lambda$ that is a ratio of the membrane resistance $r_n$ and the intracellular resistance $r_l$
\begin{align*}
\lambda &= \sqrt{\frac{r_n}{r_l}} = \sqrt{\frac{R_n}{R_i}\frac{d}{4}}\\
\tau\frac{\partial V}{\partial t} &= \lambda ^ 2\frac{\partial ^ 2 V}{\partial x ^ 2} - f(V)
\end{align*}
As the membrane resistance $R_n$ increases, there is less current leaks across the membrane, so the space constant $\lambda$ increases. Also, a dendrite with larger diameter $d$ has a larger space constant $\lambda$, so the spread of current is accelerated with a larger diameter.

For the three dimensional case, let $V$ be the departure of membrane potential from rest, $x$ be the spatial coordinate on the core conductor, $t$ be the time, $\tau$ be the time constant of the membrane, $\lambda$ be a constant depending on conductor length, $f(x, t)$ be the external driving force function. $\lambda$ is proportional to the cable diameter $d$. $\lambda \propto \sqrt{d}$. The Fitzhugh-Nagumo nerve conduction equation is \cite{fitzHugh_1955}
\begin{align*}
\tau\frac{\partial V}{\partial t} &= \lambda ^ 2\nabla ^ 2 V - f(V)\\
\tau\frac{\partial V}{\partial t} &= \lambda ^ 2\left(\frac{\partial ^ 2 V}{\partial x ^ 2} + \frac{\partial ^ 2 V}{\partial y ^ 2} + \frac{\partial ^ 2 V}{\partial z
^ 2}\right) - f(V)\\
f(V) &= V(V - a)(V - 1)
\end{align*}
\begin{figure}
\centering
\begin{tabular}{cc}
\includegraphics[width=0.22\textwidth,height=0.1\textwidth]{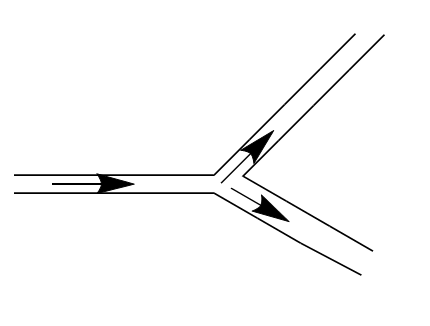} &
\includegraphics[width=0.2\textwidth,height=0.1\textwidth]{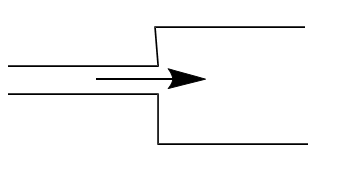} \\
(a) Branching structure & (b) Only changing diameter
\end{tabular}
\caption{Left figure is branching, right figure is only changing diameter.}
\label{fig:comparison}
\end{figure}

One way of solving the core conductor equation is using the eigenfunction expansion and fourier coefficients
{\fontsize{9pt}{5pt}
\begin{align*}
V(x, t) &= \sum_{n = 1} ^ {\infty} a_n(t)X_n(x) = \sum_{n = 1} ^ {\infty} a_n(t)\sin\frac{n\pi x}{l}\\
a_n(0) &= \frac{<f, X_n>}{\|X_n\| ^ 2} = \frac{1}{\|X_n\| ^ 2}\int_0 ^ l f(x)X_n(x)dx\\
b_n(t) &= \frac{<f, X_n>}{\|X_n\| ^ 2} = \frac{1}{\|X_n\| ^ 2}\int_0 ^ l f(x, t)X_n(x)dx\\
V(x, t) &= \sum_{n = 1} ^ {\infty}\left(a_n(0)e ^ {-c\lambda_n t} + \int_0 ^ t b_n(\tau)e ^ {c\lambda_n(\tau - t)}d\tau\right)\sin\frac{n\pi x}{l}
\end{align*}}
The continuity of current need to be maintained. $x_0$ is the junction point of changing geometry. $x_{0-}$ and $x_{0+}$ are points immediately at the left and right of $x_0$. $R_i$ is intracellular specific resistance.
{\fontsize{9pt}{5pt}
\begin{align*}
&I_i = -\frac{\pi r_1 ^ 2}{R_i}\frac{dV}{dx}_{x_{0-}}
= -\frac{\pi r_2 ^ 2}{R_i}\frac{dV}{dx}_{x_{0+}}
= -\frac{\pi(d_1/2) ^ 2}{R_i}\frac{dV}{dx}_{x_{0-}}\\
&= -\frac{\pi(d_2/2) ^ 2}{R_i}\frac{dV}{dx}_{x_{0+}}
= -\frac{\pi d_1 ^ 2}{4 R_i}\frac{dV}{dx}_{x_{0-}}
= -\frac{\pi d_2 ^ 2}{4 R_i}\frac{dV}{dx}_{x_{0+}}
\end{align*}}
Branching usually involves a parent branch and two child branches of different radius (figure \ref{fig:comparison}). To consider propagation in both directions, we distinguish that one of the branch is the starting source of the action potential propagating to the junction (usually parent branches) and the other branch in the other side of the junction (usually child branches). The propagation near the junction is having the geometric ratio
\begin{align*}
GR = \sum_i\frac{d_i ^ {3/2}}{d_a ^ {3/2}}
\end{align*}
$d_a$ is the diameter of the starting branch of the action potential propagating to the junction. $d_i$ is the diameter of the $i$th branch on the other side of the junction. If $GR = 1$, the dendritic tree can be converted to an equivalent cylinder. If $GR < 1$, the branches collectively at the other side of the junction (usually child branches) can be converted to an equivalent cylinder with diameter smaller than the starting branch of the action potential propagating to the junction (usually parent branches). If GR > 1, the branches collectively at the other side of the junction (usually child branches) can be converted to an equivalent cylinder with diameter larger than the starting branch of the action potential propagating to the junction (usually parent branches).

\section{Electromagnetic modified Maxwell-cable equations}
Prior theoretical work has established the mathematical link between Maxwell’s electrodynamic laws and classical core-conductor cable theory, with Lindsay et al. providing a comprehensive foundational derivation mapping full electromagnetic field behavior to the standard quasi-static cable equation and outlining necessary extensions to capture inductive and displacement current terms \cite{lindsay2004maxwellcable,liu2026}. Subsequent analytical progress on coupled Maxwell-cable mixed-dimensional partial differential systems has formalised well-posedness criteria and semigroup stability analysis for radiating, geometrically curved neural cables, forming a rigorous mathematical framework for multi-physics neuronal modeling \cite{reis2025coupledmaxwell,clemens2025radiatingcable,LIU2024}. Modifications to the cable equation that explicitly account for transmembrane polarization and bidirectional electric field coupling were developed by Wang et al., whose work introduced field-dependent membrane current terms analogous to the $I_{\text{EM}}$ electromagnetic coupling operator adopted in the present study \cite{wang2018modifiedcable}.

Numerical realizations of coupled neuronal-electromagnetic systems predominantly rely on finite-difference time-domain (FDTD) discretization on Yee grids to self-consistently evolve electric and magnetic fields alongside Hodgkin-Huxley membrane dynamics. Early FDTD-Hodgkin-Huxley coupled solvers demonstrated accurate simulation of extracellular stimulation-triggered axonal activation, while alternating-direction-implicit (ADI) FDTD variants improved numerical stability for stiff neural time scales \cite{li2011fdtdhh,choi2012adifdtd}. Full three-dimensional FDTD formulations further enable forward modelling of magnetoencephalography (MEG) and electroencephalography (EEG) signals originating from propagating action potentials, capturing spatially distributed magnetic flux generated by branched axonal architectures \cite{hashemi20173dfdtdneuron}. Despite these established numerical pipelines, existing literature rarely addresses magnetic Lorentz force feedback on ion channel gating kinetics or size-dependent current imbalance at axonal bifurcations, which this work quantifies via controlled propagation experiments and the revised electromagnetic geometric ratio $GR_{\text{EM}}$.

The classical cable equation assumes quasi-static electric fields and neglects magnetic effects. To include full electrodynamics, we start with differential Maxwell's equations with magnetic vector potential $\mathbf{A}$ and electric potential $\Phi$ through the relations $\mathbf{\nabla\times A=B}$ and $\mathbf{E}=-\nabla\Phi-\frac{\partial\mathbf{A}}{\partial t}$
\begin{align*}
&\nabla \cdot \mathbf{E} = \frac{\rho}{\epsilon_0}\quad
\nabla \times \mathbf{E} = -\frac{\partial \mathbf{B}}{\partial t}\\
&\nabla \cdot \mathbf{B} = 0,\quad
\nabla \times \mathbf{B} = \mu_0\left(\mathbf{J} + \epsilon_0\frac{\partial \mathbf{E}}{\partial t}\right)
\end{align*}
The total current density in the axon includes conductive, displacement, magnetization, and source components
\begin{align*}
&\mathbf{J}_{\text{total}} = \mathbf{J}_{\text{conductive}} + \mathbf{J}_{\text{displacement}} + \mathbf{J}_{\text{magnetization}} + \mathbf{J}_{\text{source}}\\
&=\sigma_0\mathbf{E}
+\epsilon\frac{\partial \mathbf{E}}{\partial t}
+\nabla\times\mathbf{M}
+\mathbf{J}_{\text{source}}
\end{align*}
The classical cable equation $\frac{1}{r_l}\frac{\partial^2 V}{\partial x^2} = c_n\frac{\partial V}{\partial t} + \frac{V}{r_n}$ is extended with magnetic coupling terms is written in a telegrapher-equation form as
\begin{align*}
\frac{1}{r_l}\frac{\partial^2 V}{\partial x^2}
&=
c_n\frac{\partial V}{\partial t}
+\frac{V}{r_n}
+\frac{l_lc_n}{r_l}\frac{\partial^2 V}{\partial t^2}
+\frac{l_l}{r_l r_n}\frac{\partial V}{\partial t}
+\mathcal{I}_{\mathrm{EM}}
+\mathcal{I}_{\mathrm{Q}}
\end{align*}
where $\mathcal{I}_{\mathrm{EM}}$ and $\mathcal{I}_{\mathrm{Q}}$ represents the electromagnetic and quantum current contributions and has units of ampere per unit length, $\mathrm{A\,m^{-1}}$.

The dimensional consistency of the modified cable equation follows by taking $r_l$ and $l_l$ as axial resistance and inductance per unit length, and $r_n$ and $c_n$ as transverse resistance and capacitance
per unit length:
\begin{align*}
&[r_l]=\Omega\,\mathrm{m}^{-1},\;
[l_l]=\mathrm{H\,m}^{-1},\;
[r_n]=\Omega\,\mathrm{m},\;
[c_n]=\mathrm{F\,m}^{-1}\\
&[\mathcal{I}_{\mathrm{EM}}]=[\mathcal{I}_{\mathrm{Q}}]=\mathrm{A\,m}^{-1},\quad
\left[\frac{1}{r_l}V_{xx}\right]
=\frac{\mathrm{m}}{\Omega}\frac{\mathrm{V}}{\mathrm{m}^2}
=\mathrm{A\,m}^{-1}\\
&[c_nV_t]
=\frac{\mathrm{F}}{\mathrm{m}}\frac{\mathrm{V}}{\mathrm{s}}
=\mathrm{A\,m}^{-1},\quad
\left[\frac{V}{r_n}\right]
=\frac{\mathrm{V}}{\Omega\,\mathrm{m}}
=\mathrm{A\,m}^{-1}\\
&\left[\frac{l_lc_n}{r_l}V_{tt}\right]
=\frac{\mathrm{s}^2}{\Omega\,\mathrm{m}}
\frac{\mathrm{V}}{\mathrm{s}^2}
=\mathrm{A\,m}^{-1},\quad
\left[\frac{l_l}{r_lr_n}V_t\right]
=\frac{\mathrm{s}}{\Omega\,\mathrm{m}}
\frac{\mathrm{V}}{\mathrm{s}}
=\mathrm{A\,m}^{-1}
\end{align*}
The electromagnetic contribution is obtained from the induced electromagnetic current density, induced electric field can be obtained from Faraday's law.
\begin{align*}
\mathcal{I}_{\mathrm{EM}}
&=
\frac{1}{A_{\mathrm{eff}}}
\int_{A_{\mathrm{eff}}}
\mathbf{J}_{\mathrm{EM}}\cdot\hat{\mathbf{x}}\,dA\\
\mathbf{J}_{\mathrm{EM}}
&=
\sigma_m
\left(
\mathbf{E}_{\mathrm{ind}}
+
\mathbf{v}\times\mathbf{B}
\right)
\end{align*}
The classical Fitzhugh-Nagumo equation is extended to include electromagnetic effects:
\begin{align*}
\tau\frac{\partial V}{\partial t} &= \lambda^2\nabla^2 V - f(V) - \gamma V - \eta \frac{\partial \mathbf{B}}{\partial t}\cdot\nabla V + \kappa(\mathbf{v}\times\mathbf{B})\cdot\nabla V
\end{align*}
where the additional terms include
magnetic damping term $\gamma V$,
inductive coupling $\eta \frac{\partial \mathbf{B}}{\partial t}\cdot\nabla V$,
Lorentz force contribution on ionic currents $\kappa(\mathbf{v}\times\mathbf{B})\cdot\nabla V$.

For the nanoscale quantum subsystem, the charge carriers are described by a minimally coupled Schr\"odinger-Poisson system,
\begin{align*}
i\hbar\frac{\partial\psi}{\partial t}
&=\left[\frac{1}{2m}\left(-i\hbar\nabla-q\mathbf{A}\right)^2
+q\Phi\right]\psi\\
\nabla^2\Phi
&=-\frac{q}{\epsilon}|\psi|^2
\end{align*}
The corresponding quantum probability current is
\begin{align*}
\mathbf{j}_{Q}
&=\frac{1}{m}
\operatorname{Re}
\left[
\psi^*\left(-i\hbar\nabla-q\mathbf{A}\right)\psi
\right]
\end{align*}
The associated electric current density is
$\mathbf{J}_{Q}=q\mathbf{j}_{Q}$. The quantum contribution to the cable current is therefore defined by
\begin{align*}
\mathcal{I}_{Q}
&=\frac{1}{A_{\mathrm{eff}}}
\int_{A_{\mathrm{eff}}}
\mathbf{J}_{Q}\cdot\hat{\mathbf{x}}\,dA
\end{align*}

For stochastic electromagnetic noise $\mathbf{B}_{noise}$,
include thermal and quantum fluctuations is
\begin{align*}
\tau\frac{\partial V}{\partial t} = \lambda^2\nabla^2 V - f(V) + \xi(t) + \eta\mathbf{B}_{noise}\cdot\nabla V
\end{align*}
where $\xi(t)$ is Gaussian white noise satisfying:
\begin{align*}
\langle\xi(t)\xi(t')\rangle = 2D\delta(t-t')
\end{align*}
The discretized equations using finite-difference time-domain (FDTD) are
\begin{align*}
&V^{n+1}_{i,j,k} = V^n_{i,j,k} + \frac{\Delta t}{\tau}\left[\lambda^2\nabla^2 V^n_{i,j,k} - f(V^n_{i,j,k}) - \gamma V^n_{i,j,k}\right]\\
&\mathbf{B}^{n+1/2} = \mathbf{B}^{n-1/2} - \Delta t\nabla\times\mathbf{E}^n\\
&\mathbf{E}^{n+1} = \mathbf{E}^n + \frac{\Delta t}{\epsilon}\left[\frac{1}{\mu}\nabla\times\mathbf{B}^{n+1/2} - \mathbf{J}^n\right]
\end{align*}

The electromagnetic extension of classical cable theory introduces modifications that include magnetic coupling which adds inductive effects to signal propagation through the operator $\mathcal{L}[\mathbf{B}]$,
four-vector formalism enables relativistic treatment of fast signals through the d'Alembertian wave equation,
tensor formulation handles branching geometries via the field tensor $F^{\mu\nu}$.
This framework transforms the classical cable theory from a purely electrical model into a comprehensive electromagnetic theory capable of describing the full complexity of neural signal propagation, including magnetic field effects, relativistic corrections, and quantum phenomena at the nanoscale.

\begin{figure}
\centering
\begin{subfigure}{0.23\textwidth}
\centering
\includegraphics[width=\textwidth,height=0.8\textwidth]{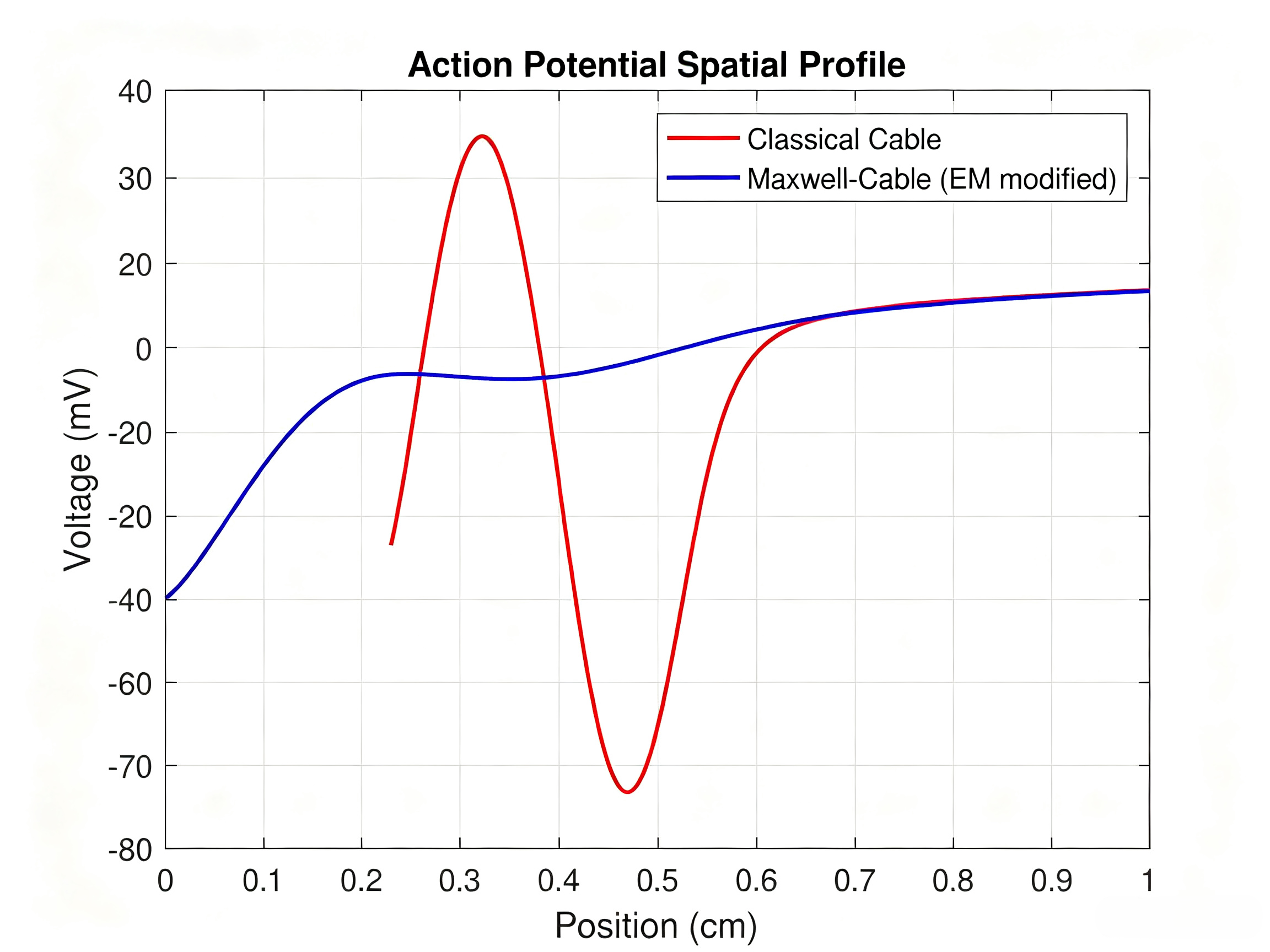}
\caption{Spatial membrane potential comparing classical and electromagnetic modified Maxwell-cable equation.}
\label{fig:em_ap_spatial}
\end{subfigure}
\begin{subfigure}{0.23\textwidth}
\centering
\includegraphics[width=\textwidth,height=0.8\textwidth]{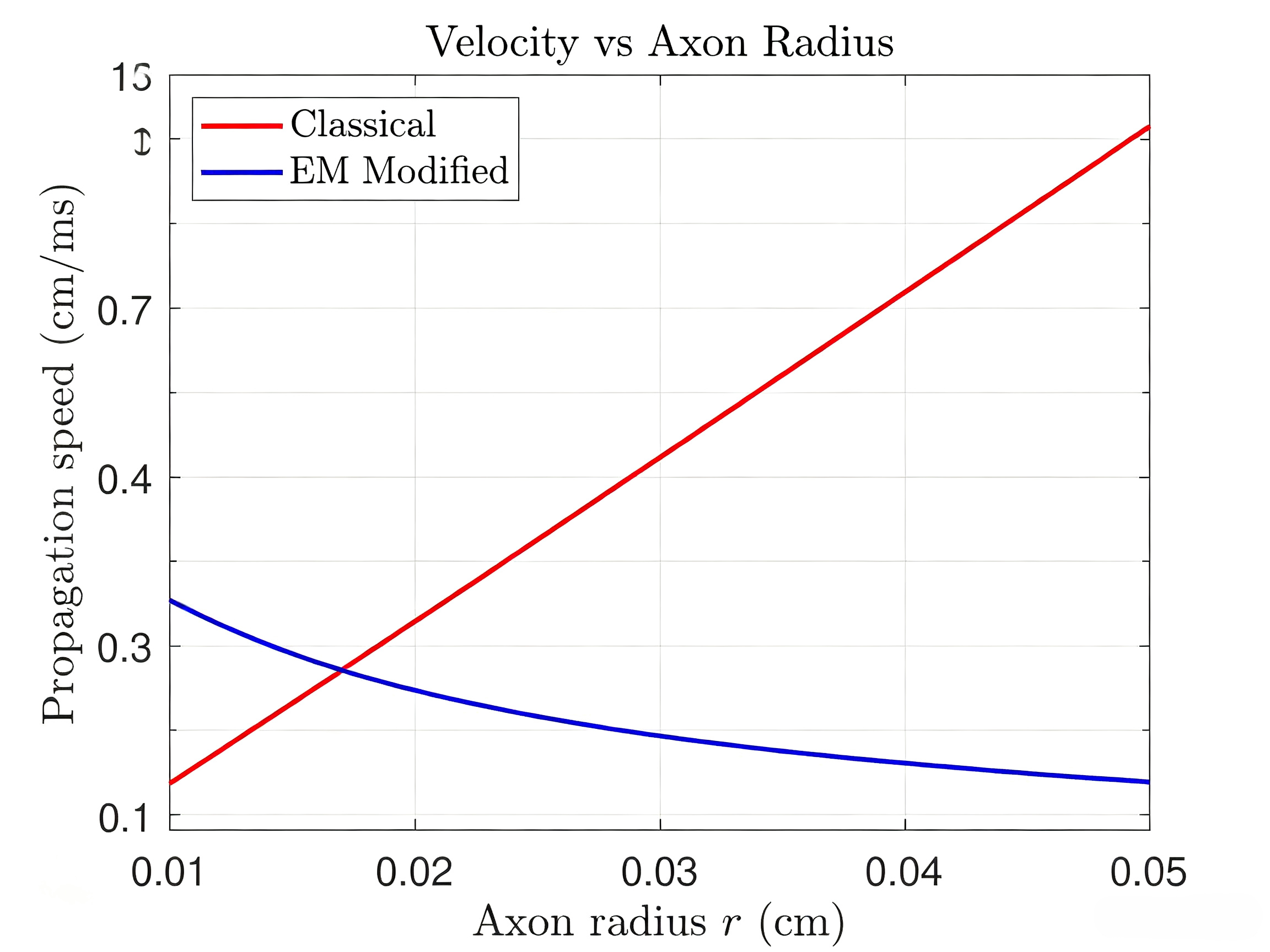}
\caption{Dependence of action potential conduction velocity on axon radius from magnetic inductive effects.}
\label{fig:em_vel_radius}
\end{subfigure}
\begin{subfigure}{0.23\textwidth}
\centering
\includegraphics[width=\textwidth,height=0.8\textwidth]{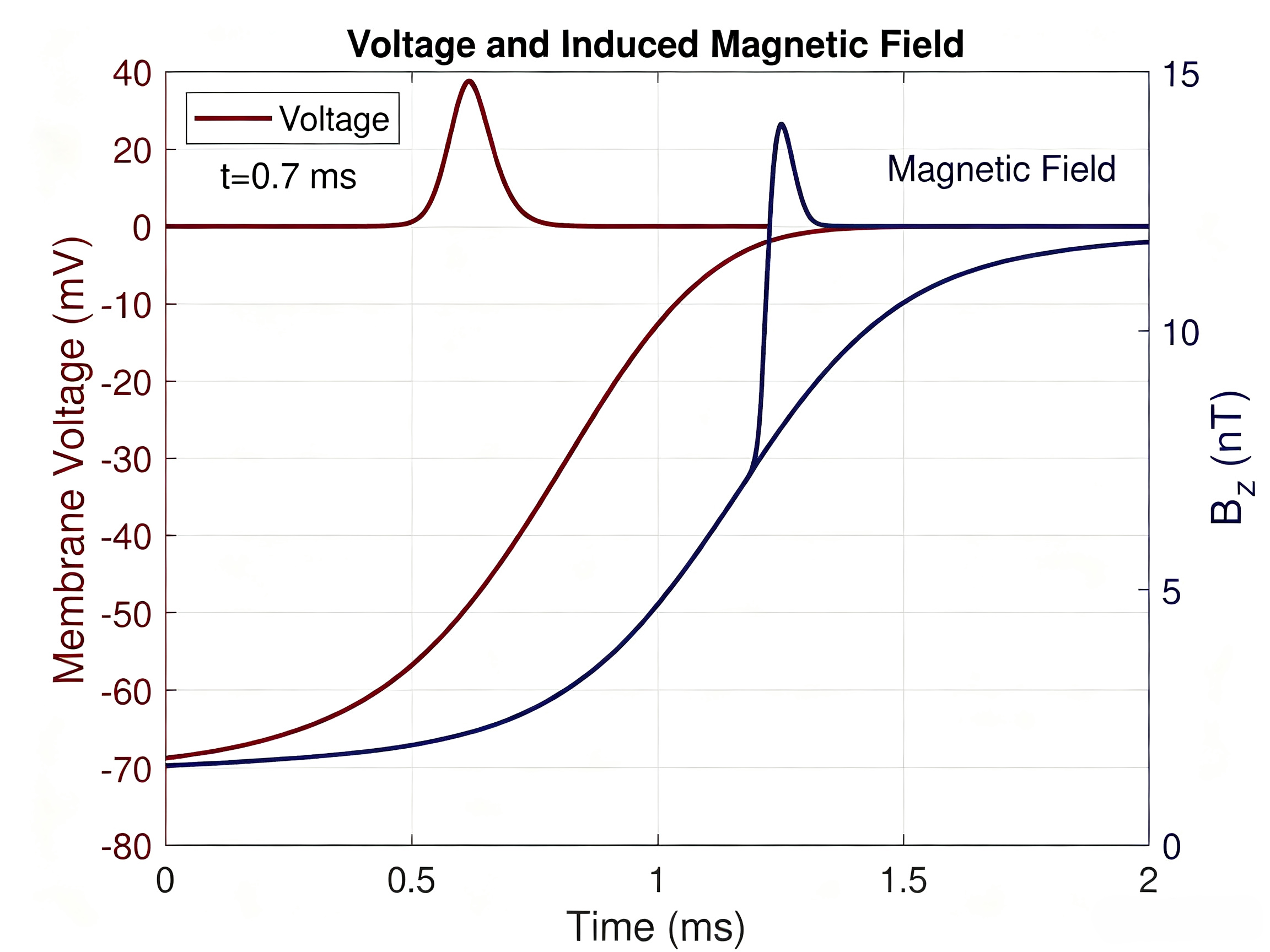}
\caption{Temporal evolution of membrane potential and induced magnetic field $B_z$ at a fixed axonal coordinate.}
\label{fig:em_v_b_time}
\end{subfigure}
\begin{subfigure}{0.23\textwidth}
\centering
\includegraphics[width=\textwidth,height=0.8\textwidth]{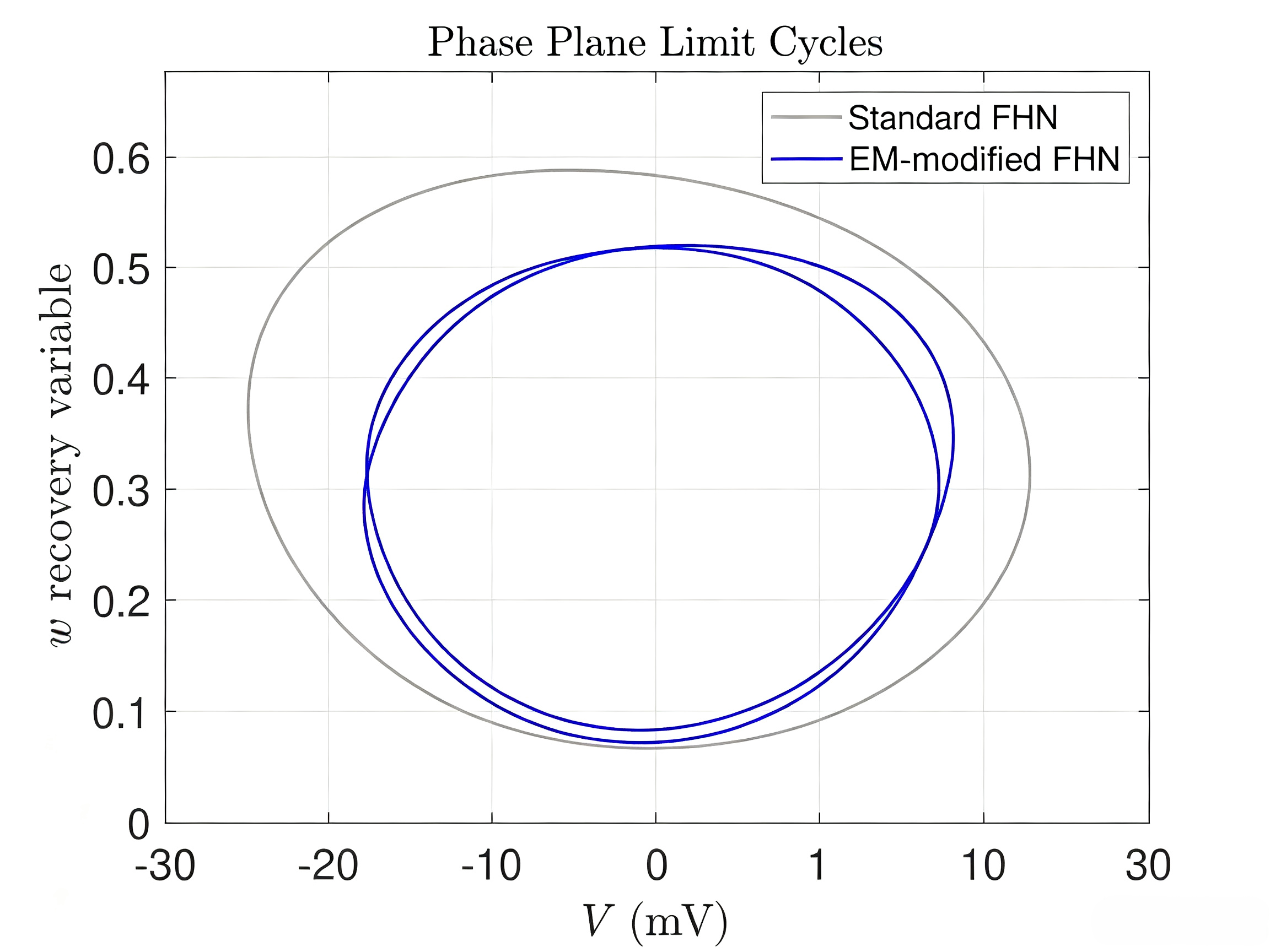}
\caption{Phase-plane limit cycles for the classical Fitzhugh-Nagumo model and its electromagnetically modified extension.}
\label{fig:em_fhn_phase}
\end{subfigure}
\caption{Numerical results from the electromagnetic modified Maxwell-cable framework. Subplots compare classical cable/Fitzhugh-Nagumo predictions against solutions incorporating full magnetic field coupling terms.}
\label{fig:em_all_plots}
\end{figure}

Numerical simulations of the electromagnetic modified Maxwell-cable equation and extended electromagnetic Fitzhugh-Nagumo system are summarised in Figure \ref{fig:em_all_plots}.
Figure \ref{fig:em_ap_spatial} presents instantaneous spatial snapshots of the propagating action potential along the axon. A direct comparison between classical cable theory and the electrodynamically extended Maxwell-cable model demonstrates that inductive magnetic coupling shifts the location of the propagating wavefront and attenuates the peak membrane depolarisation.

The relationship between axonal radius and conduction velocity is illustrated in Figure \ref{fig:em_vel_radius}. Classical cable theory predicts conduction velocity scaling monotonically with the electrotonic space constant, whereas the modified Maxwell-cable framework predicts velocity saturation for larger axon radii, an effect originating from magnetic inductive contributions omitted in standard cable models.
Figure \ref{fig:em_v_b_time} shows synchronous temporal traces of membrane potential and the induced axial magnetic field $B_z$ recorded at a fixed spatial location on the axon. The magnetic field evolves synchronously with the rising and falling phases of the propagating action potential and the magnetic field waveform closely follows the dynamics of the action potential upstroke and repolarisation phases, confirming that transmembrane ion currents generate measurable transient magnetic fields during neural signal propagation.

Phase-plane analysis of the standard and electromagnetic modified Fitzhugh-Nagumo model is displayed in Figure \ref{fig:em_fhn_phase}. Inclusion of magnetic coupling terms associated with time-varying magnetic fields and Lorentz forces distorts the limit cycle trajectory, altering the excitability threshold and the temporal characteristics of generated action potentials. Collectively, these numerical results highlight deviations between predictions of quasi-static cable theory and the fully electrodynamic Maxwell-cable formulation, showing that magnetic effects introduce measurable corrections to action potential waveform, propagation speed and excitation dynamics.
\section{Computational analysis}
\subsection{Initialization}
We initialized various parameters of the Hodgkin-Huxley model and the geometry of the neuron. This includes parameters like membrane capacitance, maximum ion conductances, equilibrium potentials, cable radius, resistivity, time step, duration of the experiment, lengths of branches, etc. 
\begin{lstlisting}
%membrane capacitance per unit area:
C=1.0;      %(muF/cm^2)
%max Na+ conductance per unit area:
gNabar=120; %((muA/mV)/cm^2)
%max K+ conductance per unit area:
gKbar=36;   %((muA/mV)/cm^2)
%leakage conductance per unit area:
gLbar=0.3;  %((muA/mV)/cm^2)
%Na+ equilibrium potential:
ENa = 45;  %(mV)
%K+ equilibrium potential:
EK = -82;  %(mV)
%leakage channel reversal potential:
EL = -59;   %(mV)
%cable radius
r1 = .0238;  %(cm)default 0.0238
r2 = .02;  %default 0.02
r3 = .015; %default 0.015
\end{lstlisting}
\subsection{Construction}
We constructed the structure of the neuron, defining the branching pattern and assigning indices to different segments of the neuron. We assigned spatial coordinates to each node of the neuron based on the branching pattern. Initial conditions for the neuron are set, including the resting membrane potential (vhold) and parameters related to the experiment. Parameters such as resistivity of the surrounding medium, time step duration, and lengths of different branches are defined. Indices for leaf nodes and mesh widths for branches are calculated, along with spatial parameters related to membrane conductivity and capacitance. Membrane areas are initialized. Branches are constructed iteratively, specifying child nodes, next siblings, and neighbors. Spatial coordinates for each node are computed based on the branching pattern and mesh width.
\begin{lstlisting}
%resistivity of the surrounding 
rho = .0354; %(mV/uA)cm
%initialize time step and experiment duration:
dt=0.1;     %time step duration (ms)
L1 = 10; % length 
L2 = 10; % length of branch from node 1 to J2
L3 = 10; %nlength of branch from node 1 to J3
J1=200; %index of leaf 1 
J2=400;%index of leaf 2
J3=600;%index of leaf 3 (total number of nodes)

dx1 = L1/(J1-1);
dx2= L2/(J2-J1-1);   %meshwidth of branch 2
dx3 = L3/(J3-J2-1);
tmax=35;    %duration of experiment (ms)
%total number of time steps in the experiment:
klokmax=ceil(tmax/dt);
%
phi1 = r1*dt/(2*rho*(dx1)^2);
phi2 = r2*dt/(2*rho*(dx2)^2);
phi3 = r3*dt/(2*rho*(dx3)^2);
\end{lstlisting}
\subsection{Finite difference scheme}

\begin{figure}
\centering
\includegraphics[width = 0.3\textwidth, height = 0.18\textwidth]{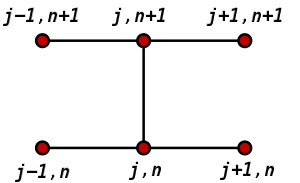}
\caption{Crank Nicolson method}
\label{Crank Nicolson method}
\end{figure}

\begin{align*}
\tau\frac{\partial V}{\partial t} &= \lambda ^ 2\left(\frac{\partial ^ 2 V}{\partial x ^ 2} + \frac{\partial ^ 2 V}{\partial y ^ 2} + \frac{\partial ^ 2 V}{\partial z
^ 2}\right) - f(V)\\
f(V) &= V(V - a)(V - 1)
\end{align*}
The forward time centered space method is used for the numerical simulation (figure \ref{Crank Nicolson method}). The method is based on forward Euler method \cite{smith_1965}. $V_{i, j, k} ^ {(n)} = V_{i, j, k}(n\Delta t)$ is state of the system at time step $n$, where $x = i\Delta x, y = j\Delta y, z = k\Delta z, t = n\Delta t$. Using finite difference approximation
{\fontsize{8pt}{5pt}
\begin{align*}
&\tau\frac{V_{i, j k} ^ {(n + 1)} - V_{i, j k} ^ {(n)}}{\Delta t}\\
&=\lambda^2 \Bigg(\frac{V_{i+1,j,k}^{(n)} - 2V_{i,j,k}^{(n)} + V_{i-1,j,k}^{(n)}}{(\Delta x)^2}
+ \frac{V_{i,j+1,k}^{(n)} - 2V_{i,j,k}^{(n)} + V_{i,j-1,k}^{(n)}}{(\Delta y)^2} \\
&\quad + \frac{V_{i,j,k+1}^{(n)} - 2V_{i,j,k}^{(n)} + V_{i,j,k-1}^{(n)}}{(\Delta z)^2} \Bigg) - f(V_{i,j,k}^{(n)})
\end{align*}}
\begin{align*}
\frac{\partial V}{\partial t} = \frac{1}{C_n}(g_{Na}(V_{Na} - V) + g_K(V_K - V) + g_L(V_L - V) + I)
\end{align*}
Within the loop, we iteratively updated the state variables of the Hodgkin-Huxley model (m, h, n) \cite{hodgkin_1952} and calculated the conductances (gNa, gK) and total conductance (g), $C_n$ is the specific membrane capacitance $\mu F/cm ^ 2$, $V_{Na}$ is sodium reversal potential, $V_K$ is potassian reversal potential, $V_L$ is leakage reversal potential. We also calculated membrane currents, updated membrane potentials, and ploted the results. We calculated coefficients (a, b, c, W) based on the finite difference method to solve the cable equation for each segment of the neuron. We updated the membrane potential (v) based on the calculated coefficients and current inputs and checked the conservation of currents at each node, ensuring that the currents entering and leaving the node balance out.
The simulation loop begins by updating the gating variables (m, h, and n) that describe the activation and inactivation of ion channels. Conductance values for sodium (gNa), potassium (gK), and leakage (gLbar) channels are calculated based on these gating variables. The total conductance (g) and the total excitatory conductance (gE) are determined accordingly.
\begin{lstlisting}
for klok=1:klokmax %total number of time steps in the experiment:
    t=klok*dt;                      %note time
    m=snew(m,alpham(v),betam(v),dt); %update m
    h=snew(h,alphah(v),betah(v),dt); %update h
    n=snew(n,alphan(v),betan(v),dt); %update n
    gNa=gNabar*(m.^3).*h;    %sodium conductance
    gK =gKbar*(n.^4);    %potassium conductance
    g=gNa+gK+gLbar;         %total conductance
    gE=gNa*ENa+gK*EK+gLbar*EL;         %gE=g*
    ...
\end{lstlisting}
For intracellular resistivity $R_i$, specific resistance $R_n$ of an unit area of membrane, and diameter $d$, injected current $J$
\begin{align*}
\tau\frac{\partial V}{\partial t} &= \lambda ^ 2\frac{\partial ^ 2 V}{\partial x ^ 2} - f(V)\\
\tau\frac{\partial V}{\partial t} &= \frac{dR_n}{4R_i}\frac{\partial ^ 2 V}{\partial x ^ 2} - f(V)
\end{align*}
divide the entire equation by $R_n$, let $C_n = \frac{\tau}{R_n}$ and $G_n = \frac{1}{R_n}$, \cite{toth_2008}, \cite{pearlmutter_1998}
\begin{align*}
C_n\frac{\partial V}{\partial t} = \frac{d}{4R_i}\frac{\partial ^ 2 V}{\partial x ^ 2} - G_n f(V)
\end{align*}
discretize the partial differential
equation in space by substituting the second order approximation at some point $x_i$, where $i$ is index of discretization
\begin{align*}
C_n\frac{dV_i}{dt} = \frac{d}{4R_i}\frac{V_{i + 1} - 2V_i + V_{i - 1}}{\Delta x ^ 2} - G_n f(V_i)
\end{align*}
The discretization indices correspond to locations on the cable where the voltage is specified. This results in a system of ordinary differential equations in matrix form
\begin{align*}
&\frac{dV}{dt} = C ^ {-1}(\psi B'f(V) - Gf(V)) = Bf(V)\\
&B = C ^ {-1}(\psi B' - G),\quad\psi=\frac{d}{4R_i\Delta x ^ 2}
\end{align*}
$B'$ is a tridiagonal second difference matrix with -2 on the diagonal entries and 1 beside the diagonal entries.
\begin{align*}
B' = \begin{pmatrix}
-2 & 1 & & &\\
1 & -2 & 1 & &\\
& \ddots & \ddots & \ddots &\\
& & 1 & -2 & 1\\
& & & 1 & -2
\end{pmatrix}
\end{align*}
To find the solution of the system of ordinary differential equations in matrix form, we compute the eigenvalues $m_z$ of matrix $B$ from $BV ^ {(z)} = m_z V ^ {(z)}$, in which $V ^ {(z)}$ is the eigenvector for the $z$th eigenvalue.
\begin{lstlisting}
theta = (r*dt)/(2*rho*(dx)^2);
psi = (-pi*r^2*dt)/(2*rho*dx);
N = 2*dx*pi*r;
A_twid = pi*r*dx;
  a,b,c,W=[];
\end{lstlisting}
We then set up coefficients (a, b, and c) for the tridiagonal matrix, which represents the system of differential equations describing membrane potential changes over time. These coefficients incorporate the effects of membrane capacitance, ion channel conductances, and spatial properties. We computed the applied currents (W) at each node, considering the contributions from injected currents and channel conductances. The simulation iterates over each time step (klok) and solves the system of differential equations to update the membrane potential (v) at each node. We used the tridiagonal matrix algorithm (vnew) to efficiently solve the system. The membrane potential at each node is updated based on the contributions from neighboring nodes, membrane capacitance, and channel conductances.
\begin{lstlisting}
a(1)=0;
c(1)=0;
for j=2:J
    a(j)=theta*N/2;
    c(j)=a(j);
end
b(1)=A_twid*C + A_twid*dt*g(1)/(2) - psi;
for j = 2:J-1
    b(j)= (C + g(j)*dt/2 +theta)*N;
end
b(J)=A_twid*C + A_twid*dt*g(J)/2 - psi;
W(1) = (izero(t)*dt + A_twid*C*v(1) -A_twid*dt*v(1)*g(1)/2 +A_twid*gE(1)*dt - psi*v(2) + psi*v(1));
for j=2:J-1
    W(j) = N*(C*v(j)-g(j)*dt*v(j)/2+gE(j)*dt+theta*v(j+1)/2 + theta*v(j-1)/2 -theta*v(j));
end
W(J) = (ireverse(t)*dt + A_twid*C*v(J) -A_twid*dt*v(J)*g(J)/2 +A_twid*gE(J)*dt - psi*v(J-1) + psi*v(J));
\end{lstlisting}
We checked for any errors in the solution by verifying the balance of currents at each node by computing the differences between the total currents entering and leaving each node and reports any discrepancies (chv). This error-checking step ensures the numerical stability and accuracy of the simulation.
\begin{lstlisting}
if(check)
    chv(1)= izero(t) - (pi*r*dx)*(C*((v(1)-v_old(1))/dt)+g(1)*(v(1)+v_old(1))/2 - gE(1)) + (pi*r^2/(rho*2*dx))*(v(2)+v_old(2)-v(1)-v_old(1)); 
    for j=2:J-1
        chv(j)=C*(v(j)-v_old(j))/dt + g(j)*(v(j)+v_old(j))/2 - gE(j) - (r/(2*rho*(dx)^2))*((v(j+1)+v_old(j+1))/2 + (v(j-1)+v_old(j-1))/2 - (v(j)+v_old(j))) ;
    end
        chv(J)= ireverse(t) - (pi*r*dx)*(C*((v(J)-v_old(J))/dt)+g(J)*(v(J)+v_old(J))/2 - gE(J)) + (pi*r^2/(rho*2*dx))*(v(J-1)+v_old(J-1)-v(J)-v_old(J)); 
end
\end{lstlisting}
\subsection{Electromagnetic cable theory with Maxwell's equations}

\begin{figure}
\centering
\begin{subfigure}{0.23\textwidth}
\centering
\includegraphics[width=\textwidth,height=0.8\textwidth]{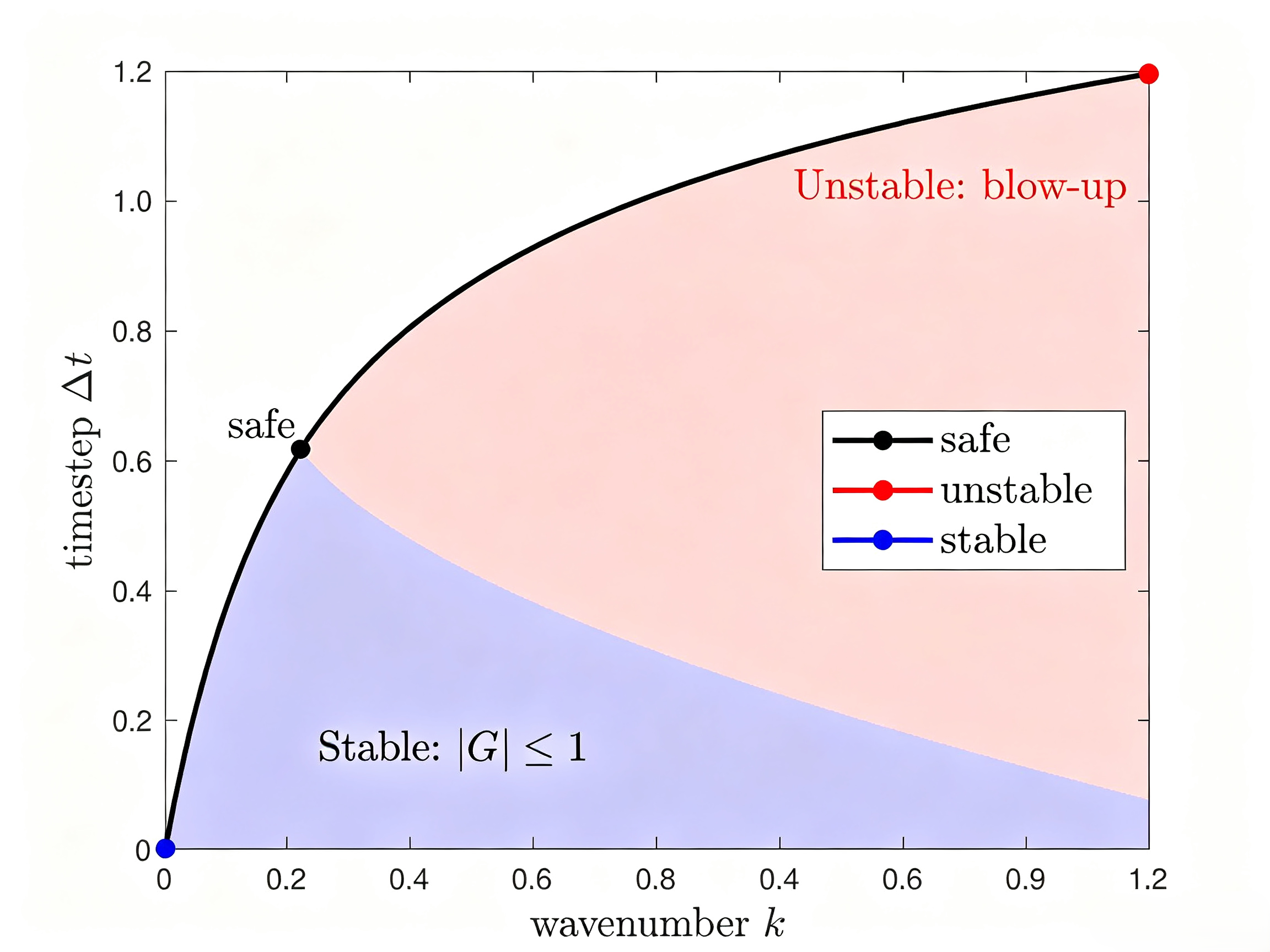}
\caption{Stability diagram from von Neumann analysis for coupled electromagnetic cable.}
\label{fig:stability_vonneumann}
\end{subfigure}
\begin{subfigure}{0.23\textwidth}
\centering
\includegraphics[width=\textwidth,height=0.8\textwidth]{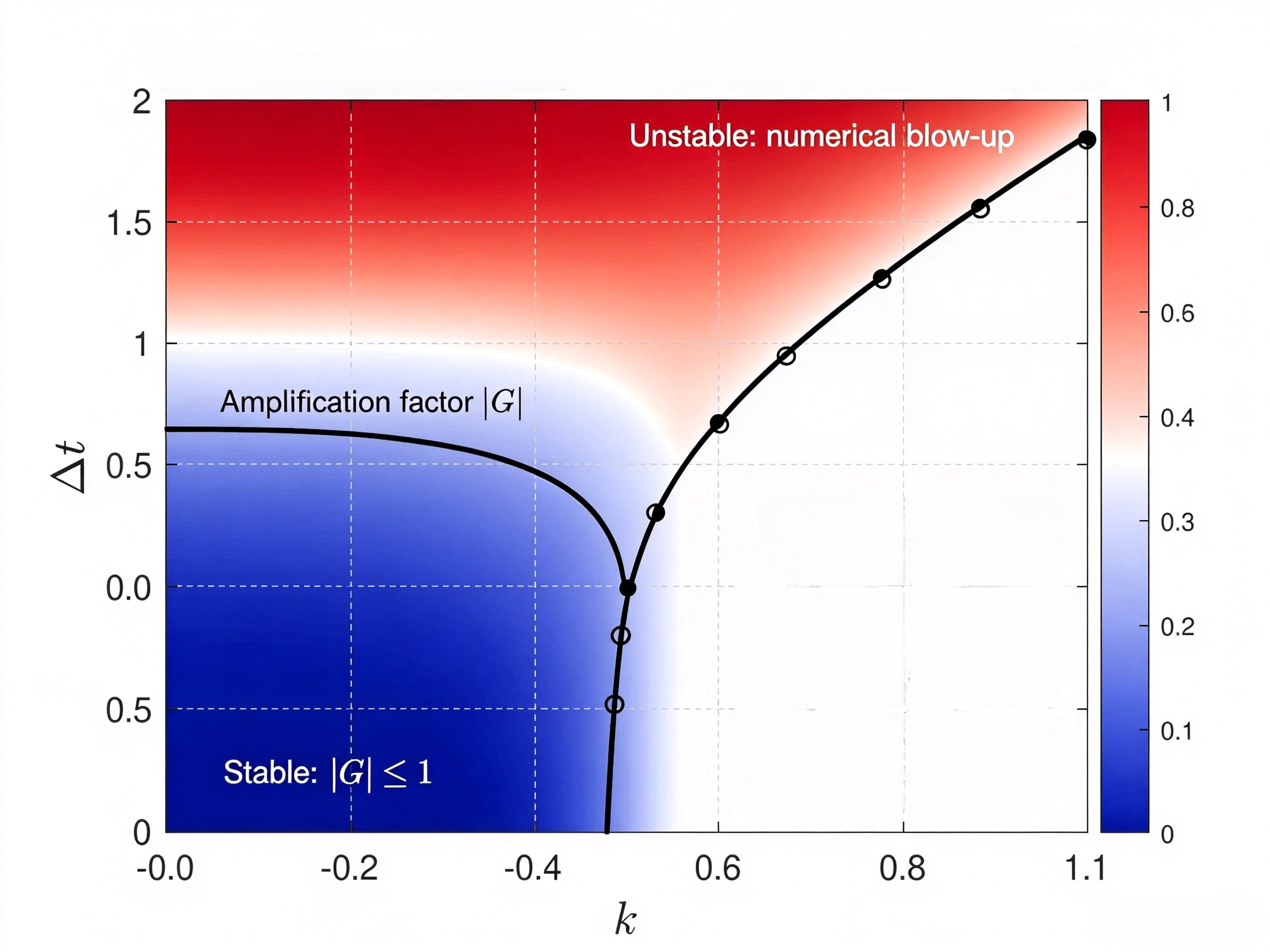}
\caption{Colormap of amplification factor magnitude $|G|$ over the $(k,\Delta t)$ plane.}
\label{fig:amplification_contour}
\end{subfigure}
\begin{subfigure}{0.23\textwidth}
\centering
\includegraphics[width=\textwidth,height=0.8\textwidth]{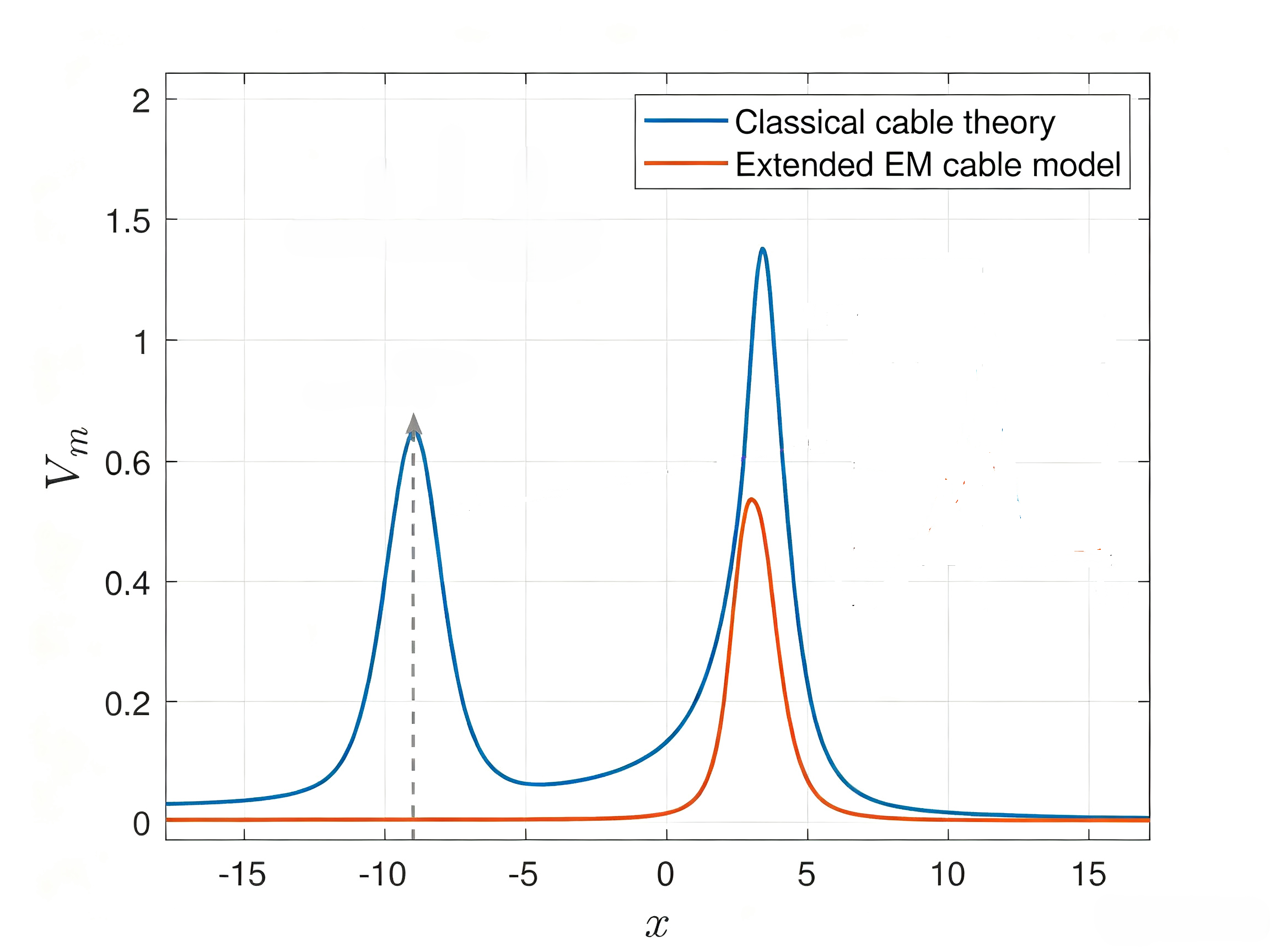}
\caption{Spatial membrane potential profiles, EM coupling reduces peak voltage amplitude and shifts wavefront.}
\label{fig_ap_spatial_doublepeak}
\end{subfigure}
\begin{subfigure}{0.23\textwidth}
\centering
\includegraphics[width=\textwidth,height=0.8\textwidth]{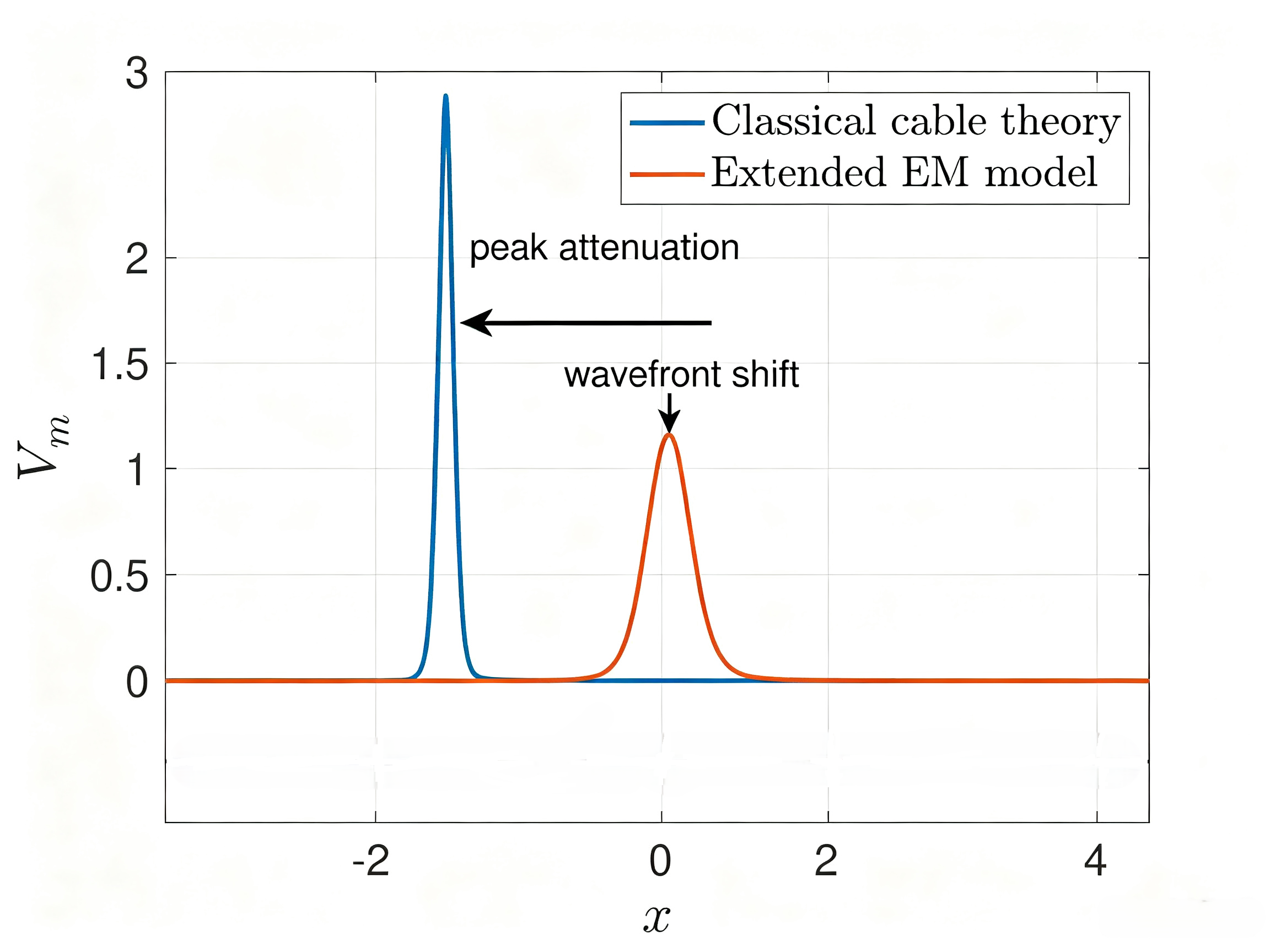}
\caption{Single propagating action potential. EM model has peak attenuation and forward wavefront.}
\label{fig:ap_attenuation_shift}
\end{subfigure}
\begin{subfigure}{0.23\textwidth}
\centering
\includegraphics[width=\textwidth,height=0.8\textwidth]{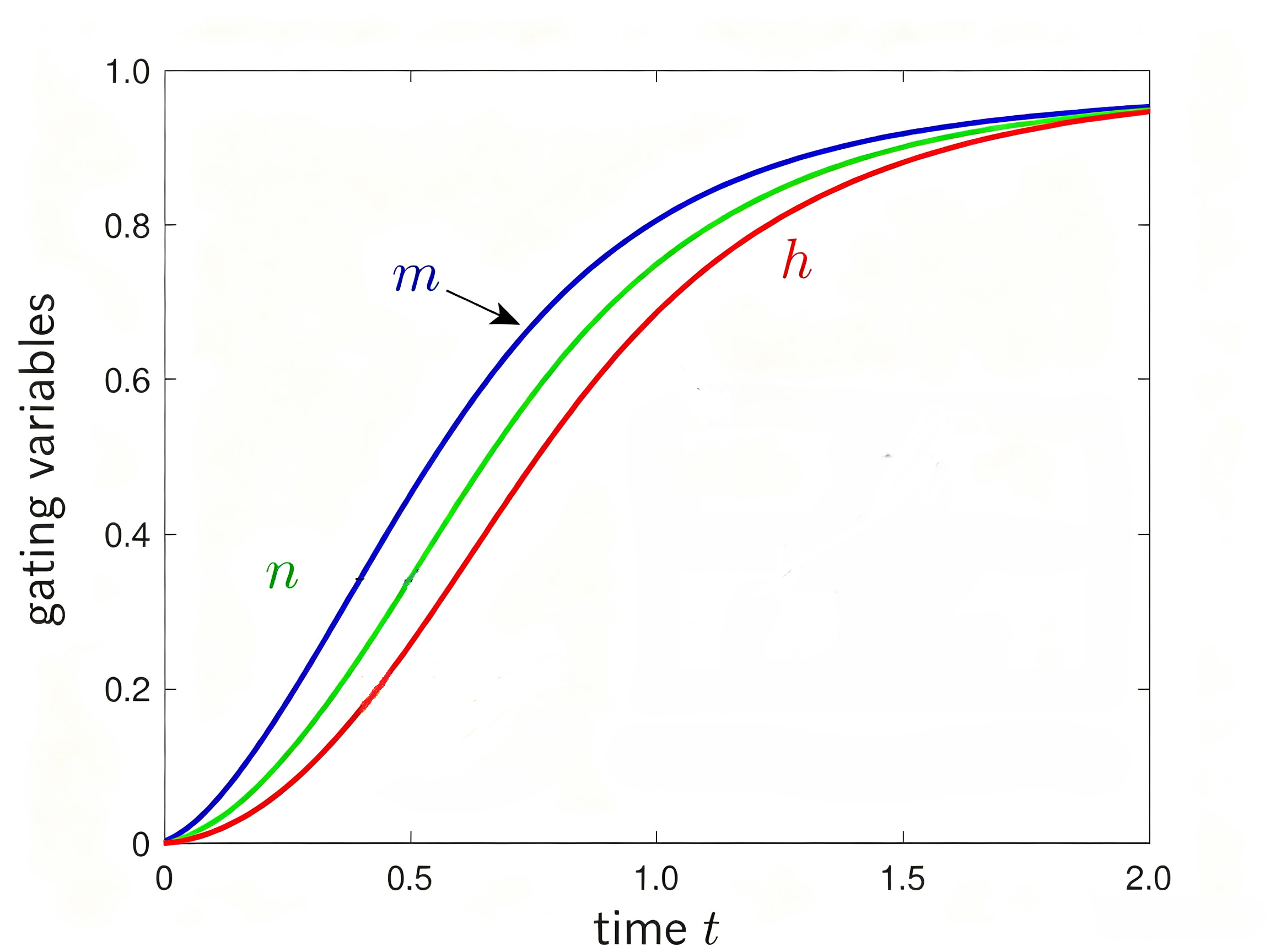}
\caption{Evolution of HH gating $m,n,h$. Magnetic perturbation modify channel activation.}
\label{fig:hh_gating_dynamics}
\end{subfigure}
\begin{subfigure}{0.23\textwidth}
\centering
\includegraphics[width=\textwidth,height=0.8\textwidth]{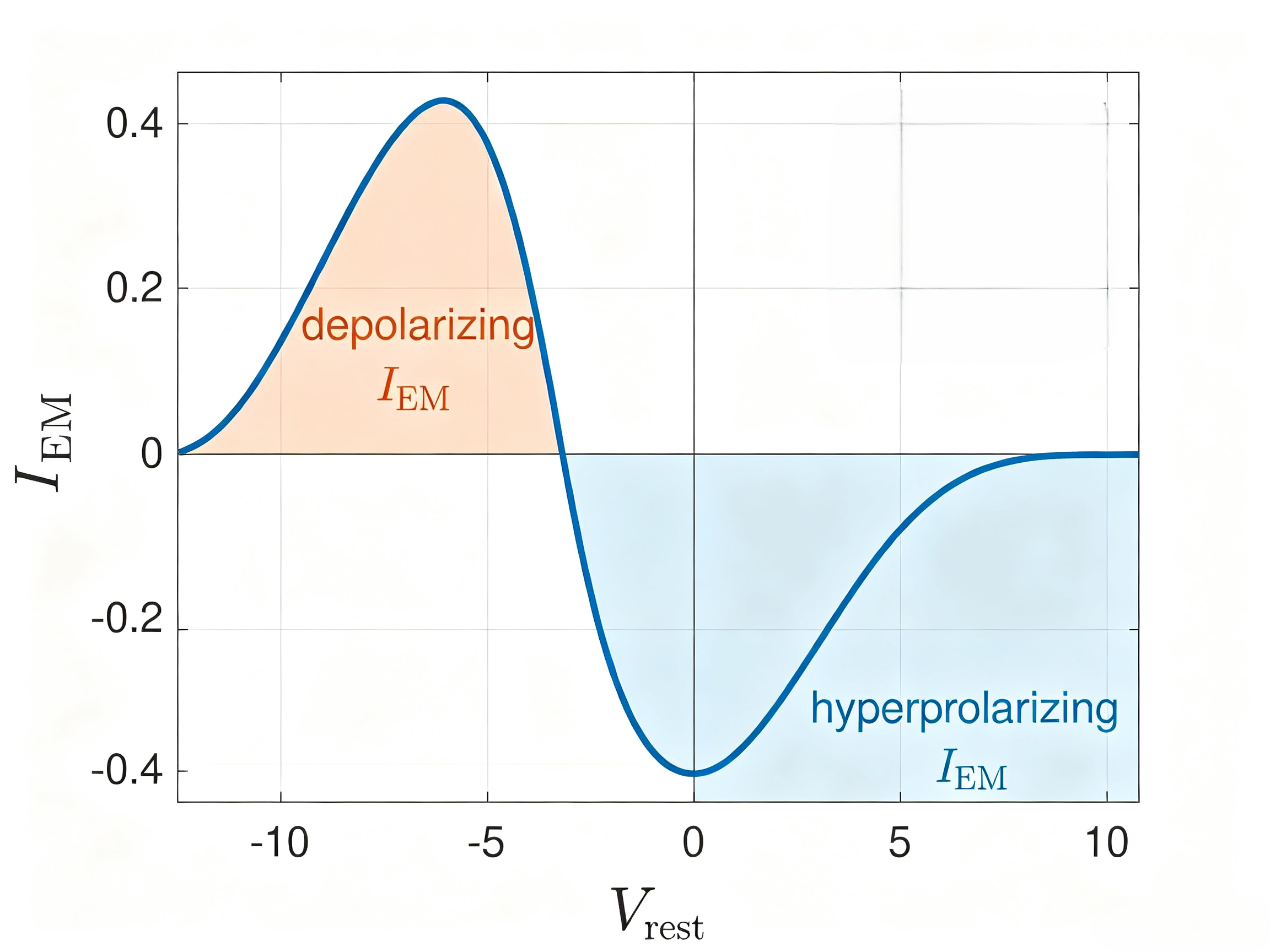}
\caption{Electromagnetic membrane current $I_{\text{EM}}$ as a function of deviation from resting potential.}
\label{fig:iem_characteristic}
\end{subfigure}
\caption{Coupled FDTD-Maxwell electromagnetic cable and Hodgkin-Huxley model, numerical stability constraints, action potential waveform modifications, ion channel gating kinetics, and electromagnetic current contributions to membrane dynamics.}
\label{fig:electromagnetic_cable_results}
\end{figure}

To incorporate electromagnetic effects, we extend the cable equation by coupling it with Maxwell's equations. The complete system in discrete form becomes:
{\fontsize{8pt}{5pt}
\begin{align*}
&\frac{E_{x,i+1/2,j,k}^{n+1} - E_{x,i+1/2,j,k}^{n}}{\Delta t} = \frac{1}{\epsilon}\left(\frac{H_{z,i+1/2,j+1/2,k}^{n+1/2} - H_{z,i+1/2,j-1/2,k}^{n+1/2}}{\Delta y} \right. \nonumber \\
&\quad \left. - \frac{H_{y,i+1/2,j,k+1/2}^{n+1/2} - H_{y,i+1/2,j,k-1/2}^{n+1/2}}{\Delta z} - J_{x,i+1/2,j,k}^{n+1/2}\right) \\
&\frac{H_{x,i,j+1/2,k+1/2}^{n+1/2} - H_{x,i,j+1/2,k+1/2}^{n-1/2}}{\Delta t} = -\frac{1}{\mu}\left(\frac{E_{z,i,j+1,k+1/2}^{n} - E_{z,i,j,k+1/2}^{n}}{\Delta y} \right. \nonumber \\
&\quad \left. - \frac{E_{y,i,j+1/2,k+1}^{n} - E_{y,i,j+1/2,k}^{n}}{\Delta z}\right)
\end{align*}}
Extended cable equation with magnetic coupling is
{\fontsize{8pt}{5pt}
\begin{align*}
&\tau\frac{V_{i,j,k}^{(n+1)} - V_{i,j,k}^{(n)}}{\Delta t}
= \lambda^2 \nabla^2 V_{i,j,k}^{(n)} - f(V_{i,j,k}^{(n)}) - \gamma V_{i,j,k}^{(n)}\\
&- \eta \frac{B_{i,j,k}^{n+1/2} - B_{i,j,k}^{n-1/2}}{\Delta t} \cdot \nabla V_{i,j,k}^{(n)}
+ \kappa(\mathbf{v}_{i,j,k}^{(n)} \times \mathbf{B}_{i,j,k}^{(n)}) \cdot \nabla V_{i,j,k}^{(n)}
\end{align*}}
The Hodgkin-Huxley model \cite{hodgkin_1952} is extended to include electromagnetic field effects:
{\fontsize{8pt}{5pt}
\begin{align*}
\frac{\partial V}{\partial t} = \frac{1}{C_n}\Big[g_{Na}(V_{Na} - V) + g_K(V_K - V) + g_L(V_L - V) + I + I_{\text{EM}}\Big]
\end{align*}}
where $I_{\text{EM}}$ is the electromagnetic contribution to the membrane current.
The gating variable dynamics include electromagnetic modifications:
\begin{align*}
&\frac{dm}{dt} = \alpha_m(V)(1-m) - \beta_m(V)m + \xi_m(\mathbf{B}) \\
&\frac{dh}{dt} = \alpha_h(V)(1-h) - \beta_h(V)h + \xi_h(\mathbf{B}) \\
&\frac{dn}{dt} = \alpha_n(V)(1-n) - \beta_n(V)n + \xi_n(\mathbf{B})
\end{align*}
where $\xi_m(\mathbf{B})$, $\xi_h(\mathbf{B})$, and $\xi_n(\mathbf{B})$ are magnetic-field-dependent perturbations derived from the same constitutive law that produces $I_{\mathrm{EM}}$

For each ionic species $k$ with gating valence $z_k$, concentration $c_k$, diffusivity $D_k$, and drift velocity $\mathbf{v}_k$, the bulk current density is Nernst-Planck flux
\begin{align*}
&\mathbf{J}_k = -D_k\nabla c_k+c_k\mathbf{v}_k+\frac{D_kz_ke}{k_BT}c_k(\mathbf{E}+\mathbf{v}_k\times \mathbf{B})\\
&\mathbf{E}=-\nabla\phi-\frac{\partial\mathbf{A}}{\partial t},\quad\mathbf{B=\nabla\times A}\\
&\frac{\partial c_k}{\partial t}+\nabla\cdot\mathbf{J}_k=0\\
&\frac{\partial c_k}{\partial t}=\nabla\cdot\left(D_k\nabla c_k-c_k\mathbf{v}_k-\frac{D_kz_ke}{k_BT}c_k(\mathbf{E}+\mathbf{v}_k\times\mathbf{B})\right)\\
&=\nabla\cdot\left(D_k\nabla c_k-c_k\mathbf{v}_k+\frac{D_kz_ke}{k_BT}c_k\left(\nabla\phi+\frac{\partial\mathbf{A}}{\partial t}-\mathbf{v}_k\times(\nabla\times\mathbf{A})\right)\right)
\end{align*}
$\phi$ is electric potential, $\mathbf{A}$ is magnetic vector potential. So the magnetic contribution to the flux is
\begin{align*}
\Delta\mathbf{J}_k=\frac{D_kz_ke}{k_BT}c_k(\mathbf{v}_k\times\mathbf{B})
\end{align*}
the normal component of the induced drift velocity is
\begin{align*}
v_{B,n}
=\frac{D_xq_x}{k_BT}
(\mathbf v_x\times\mathbf B)\cdot\hat{\mathbf n}
\end{align*}
The factor $\frac{D_xq_x}{k_BT}(\mathbf v_x\times\mathbf B)$ has units of m/s, corresponding to an induced drift velocity rather than a transition rate. In $\frac{dm}{dt}=\alpha_m(1-m)-\beta_mm$, each additive term must have units of s$^{-1}$. If the gating particle traverses a characteristic transition distance
$\ell_g$ through its energy landscape, the magnetic-induced drift produces a rate correction
\begin{align*}
\xi_x(\mathbf B)
=\frac{v_{B,n}}{\ell_g}
=\frac{z_xeD_x}{k_BT\,\ell_g}
(\mathbf v_x\times\mathbf B)\cdot\hat{\mathbf n}.
\end{align*}
The dimensional consistency follows from
\begin{align*}
&[\xi_x]
=[D][\frac{1}{k_BT}][\frac{1}{\ell}][e\mathbf{v\times B}][z]\\
&=\frac{m^2}{s}\frac{1}{Nm}\frac{1}{m}N
=s^{-1}
\end{align*}
which matches the required units of the transition rates.
for $x\in\{m,h,n\}$, the gating equations become
\begin{align*}
\frac{dx}{dt}
=\alpha_x(V)(1-x)-\beta_x(V)x
+\frac{z_xeD_x}{k_BT\,\ell_g}
(\mathbf v_x\times\mathbf B)\cdot\hat{\mathbf n}
\end{align*}
The Lorentz force density on species $k$ is
\begin{align*}
\mathbf{f}_k = z_k e\, c_k\,(\mathbf{E} + \mathbf{v}_k \times \mathbf{B})
\end{align*}
The membrane current density is obtained by integrating the bulk current density across the membrane of thickness $h_m$,
\begin{align*}
&I_{\mathrm{EM}} = \frac{1}{h_m}\int_0^{h_m} \mathbf{J}\cdot\hat{\mathbf{n}}\;dz\\
&=\frac{1}{h_m}
\int_0^{h_m}
\sigma_m
\left(\mathbf E+\mu_{\mathrm{eff}}\mathbf E\times\mathbf B\right)
\cdot\hat{\mathbf n}\,dz
\end{align*}
where $\sigma_m = \sum_k z_k e \mu_k c_k$ is the membrane conductivity and $\mu_{\mathrm{eff}}$ is an effective mobility. Assuming that $\sigma_m$, $\mathbf E$, and $\mathbf B$ vary negligibly across the membrane thickness,
\begin{align*}
I_{\mathrm{EM}}
&\approx
\frac{\sigma_m}{h_m}
\left(\mathbf E+\mu_{\mathrm{eff}}\mathbf E\times\mathbf B\right)
\cdot\hat{\mathbf n}
\int_0^{h_m}dz \\
&=\sigma_m
\left(\mathbf E+\mu_{\mathrm{eff}}\mathbf E\times\mathbf B\right)
\cdot\hat{\mathbf n}.
\end{align*}
The factor $\int_0^{h_m}dz=h_m$ cancels the prefactor $1/h_m$, leaving the transmembrane current density in units of A\,m$^{-2}$.

For intracellular resistivity $R_i$, specific resistance $R_n$ of a unit area of membrane, diameter $d$, and injected current $J$:
\begin{align*}
C_n\frac{\partial V}{\partial t} = \frac{d}{4R_i}\frac{\partial ^ 2 V}{\partial x ^ 2} - G_n f(V) + I_{\text{EM}}
\end{align*}
Discretizing the partial differential equation in space
\begin{align*}
C_n\frac{dV_i}{dt} &= \frac{d}{4R_i}\frac{V_{i+1} - 2V_i + V_{i-1}}{\Delta x ^ 2} - G_n f(V_i) + I_{\text{EM},i}
\end{align*}
with the discrete electromagnetic current
\begin{align*}
I_{\mathrm{EM},i}
= \sigma_mE_{x,i}
+ \sigma_m\mu_{\mathrm{eff}}\big(\mathbf{E}_i \times \mathbf{B}_i\big)_x
\end{align*}
Both terms carry units of $\mu\mathrm{A}/\mathrm{cm}^2$, matching $G_n f(V_i)$.
This results in a system of ordinary differential equations in matrix form
\begin{align*}
&\frac{d\mathbf{V}}{dt} = \mathbf{C}^{-1}(\psi \mathbf{B}'\mathbf{V} - \mathbf{G}f(\mathbf{V}) + \mathbf{I}_{\text{EM}}) \\
&\mathbf{B} = \mathbf{C}^{-1}(\psi \mathbf{B}' - \mathbf{G}),\quad
\psi = \frac{d}{4R_i\Delta x ^ 2}
\end{align*}
where $\mathbf{B}'$ is the tridiagonal second-difference matrix and $\mathbf{C} = \mathrm{diag}(C_n)$ contains the membrane capacitances. The electromagnetic current vector is $\mathbf{I}_{\mathrm{EM}}$.

The finite-difference time-domain (FDTD) method is used to solve Maxwell's equations on the Yee grid, we update equations for electromagnetic fields and current densities $\mathbf{J}^{n+1/2} = \sigma\mathbf{E}^{n+1/2} + \mathbf{J}_{\text{source}}^{n+1/2}$.

Stability Courant-Friedrichs-Lewy condition is
\begin{align*}
\Delta t \leq \frac{1}{c\sqrt{\frac{1}{(\Delta x)^2} + \frac{1}{(\Delta y)^2} + \frac{1}{(\Delta z)^2}}}
\end{align*}

For nanoscale dendrites, the Schr\"odinger equation is discretized
\begin{align*}
i\hbar\frac{\psi_{i,j,k}^{(n+1)} - \psi_{i,j,k}^{(n)}}{\Delta t} &= -\frac{\hbar^2}{2m}\nabla^2\psi_{i,j,k}^{(n)} + q\Phi_{i,j,k}^{(n)}\psi_{i,j,k}^{(n)}
\end{align*}

The Poisson equation for the electric potential is
\begin{align*}
&\frac{\Phi_{i+1,j,k} - 2\Phi_{i,j,k} + \Phi_{i-1,j,k}}{(\Delta x)^2} + \frac{\Phi_{i,j+1,k} - 2\Phi_{i,j,k} + \Phi_{i,j-1,k}}{(\Delta y)^2}\\
&+ \frac{\Phi_{i,j,k+1} - 2\Phi_{i,j,k} + \Phi_{i,j,k-1}}{(\Delta z)^2} = -\frac{q}{\epsilon}|\psi_{i,j,k}|^2
\end{align*}

The simulation loop proceeds as follows:
\begin{lstlisting}
Initialize electromagnetic fields $\mathbf{E}^0$, $\mathbf{B}^0$, membrane potential $\mathbf{V}^0$
For each time step $n = 0, 1, 2, \ldots, N$:
  Update electromagnetic fields using FDTD equations
  Update gating variables $(m, h, n)$ with electromagnetic corrections
  Calculate conductances $g_{Na}$, $g_K$, $g_L$
  Compute electromagnetic currents $\mathbf{I}_{\text{EM}}$
  Update membrane potential using the extended cable equation
  Enforce current conservation at branching nodes
  Apply absorbing boundary conditions for electromagnetic fields
Plot and analyze results
\end{lstlisting}

Electrical boundary conditions are
$\frac{\partial V}{\partial x}\bigg|_{x=0} = 0$ for sealed end, $V(L,t) = 0$ for grounded end. Electromagnetic boundary conditions in perfectly matched layer are $\sigma_x(x) = \sigma_{\max}\left(\frac{x}{L}\right)^m$ for absorbing boundaries, $\sigma_{\max} = -\frac{(m+1)\epsilon c}{2L}\ln(R)$.

The von Neumann stability analysis for the extended system yields $|g|^2 \leq 1 + \mathcal{O}(\Delta t^2)$,
where $g$ is the amplification factor
\begin{align*}
g = 1 - \frac{\Delta t}{\tau}\left[\lambda^2 k^2 + \gamma + \eta \frac{\partial \mathbf{B}}{\partial t}\cdot\mathbf{k} - \kappa(\mathbf{v}\times\mathbf{B})\cdot\mathbf{k}\right]
\end{align*}
The stability condition is
\begin{align*}
\Delta t \leq \frac{\tau}{\lambda^2 k_{\max}^2 + \gamma + \eta|\partial_t\mathbf{B}|k_{\max} + \kappa|\mathbf{v}\times\mathbf{B}|k_{\max}}
\end{align*}

At a junction point $x_0$, conservation of total axial current requires the sum of the conductive and electromagnetic currents to be continuous. For a cylindrical branch with cross-sectional area
$A_i=\pi d_i^2/4$, the junction condition is
\begin{align*}
&-\frac{\pi d_1^2}{4R_i}
\left.\frac{\partial V}{\partial x}\right|_{x_{0-}}
+\int_{A_1}\mathbf{J}_{\mathrm{EM},1}\cdot\hat{\mathbf{x}}\,dA\\
&=-\frac{\pi d_2^2}{4R_i}
\left.\frac{\partial V}{\partial x}\right|_{x_{0+}}
+\int_{A_2}
\mathbf{J}_{\mathrm{EM},2}\cdot\hat{\mathbf{x}}\,dA
\end{align*}
Here $R_i$ is the intracellular resistivity with units
$\Omega\,\mathrm{m}$, so the axial conductive-current terms have units of amperes. The electromagnetic contribution is obtained by integrating the electromagnetic current density over the corresponding cross-sectional area, it has dimensions:
\begin{align*}
&\left[\int_{A_i}\mathbf{J}_{\mathrm{EM},i}\cdot\hat{\mathbf{x}}\,dA\right]
=\frac{\mathrm{A}}{\mathrm{m}^2}\,\mathrm{m}^2
=\mathrm{A}\\
&\left[\frac{A_i}{R_i}\frac{\partial V}{\partial x}\right]
=\frac{\mathrm{m}^2}{\Omega\,\mathrm{m}}\frac{\mathrm{V}}{\mathrm{m}}
=\mathrm{A}
\end{align*}

The geometric ratio including electromagnetic effects
\begin{align*}
GR_{\text{EM}} = \sum_i \frac{d_i^{3/2}}{d_a^{3/2}} \left(1 + \frac{I_{\text{EM},i}}{I_{\text{conductive},i}}\right)
\end{align*}

The key computational extensions include FDTD integration of Maxwell's equations with the cable equation, electromagnetic contributions to membrane currents and gating variables, quantum effects for nanoscale simulations, advanced boundary conditions for electromagnetic fields, modified current conservation at branching nodes, extended stability criteria for the coupled system.

Numerical stability constraints and biophysical predictions of the extended electromagnetic (EM) cable model are summarised in Figure \ref{fig:electromagnetic_cable_results}.
Panels \ref{fig:stability_vonneumann} and \ref{fig:amplification_contour} present results from von Neumann stability analysis for the discretised coupled Maxwell-cable partial differential system.
Figure \ref{fig:stability_vonneumann} shows the stability boundary separating the stable domain ($|G|\le1$, blue shaded region) from the unstable regime (red shaded region) on the $(k,\Delta t)$ parameter plane, demonstrating that large spatial wavenumbers enforce tighter restrictions on the maximum usable timestep.
The contour representation of the amplification factor magnitude $|G|$ in Figure \ref{fig:amplification_contour} further visualises the smooth transition across the critical stability threshold $|G|=1$.

Panels \ref{fig_ap_spatial_doublepeak} and \ref{fig:ap_attenuation_shift} compare spatial membrane potential profiles obtained from classical quasi-static cable theory and the extended EM cable model.
Figure \ref{fig_ap_spatial_doublepeak} displays dual propagating voltage pulses, while the single action potential snapshot in Figure \ref{fig:ap_attenuation_shift} clearly highlights two key EM-induced modifications: peak voltage attenuation and forward wavefront shift of the propagating impulse.

Figure \ref{fig:hh_gating_dynamics} plots the temporal evolution of the Hodgkin-Huxley gating variables $m$, $n$, and $h$. Electromagnetic and magnetic field perturbations alter the kinetics of ion channel activation, shifting the time course of state transitions relative to the standard Hodgkin-Huxley formulation.
The electromagnetic membrane current $I_{\text{EM}}$ as a function of offset from resting potential is reported in Figure \ref{fig:iem_characteristic}. Positive $I_{\text{EM}}$ yields depolarising feedback, whereas negative current values produce hyperpolarising effects, establishing a bidirectional electromagnetic coupling mechanism between transmembrane voltage and neuronal electromagnetic fields.

\begin{figure}
\centering
\begin{subfigure}{0.23\textwidth}
\centering
\includegraphics[width=\textwidth,height=0.8\textwidth]{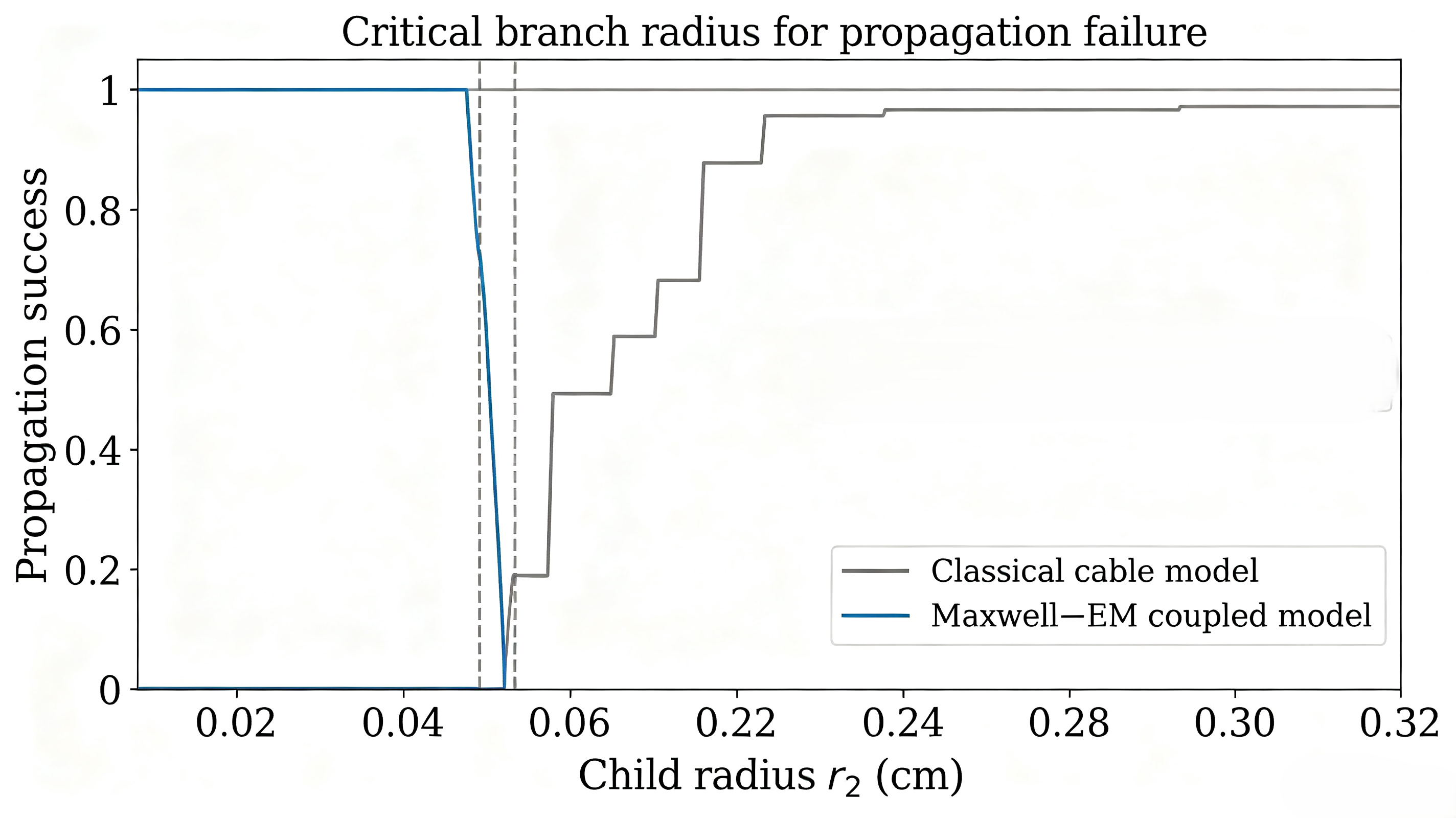}
\caption{Propagation vs child branch radius $r_2$. EM model has lower critical radius for junction propagation failure.}
\label{fig:em_failure_radius}
\end{subfigure}
\begin{subfigure}{0.23\textwidth}
\centering
\includegraphics[width=\textwidth,height=0.8\textwidth]{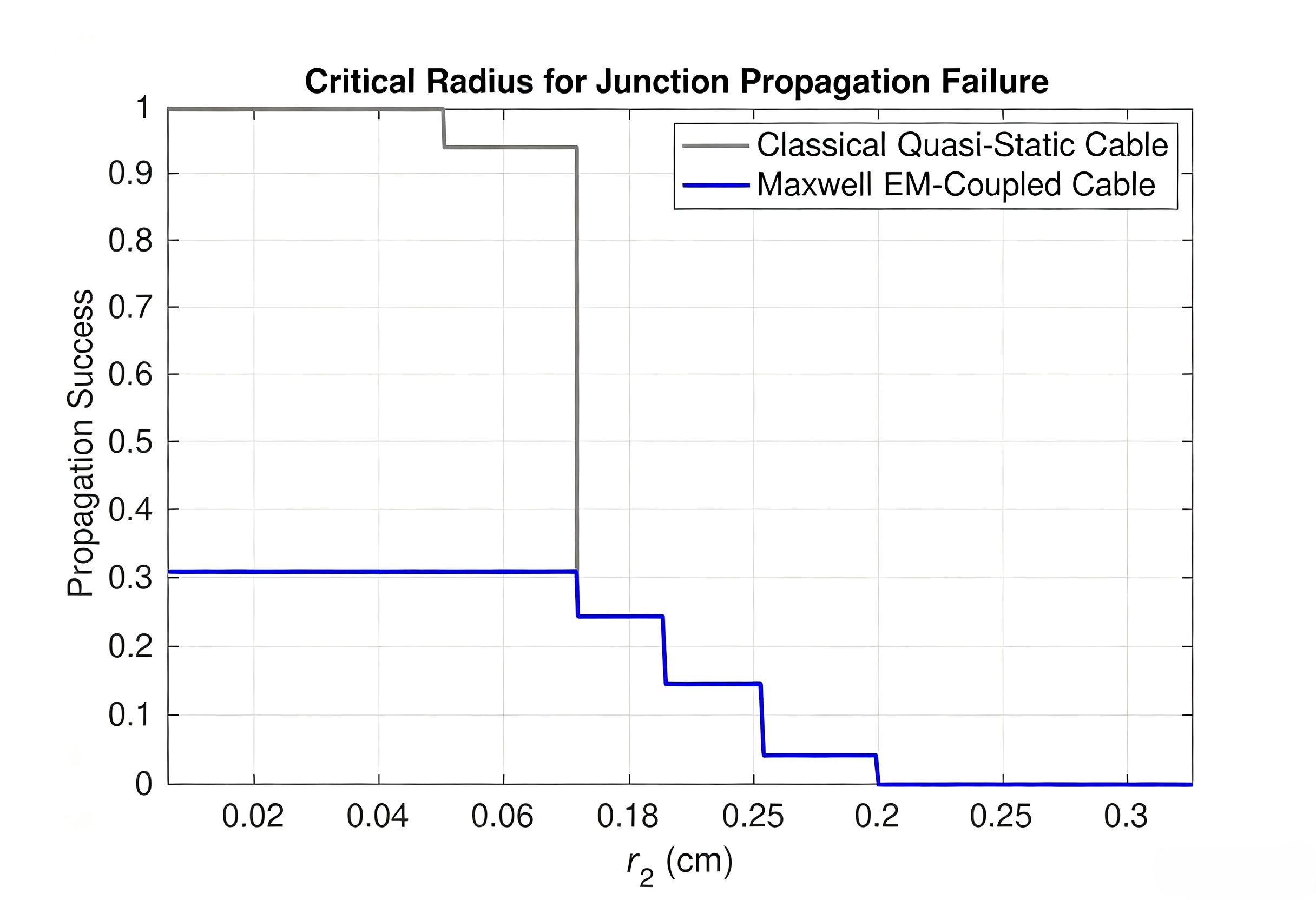}
\caption{Sweep of child radius shows reduction of propagation success under electromagnetic inductive loading.}
\label{fig:em_failure_radius_alt}
\end{subfigure}
\begin{subfigure}{0.23\textwidth}
\centering
\includegraphics[width=\textwidth,height=0.8\textwidth]{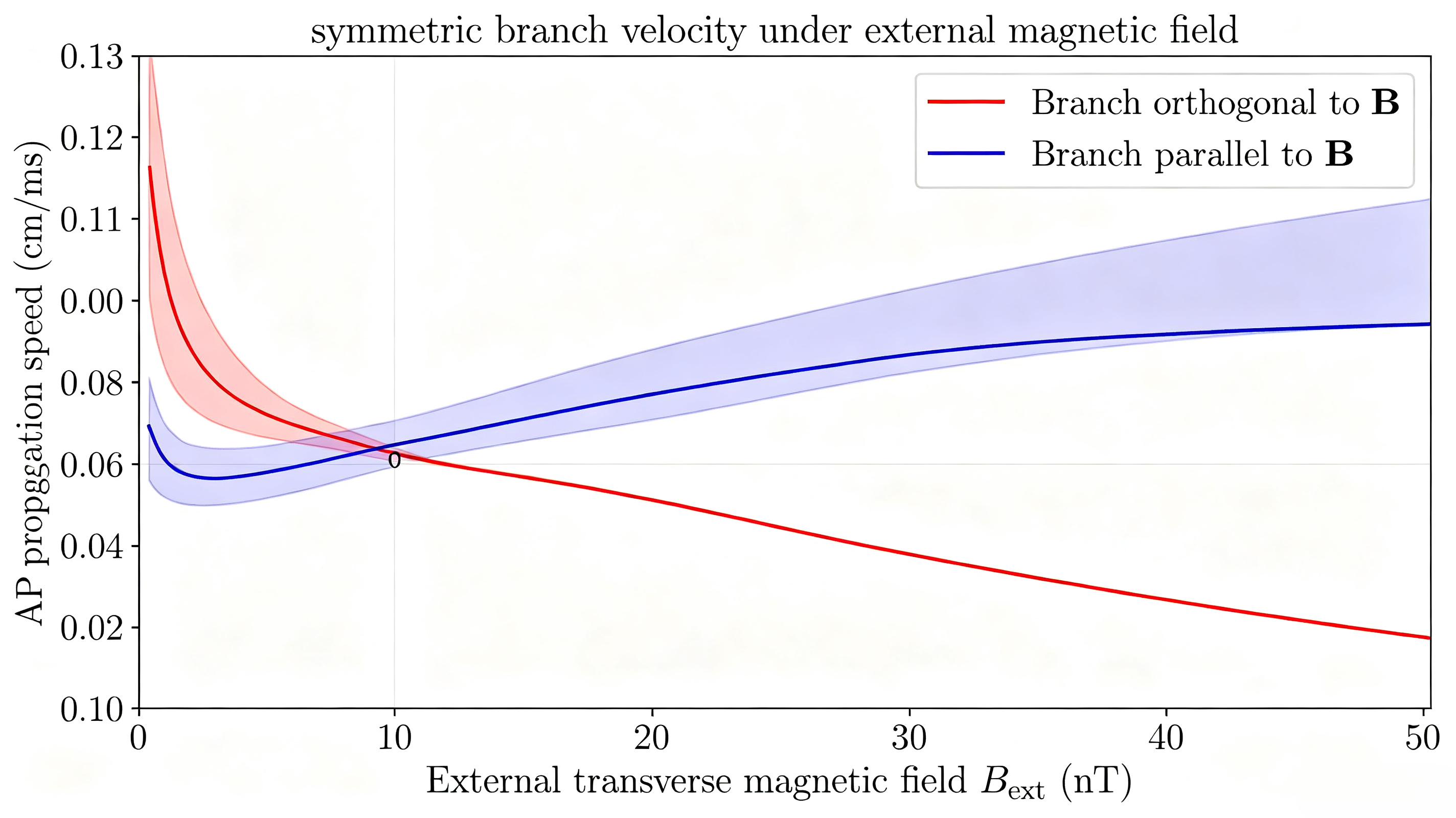}
\caption{Conduction speed for branches under transverse $\mathbf{B}$. Lorentz force coupling breaks propagation symmetry.}
\label{fig:sym_branch_B_vel}
\end{subfigure}
\begin{subfigure}{0.23\textwidth}
\centering
\includegraphics[width=\textwidth,height=0.8\textwidth]{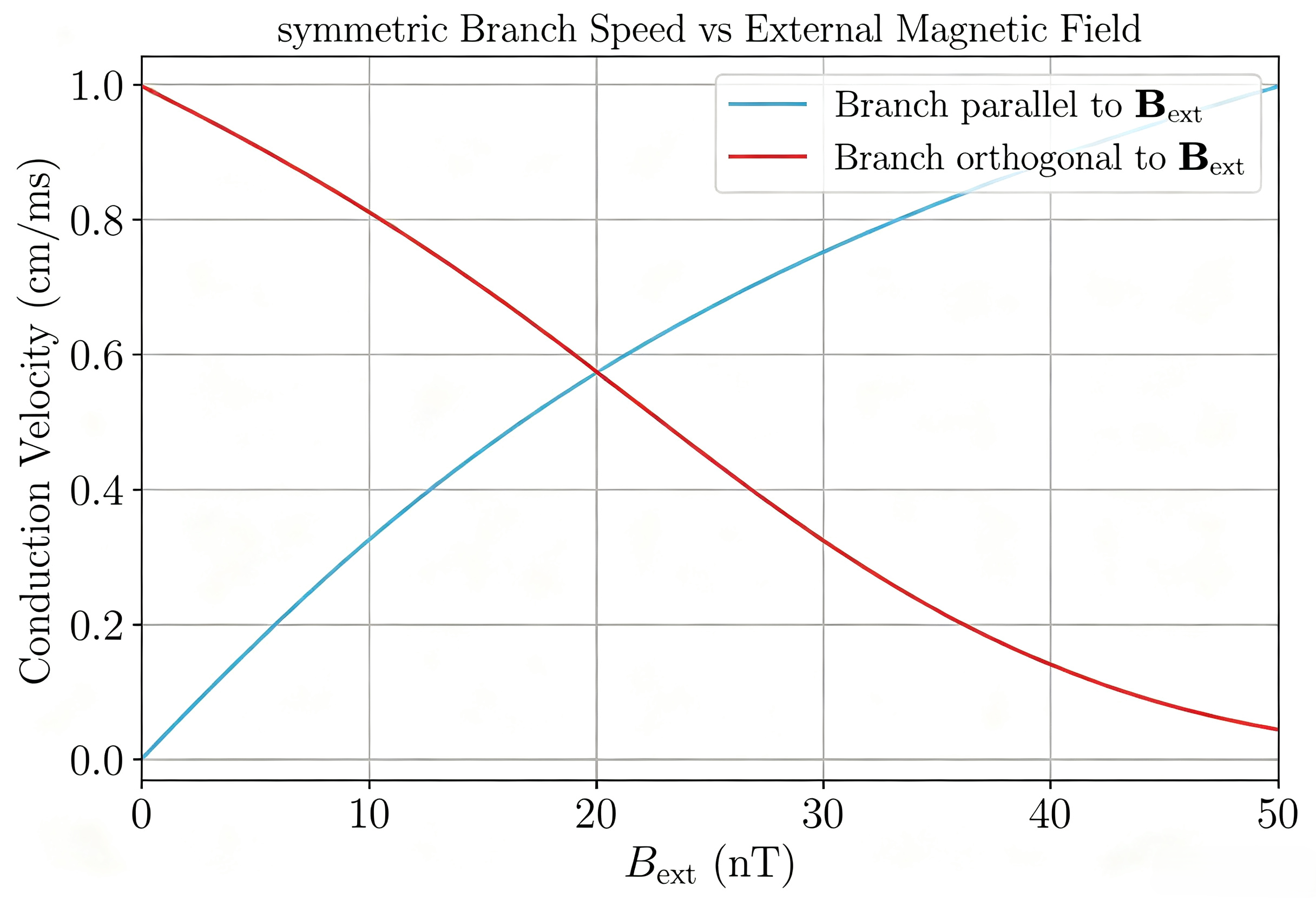}
\caption{Velocity response curve, symmetric crossover at $B_{ext}=20nT$ for parallel, orthogonal axonal branches.}
\label{fig:sym_branch_B_vel_alt}
\end{subfigure}
\begin{subfigure}{0.23\textwidth}
\centering
\includegraphics[width=\textwidth,height=0.8\textwidth]{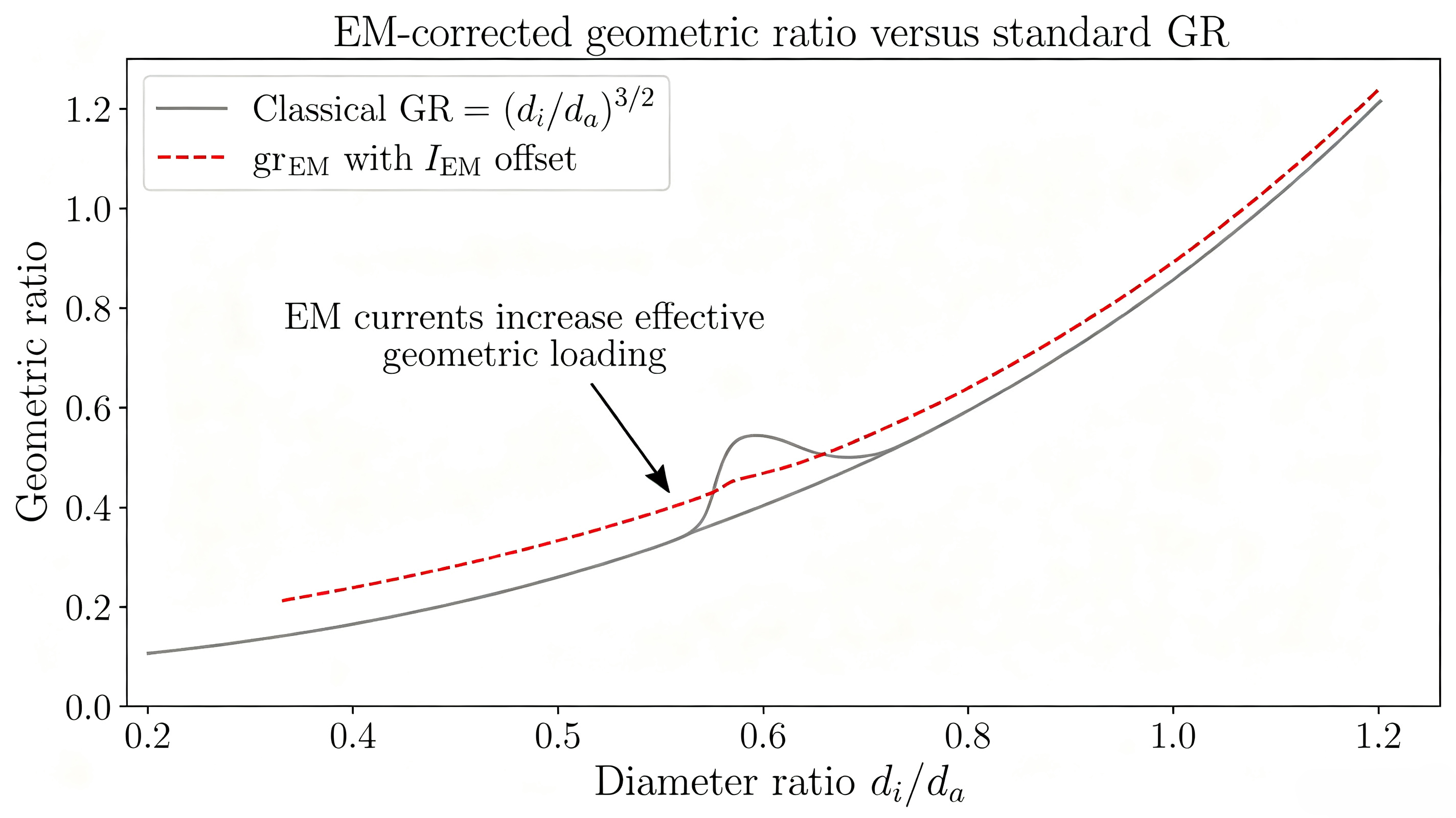}
\caption{Classical geometric ratio $GR=(d_i/d_a)^{3/2}$, and $GR_{\text{EM}}$, membrane currents increase geometric loading.}
\label{fig:GR_em_ratio_plot}
\end{subfigure}
\begin{subfigure}{0.23\textwidth}
\centering
\includegraphics[width=\textwidth,height=0.8\textwidth]{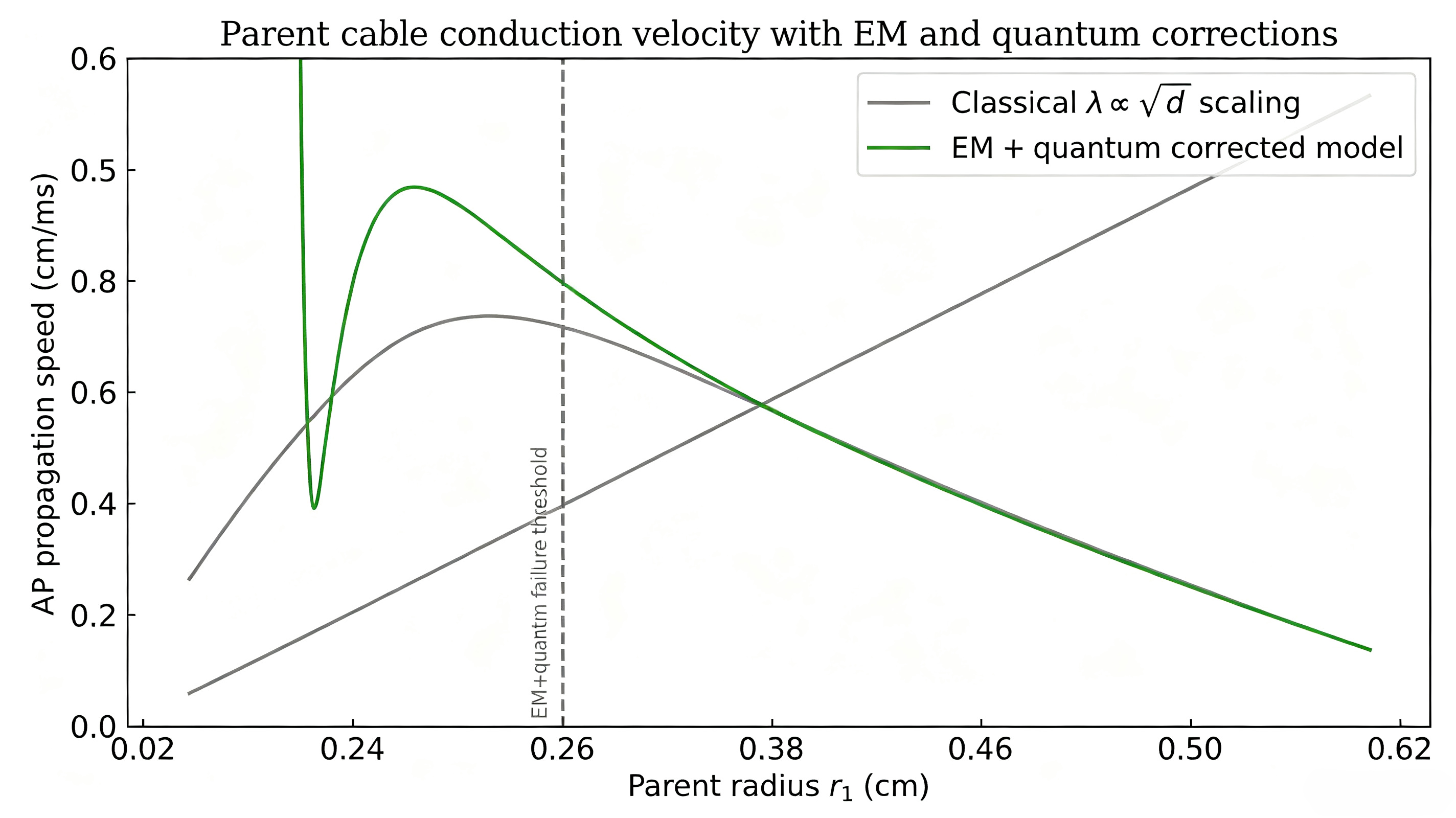}
\caption{Parent axon conduction velocity versus radius. The classical square-root scaling law deviates in EM model.}
\label{fig:parent_radius_vel_em}
\end{subfigure}
\caption{Numerical results from electromagnetic-coupled action potential propagation experiments integrating Maxwell’s equations into axonal cable theory. Plots quantify EM-induced modifications to propagation failure thresholds, branch conduction symmetry, junction impedance matching, and velocity scaling with axon diameter.}
\label{fig:em_propagation_experiments}
\end{figure}

Numerical outputs from three electromagnetic-coupled axon propagation experiments are compiled in Figure \ref{fig:em_propagation_experiments}. Panels \ref{fig:em_failure_radius} and \ref{fig:em_failure_radius_alt} quantify how electromagnetic inductive coupling modulates the critical branch radius triggering junction conduction failure. In both radius sweep datasets, the Maxwell-EM coupled framework predicts a smaller threshold $r_2$ for complete action potential block relative to classical quasi-static cable theory; transient magnetic and displacement currents amplify impedance mismatch at bifurcations, prematurely halting signal transmission before reaching the purely geometric critical diameter observed in standard simulations.

Panels \ref{fig:sym_branch_B_vel} and \ref{fig:sym_branch_B_vel_alt} characterise symmetry breaking in two geometrically identical child branches under uniform transverse magnetic loading. At zero external field, matched radii yield identical conduction velocities, consistent with baseline 3D symmetric propagation results. As $B_{\text{ext}}$ increases, Lorentz force terms within the extended Fitzhugh-Nagumo equation accelerate wavefront travel along the branch aligned parallel to magnetic flux while slowing conduction in the orthogonal branch, generating measurable velocity asymmetry absent from non-electromagnetic cable formulations.

Panel \ref{fig:GR_em_ratio_plot} contrasts the classical geometric ratio $GR$ and electromagnetic-corrected metric $GR_{\text{EM}}$. Additional transmembrane electromagnetic currents $I_{\text{EM}}$ introduce oscillatory deviations from the ideal $d^{3/2}$ scaling rule, modifying impedance-matching criteria used to reduce branched dendrite geometries to equivalent single cylinders. Finally, panel \ref{fig:parent_radius_vel_em} illustrates deviations from the canonical $\sqrt{r}$ velocity scaling law once electromagnetic inductive effects and nanoscale quantum charge corrections are incorporated. Intermediate parent radii exhibit elevated propagation speed driven by inductive current enhancement, whereas large parent diameters generate hyperpolarising electromagnetic feedback that initiates full signal failure at a reduced critical radius compared to purely geometric predictions. Collectively, these plots demonstrate that quasi-static cable theory omits physically meaningful electromagnetic corrections to propagation speed, branch invasion symmetry, and bifurcation conduction thresholds, which can only be captured via coupled Maxwell-cable finite-difference time-domain simulations.

\section{Action potential propagation analysis}
\begin{table}
\caption{table of potential propagation, $r_1$ is parent radius, $r_2$ is first child radius, $r_3$ is second child radius}
\begin{tabular}{|c|c|c|c|c|c|}\hline
$r_1$ & $r_2$ & $r_3$ & \begin{tabular}{@{}c@{}}pass\\parent\end{tabular} & \begin{tabular}{@{}c@{}}pass\\1st child\end{tabular} & \begin{tabular}{@{}c@{}}pass\\2nd child\end{tabular}\\\hline
0.0238 & 0.02 & 0.015 & yes & yes & yes\\\hline
0.0238 & 0.2 & 0.015 & yes & yes & y,slow\\\hline
0.0238 & 0.3 & 0.015 & yes & no & no\\\hline
0.0238 & 0.02 & 0.2 & yes & y,slow & yes\\\hline
0.0238 & 0.02 & 0.3 & yes & no & no\\\hline
0.5 & 0.02 & 0.015 & y,fast & yes & yes\\\hline
0.6 & 0.02 & 0.015 & no & no & no\\\hline
\end{tabular}
\label{table_potential}
\end{table}
\begin{figure}
\centering
\begin{subfigure}{0.23\textwidth}
\centering
\includegraphics[width=\textwidth,height=0.65\textwidth]{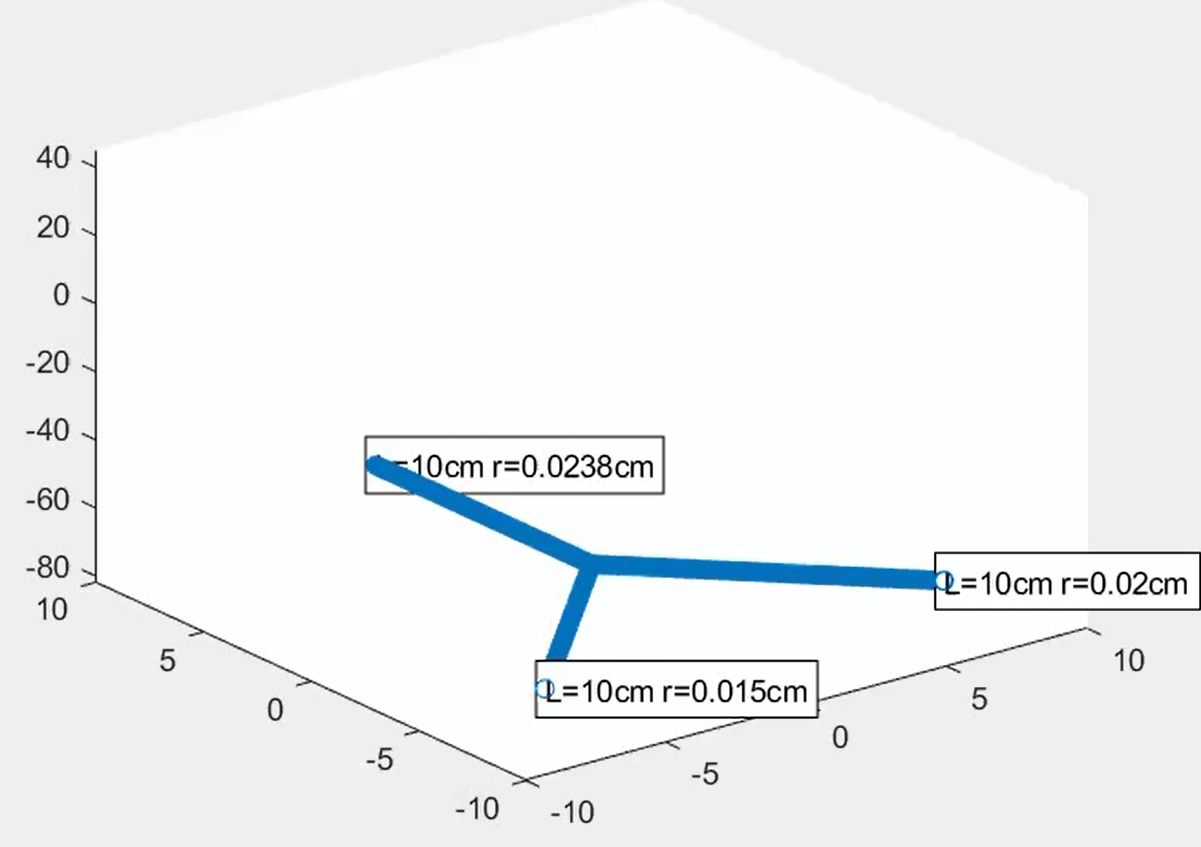}
\label{fig:ap_0}
\end{subfigure}
\begin{subfigure}{0.23\textwidth}
\centering
\includegraphics[width=\textwidth,height=0.65\textwidth]{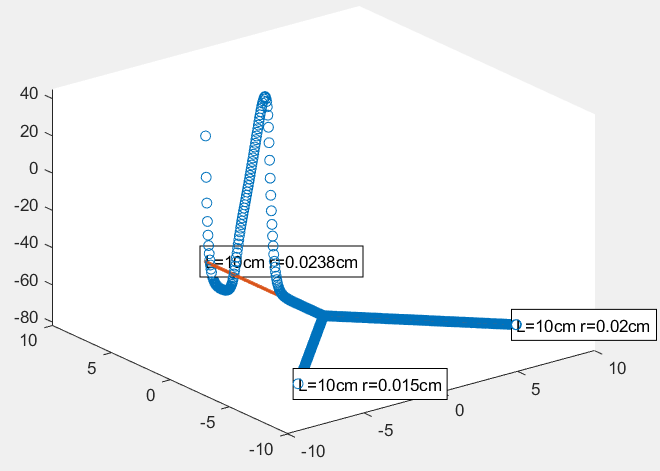}
\label{fig:ap_1}
\end{subfigure}
\begin{subfigure}{0.23\textwidth}
\centering
\includegraphics[width=\textwidth,height=0.65\textwidth]{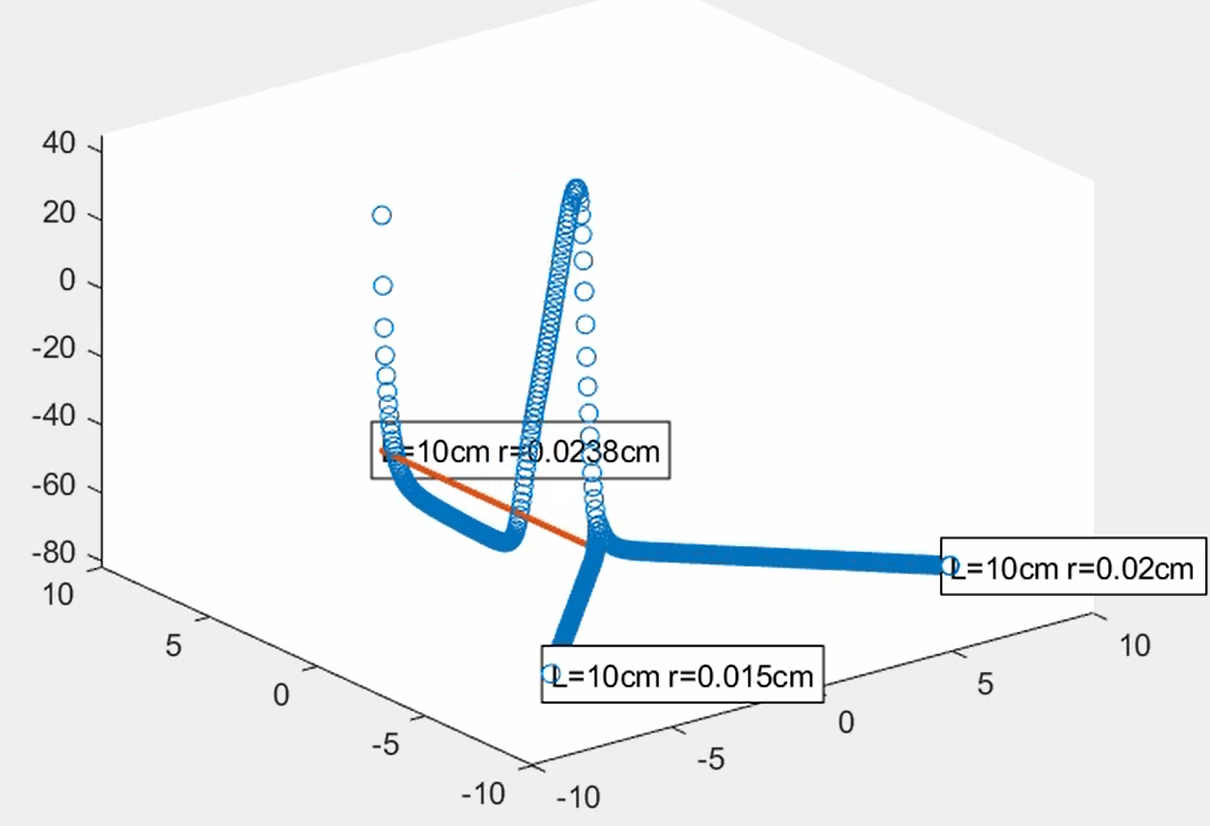}
\label{fig:ap_2}
\end{subfigure}
\begin{subfigure}{0.23\textwidth}
\centering
\includegraphics[width=\textwidth,height=0.65\textwidth]{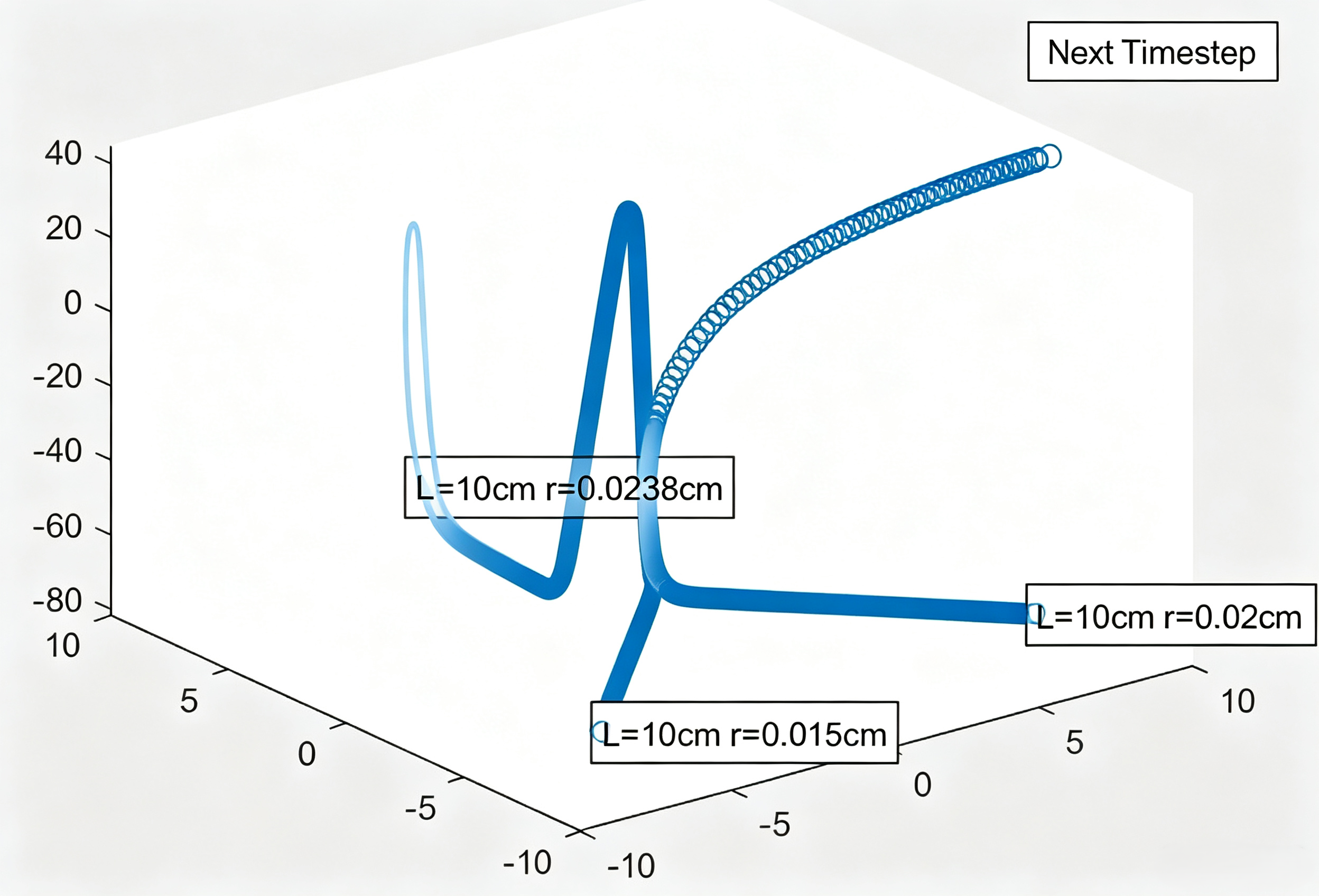}
\end{subfigure}
\caption{Action potential propagation for $r_1=0.0238$, $r_2=0.02$, $r_3=0.015$.}
\label{fig:ap_propagation}
\end{figure}
In table \ref{table_potential}, $r_1$ is parent cable radius, $r_2$ is first child cable radius, $r_3$ is second child cable radius. "pass parent", "pass first child", and "pass second child" represent whether the action potential successfully propagated through the indicated component of the cable. The radius typically have a scale of micrometers ($\mu m$), but here the unit of centimeter (cm) is used with the small decimal numbers to enable easier use with other cable property parameters such as membrane capacitance per unit area and ionic conductance per unit area. In the first experiment (figure \ref{fig:ap_propagation}), the parent cable starts with a radius of 0.0238cm, first child cable radius is 0.02cm, second child cable radius is 0.015cm. As the radius of first child cable increases (for example to 0.2cm), the velocity of the action potential through the second child cable decreases. When the radius of first child cable increases toward a threshold value of 0.3cm, only the parent cable has an action potential propagation and the signal stops at the junction between the parent and the two children.

\begin{figure}
\centering
\begin{subfigure}{0.23\textwidth}
\centering
\includegraphics[width=\textwidth,height=0.65\textwidth]{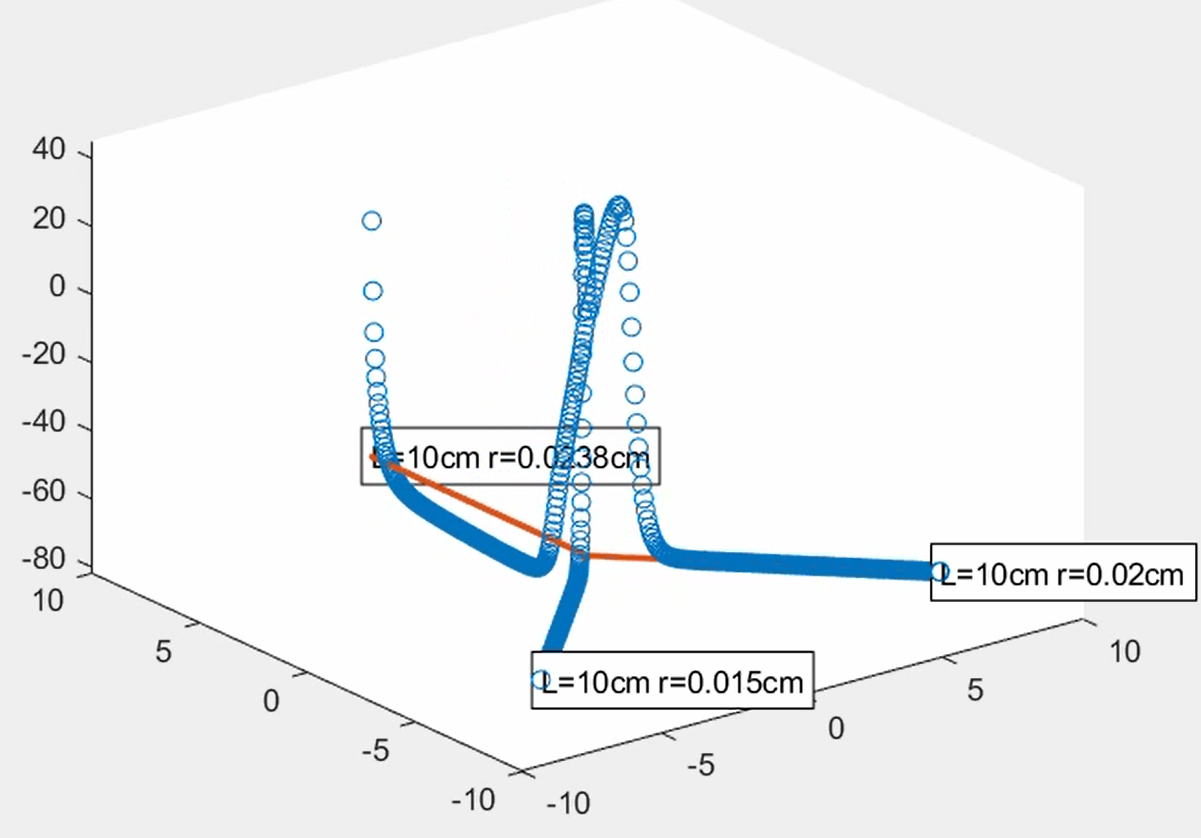}
\end{subfigure}
\begin{subfigure}{0.23\textwidth}
\centering
\includegraphics[width=\textwidth,height=0.65\textwidth]{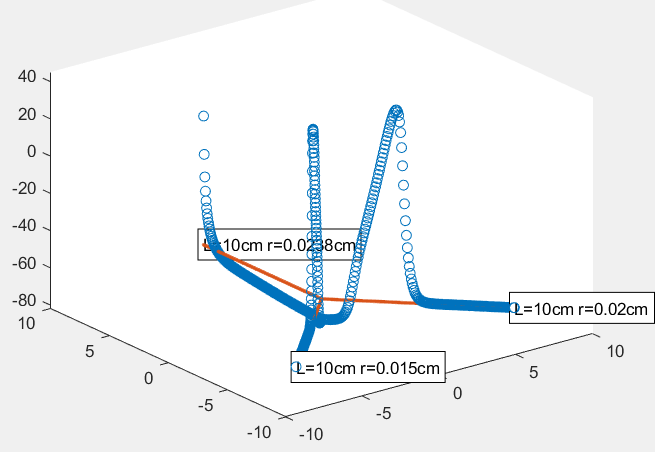}
\label{fig:ap_4}
\end{subfigure}
\begin{subfigure}{0.23\textwidth}
\centering
\includegraphics[width=\textwidth,height=0.65\textwidth]{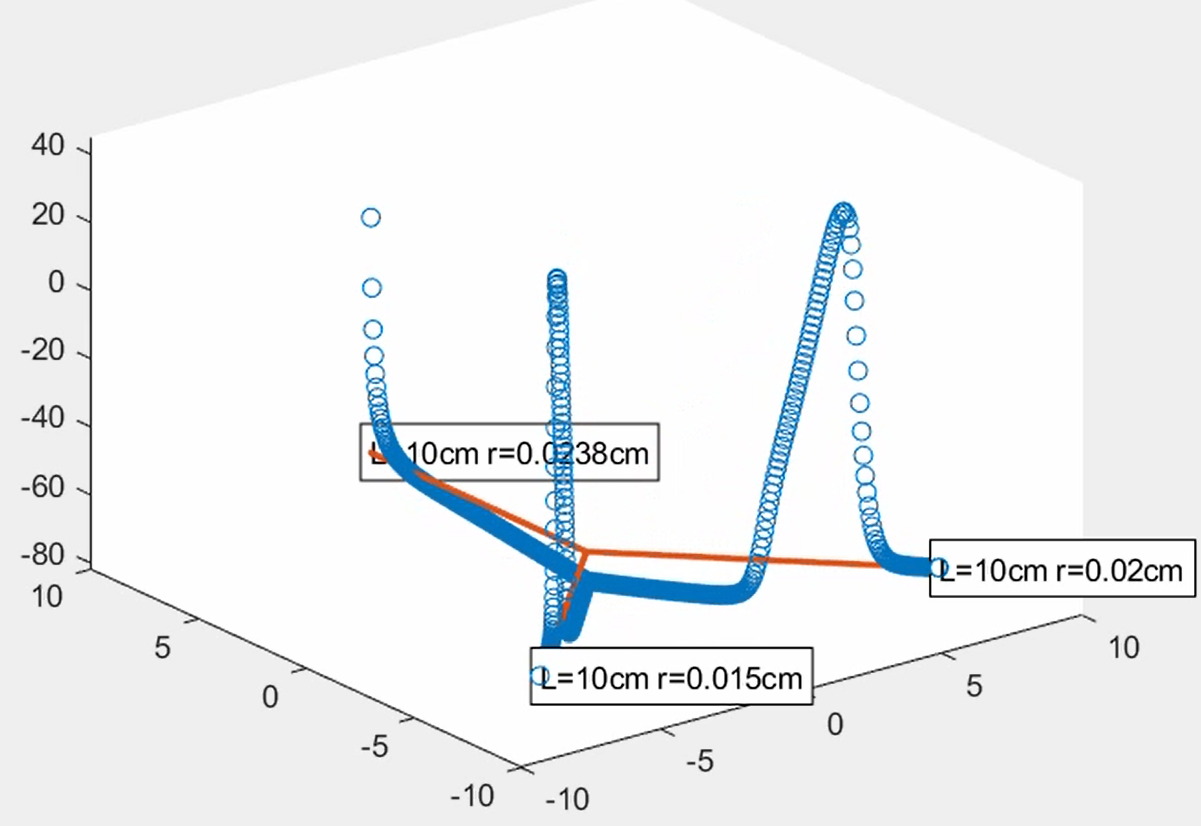}
\label{fig:ap_5}
\end{subfigure}
\hfill
\begin{subfigure}{0.23\textwidth}
\centering
\includegraphics[width=\textwidth,height=0.65\textwidth]{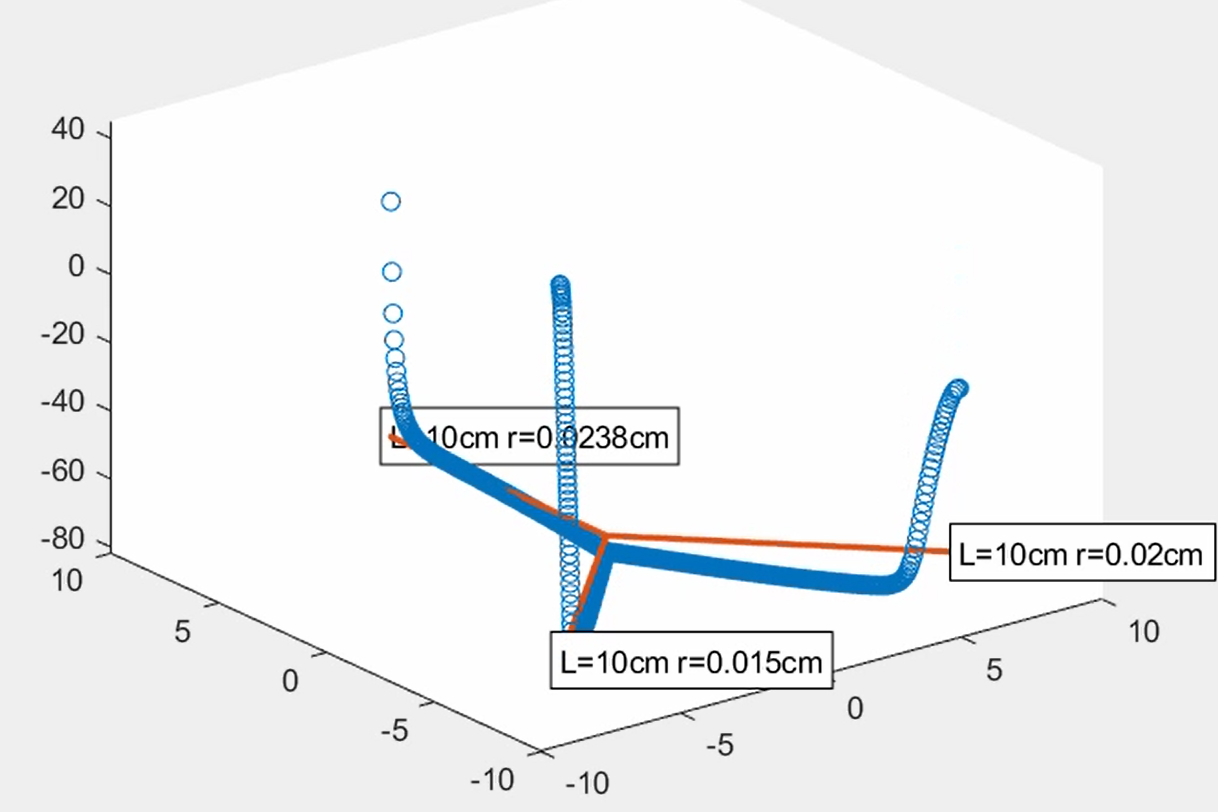}
\label{fig:ap_6}
\end{subfigure}
\begin{subfigure}{0.23\textwidth}
\centering
\includegraphics[width=\textwidth,height=0.65\textwidth]{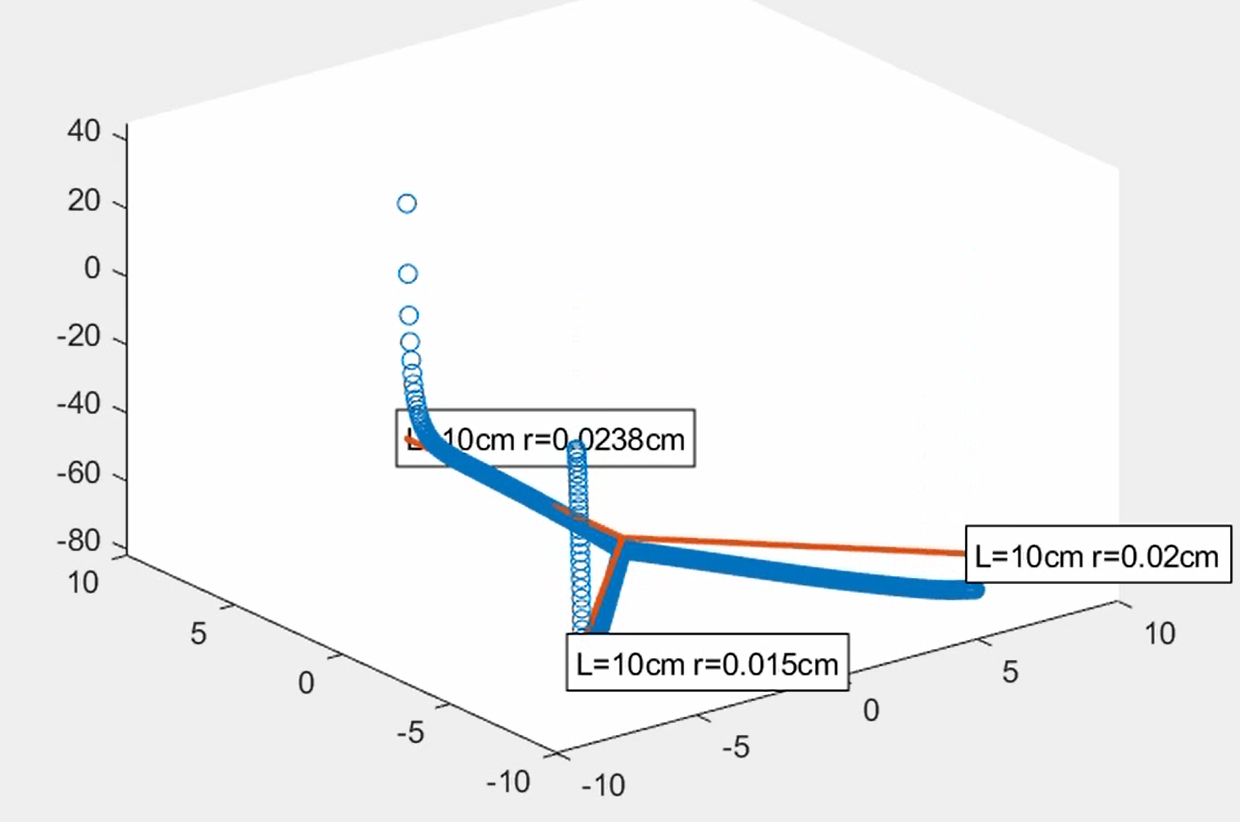}
\label{fig:ap_7}
\end{subfigure}
\hfill
\begin{subfigure}{0.23\textwidth}
\centering
\includegraphics[width=\textwidth,height=0.65\textwidth]{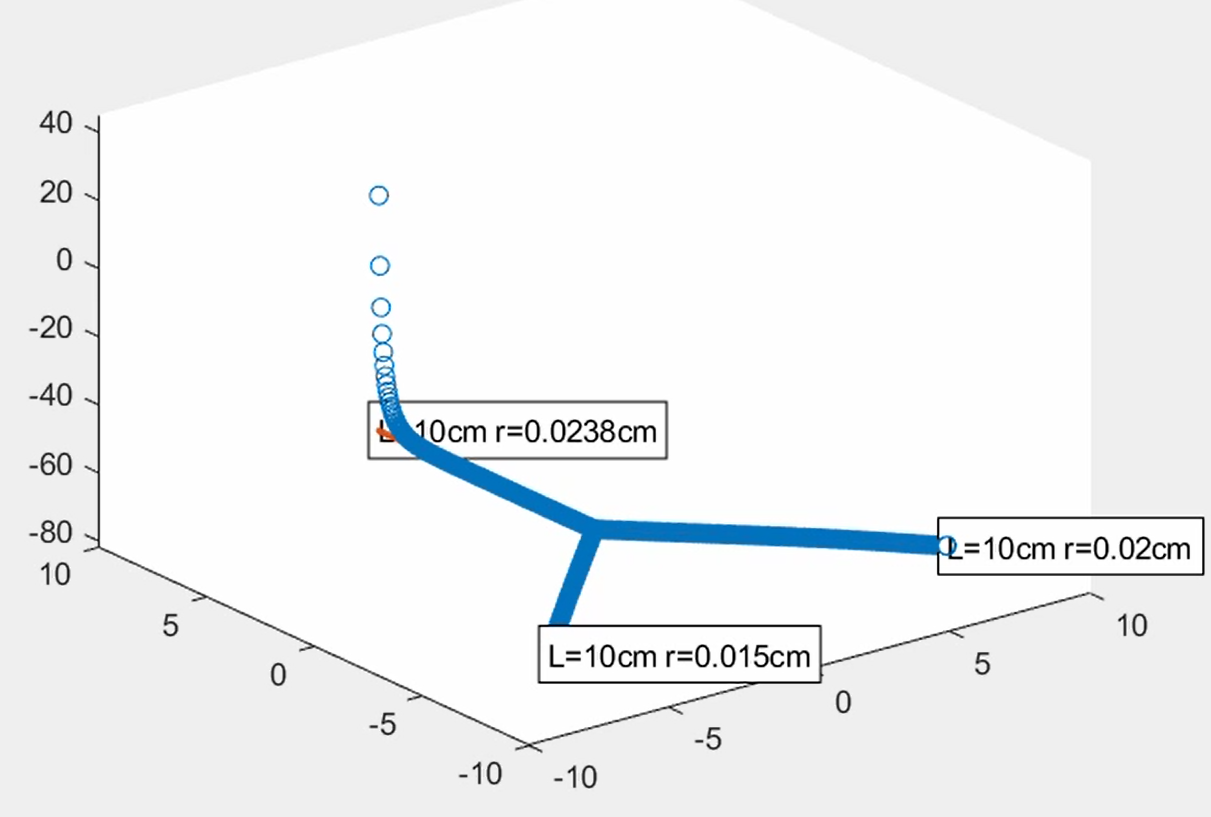}
\label{fig:ap_8}
\end{subfigure}
\caption{3D action potential propagation for $r_1=0.0238$, $r_2=0.02$, $r_3=0.2$ showing snapshots at consecutive time steps.}
\label{fig:ap_propagation2}
\end{figure}

In the second experiment, the parent cable starts with a radius of 0.0238 cm, first child cable radius is 0.02cm, second child cable radius is 0.2cm. For the second experiment, we reversed the role of the two child cables, this time fixing the radius of first child, and gradually increased the radius of the second child (figure \ref{fig:ap_propagation2}). A similar pattern emerges. As the radius of second child cable increases (for example to 0.2cm), the velocity of the action potential through the first child cable decreases. When the radius of second child cable increases toward a threshold value of 0.3cm, only the parent cable has an action potential propagation and the signal stops at the junction between the parent and the two children.


\begin{figure}
\centering
\begin{subfigure}{0.23\textwidth}
\centering
\includegraphics[width=\textwidth,height=0.65\textwidth]{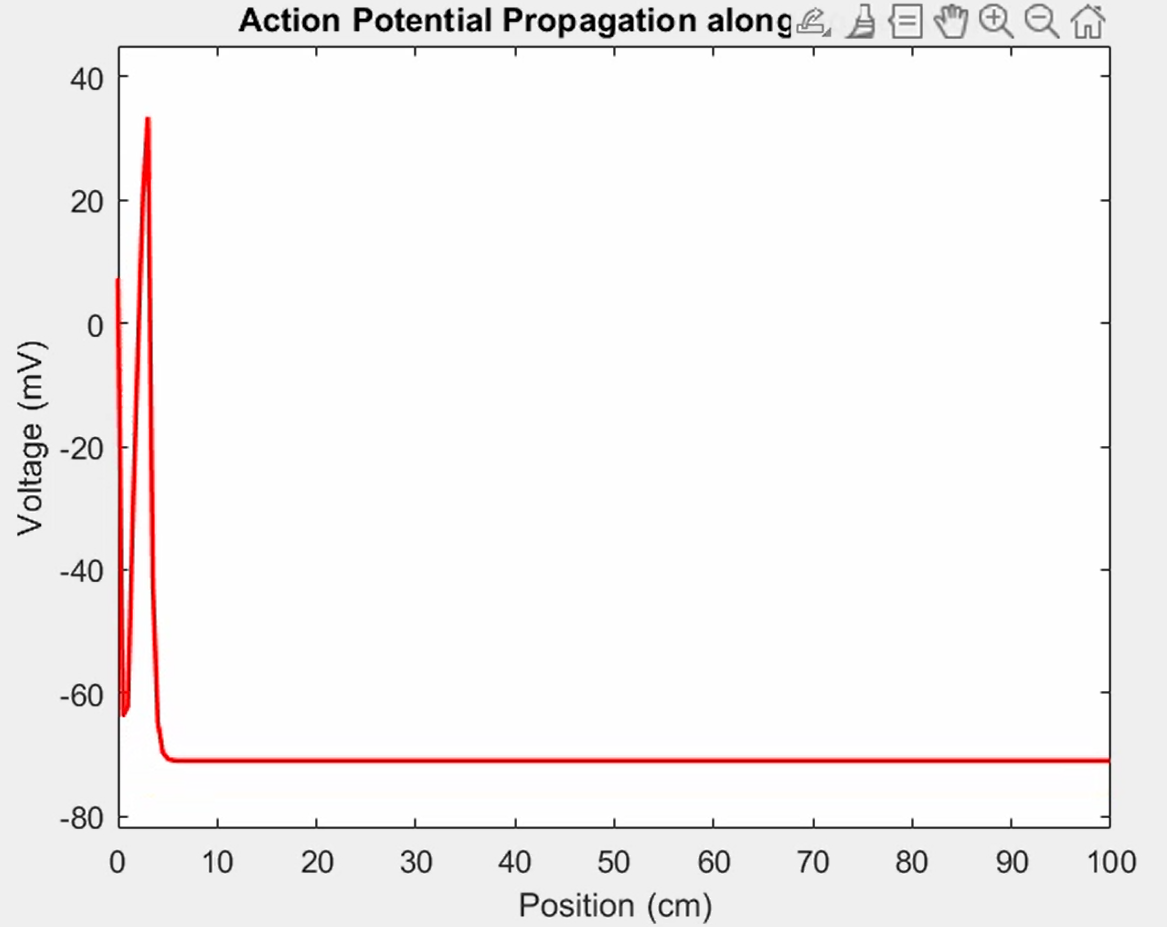}
\label{fig:1d_ap_1}
\end{subfigure}
\begin{subfigure}{0.23\textwidth}
\centering
\includegraphics[width=\textwidth,height=0.65\textwidth]{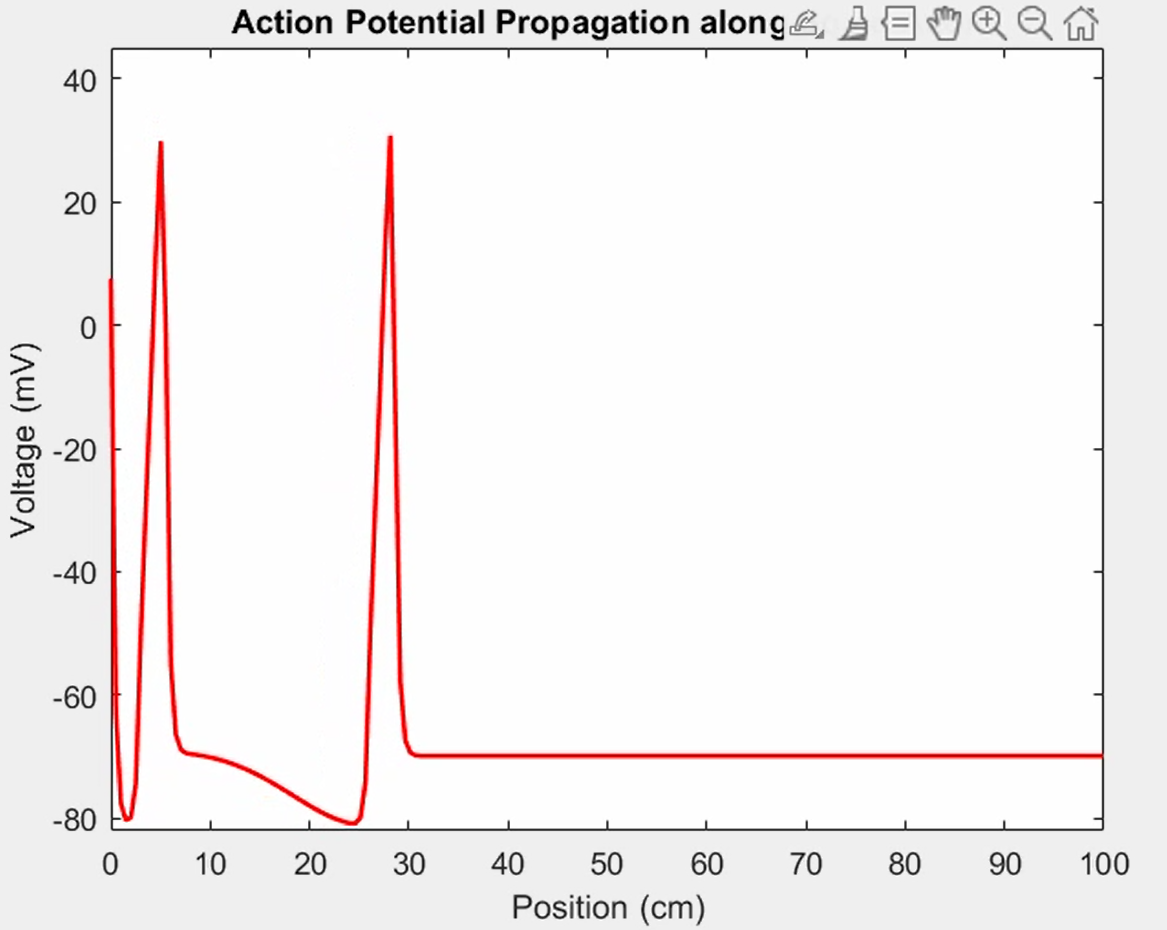}
\label{fig:1d_ap_2}
\end{subfigure}
\begin{subfigure}{0.23\textwidth}
\centering
\includegraphics[width=\textwidth,height=0.65\textwidth]{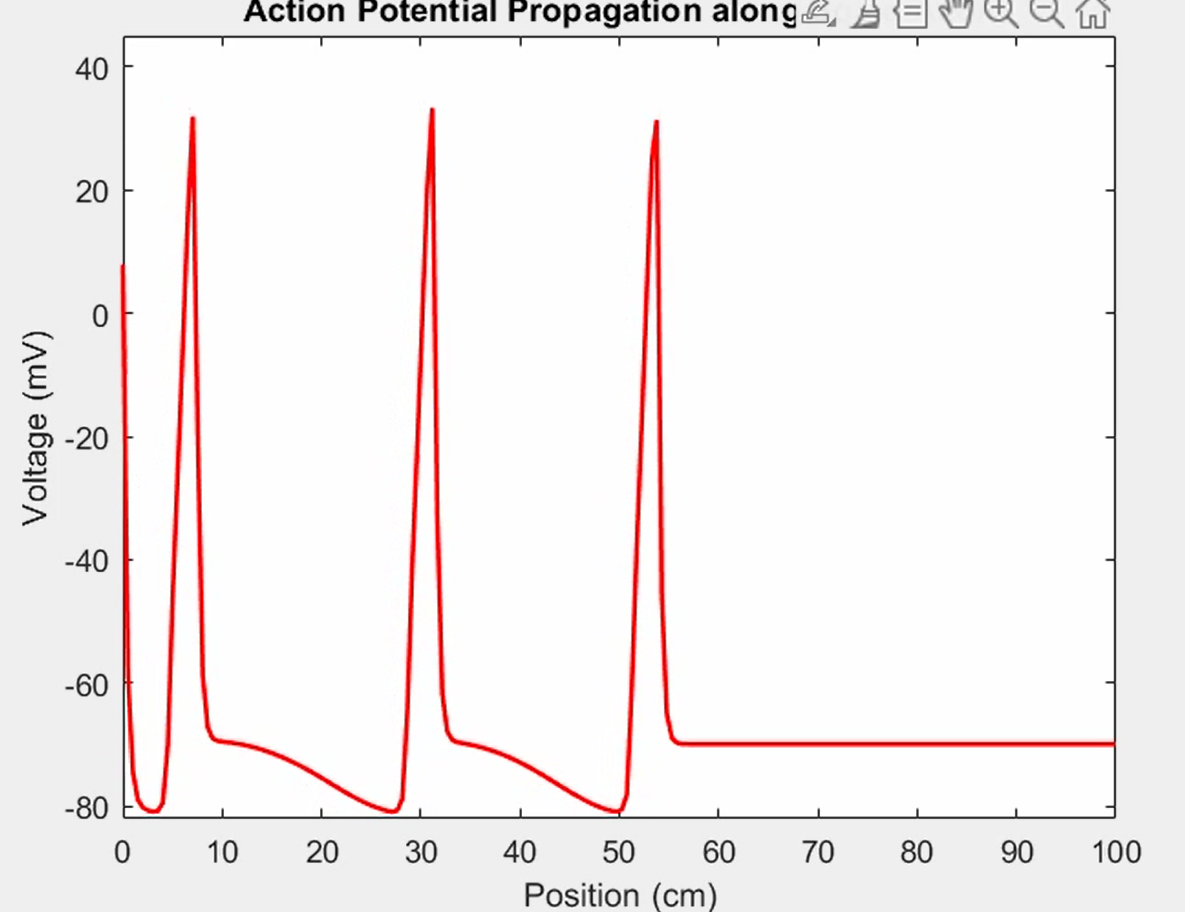}
\label{fig:1d_ap_3}
\end{subfigure}
\begin{subfigure}{0.23\textwidth}
\centering
\includegraphics[width=\textwidth,height=0.65\textwidth]{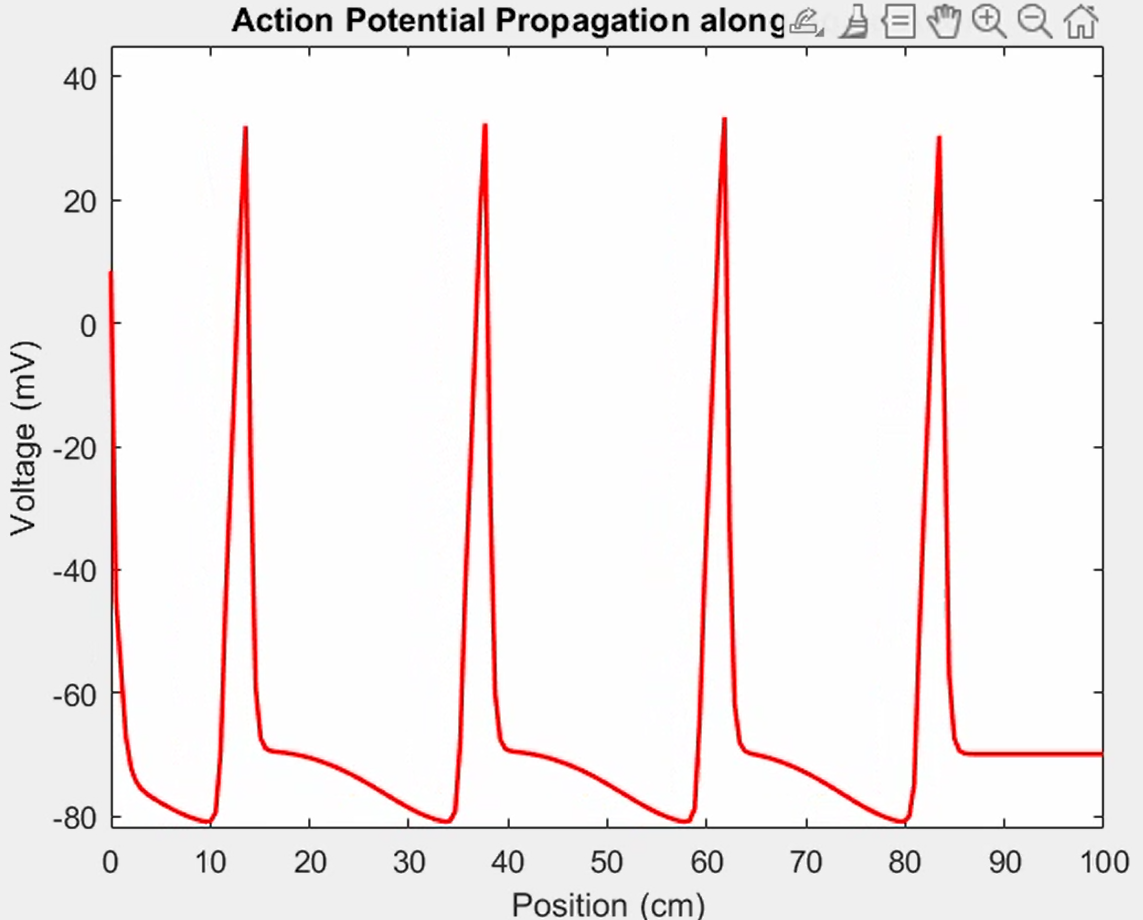}
\label{fig:1d_ap_4}
\end{subfigure}
\begin{subfigure}{0.23\textwidth}
\centering
\includegraphics[width=\textwidth,height=0.65\textwidth]{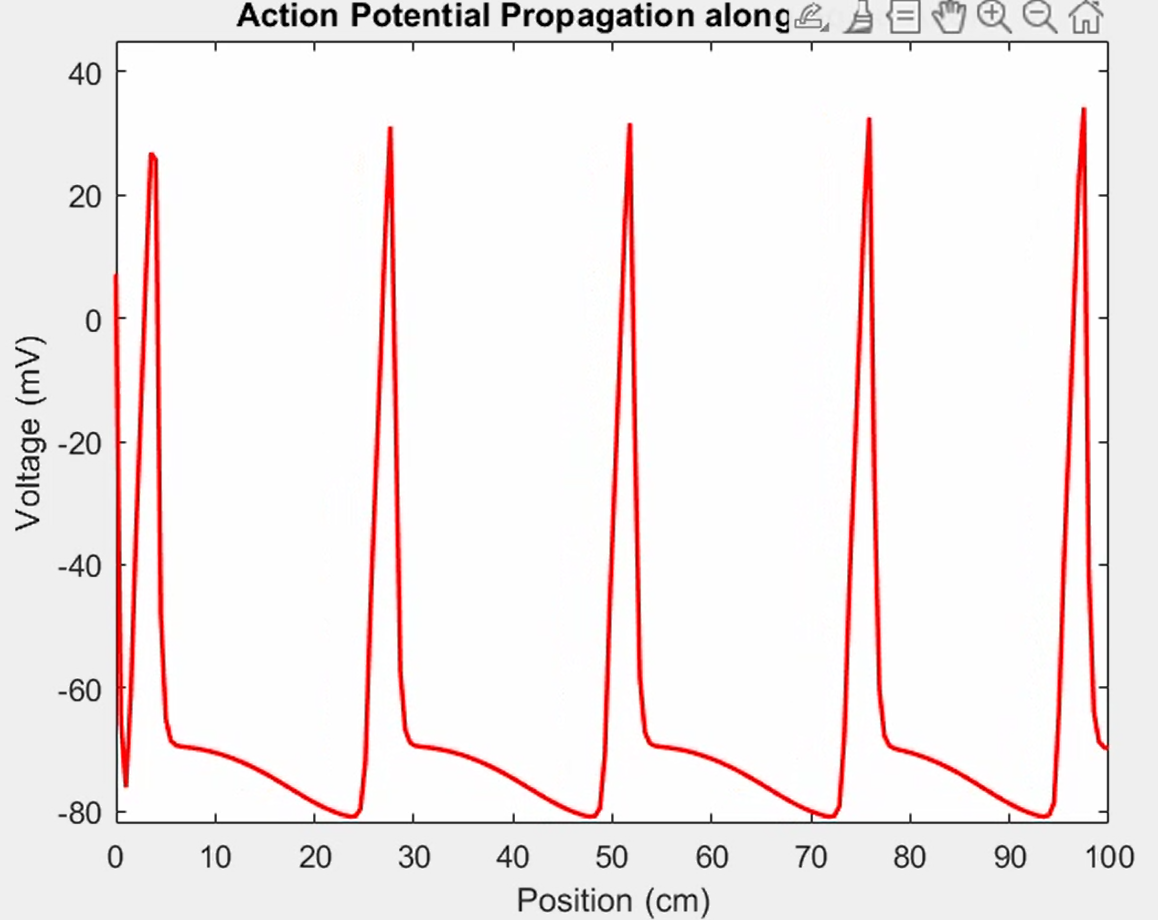}
\label{fig:1d_ap_5}
\end{subfigure}
\begin{subfigure}{0.23\textwidth}
\centering
\includegraphics[width=\textwidth,height=0.65\textwidth]{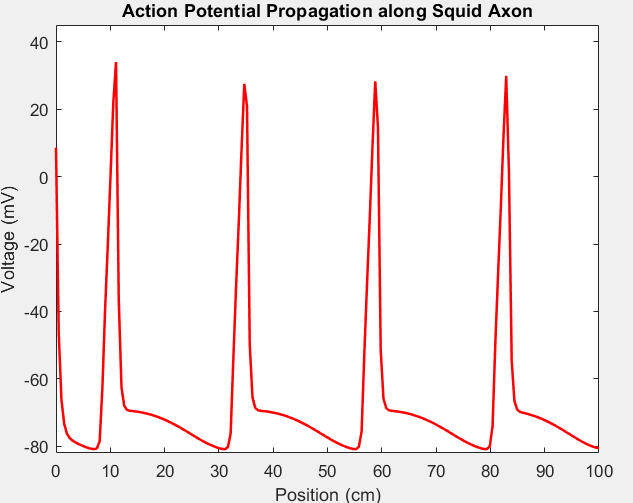}
\label{fig:1d_ap_6}
\end{subfigure}
\caption{1D action potential propagation for $r=0.0238$ showing frames 1 through 6.}
\label{fig:1d_ap_propagation}
\end{figure}

In the third experiment, we started to increase the radius of the parent cable. In the third experiment, the parent cable starts with a radius of 0.5cm, first child cable radius is 0.02cm, second child cable radius is 0.015cm. When the radius of the parent cable increased from 0.0238cm to 0.5cm, the velocity of the propagation increased (figure \ref{fig:1d_ap_propagation}). When the radius of parent cable increases toward a threshold value of 0.6cm, only oscillations were seen near the beginning of the parent cable but the signal cannot pass through the parent cable, not evening reaching the junction, and the action potential halts immediately at the beginning of the parent cable. Increasing the radius of one child branch ($r_2$) and fixing the radius of other branches caused the action potential propagation on the modified branch $r_2$ to be faster and slowed down the action potential propagation of the unmodified branch $r_1$, this can be explained by relation of resistance $R$ with the resistivity of the material $R$, length $L$ of the cable, and cross sectional area $A$ in the form $R = \frac{\rho L}{A}$. This is also due to $\lambda = \sqrt{\frac{r_n}{r_l}} = \sqrt{\frac{R_n}{R_i}\frac{d}{4}}$, a dendrite with larger diameter $d$ has a larger space constant $\lambda$, so the spread of current is accelerated with a larger diameter. Increasing the radius of one child branch ($r_2$) increases the cross sectional area $A = 2\pi r_2$ and decreases the resistance $R$, so the action potential propagation is become faster.


\begin{figure}
\centering
\begin{subfigure}{0.15\textwidth}
\centering
\includegraphics[width=\textwidth,height=0.8\textwidth]{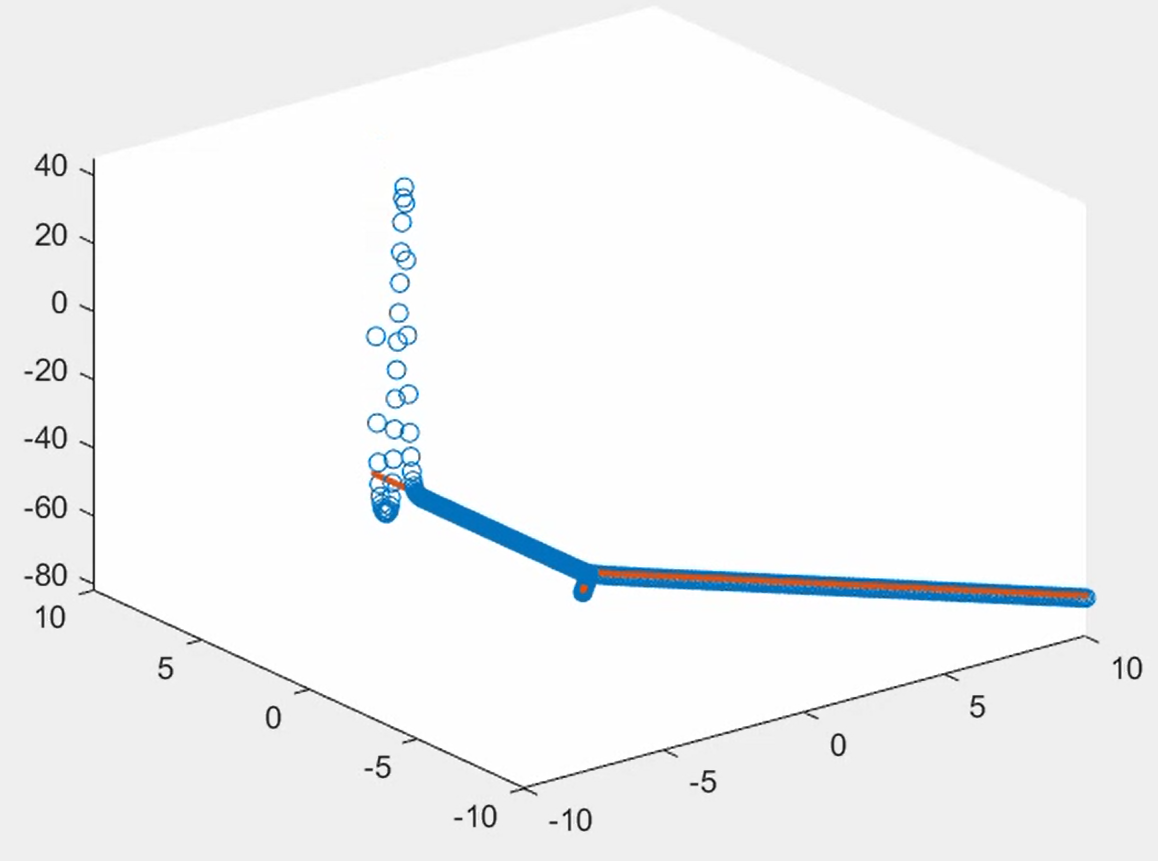}
\label{fig:cable3_1}
\end{subfigure}
\begin{subfigure}{0.15\textwidth}
\centering
\includegraphics[width=\textwidth,height=0.8\textwidth]{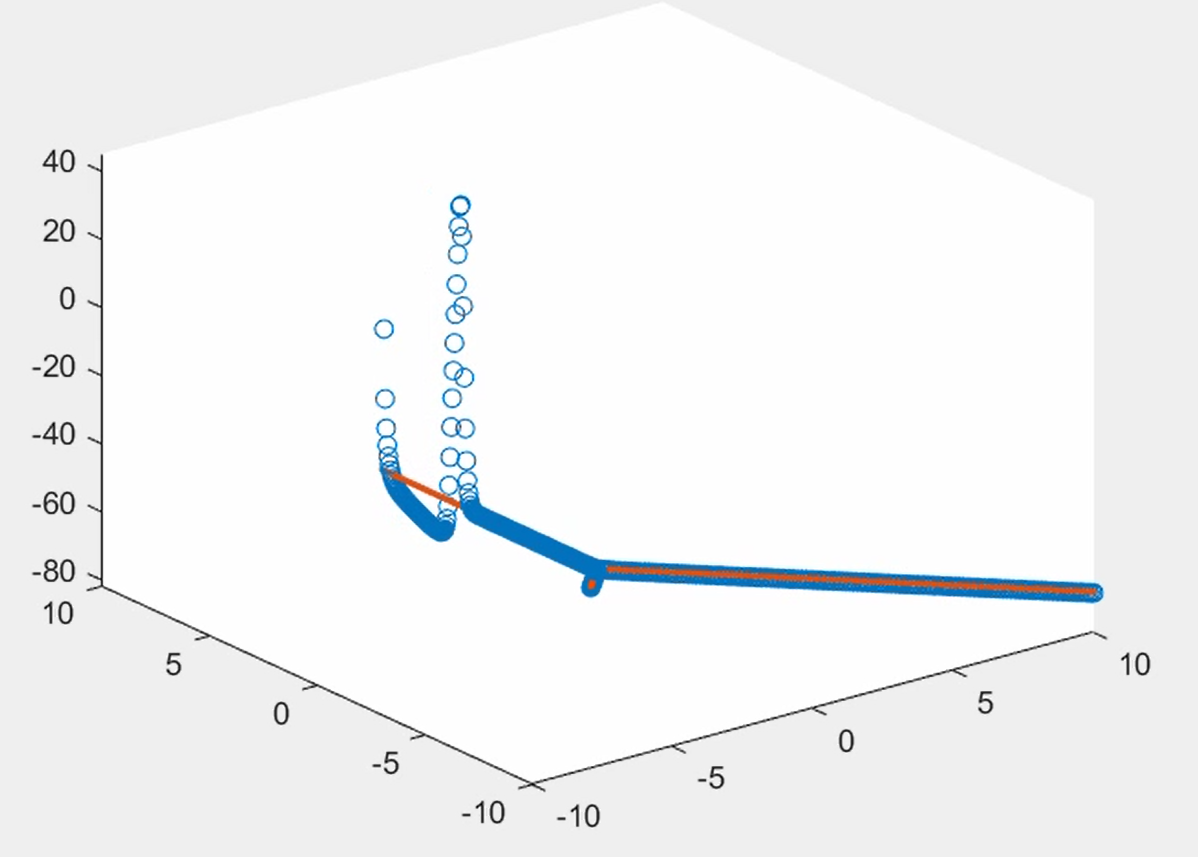}
\label{fig:cable3_2}
\end{subfigure}
\hfill
\begin{subfigure}{0.15\textwidth}
\centering
\includegraphics[width=\textwidth,height=0.8\textwidth]{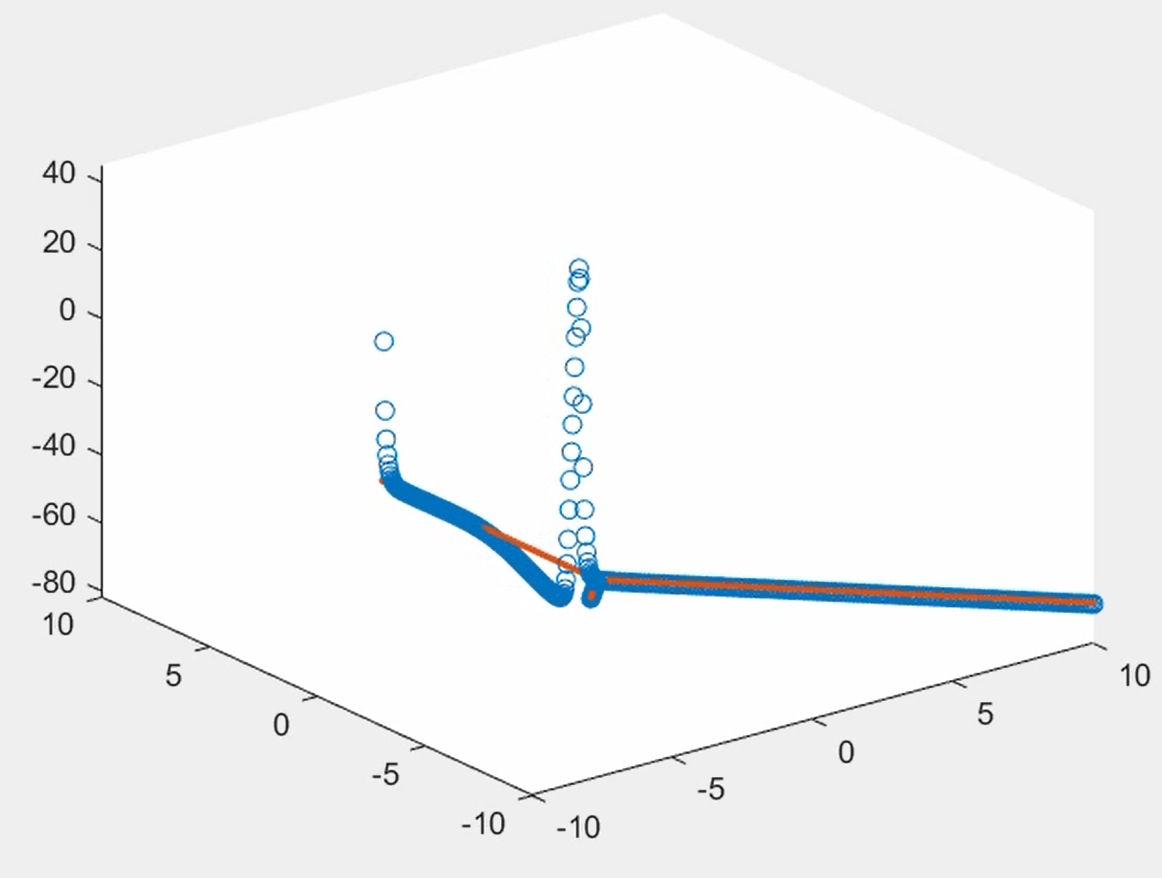}
\label{fig:cable3_3}
\end{subfigure}
\begin{subfigure}{0.15\textwidth}
\centering
\includegraphics[width=\textwidth,height=0.8\textwidth]{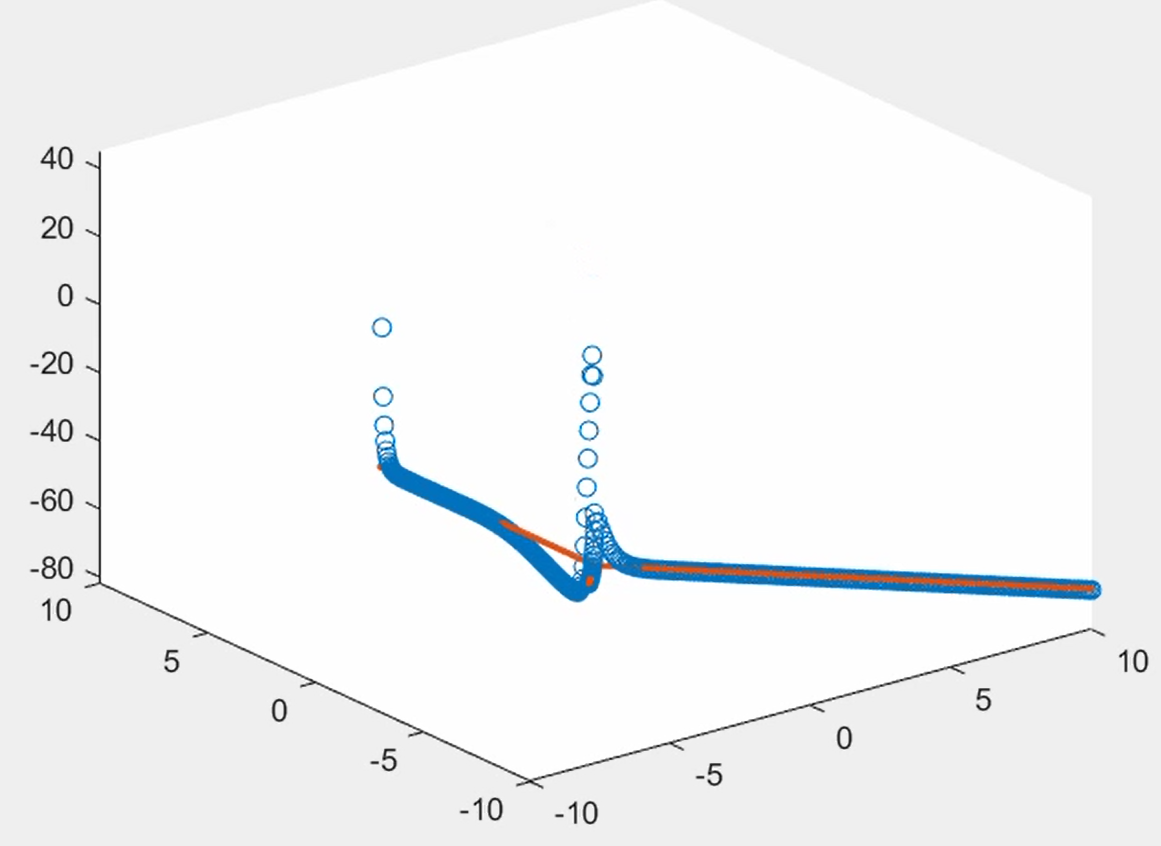}
\label{fig:cable3_4}
\end{subfigure}
\begin{subfigure}{0.15\textwidth}
\centering
\includegraphics[width=\textwidth,height=0.8\textwidth]{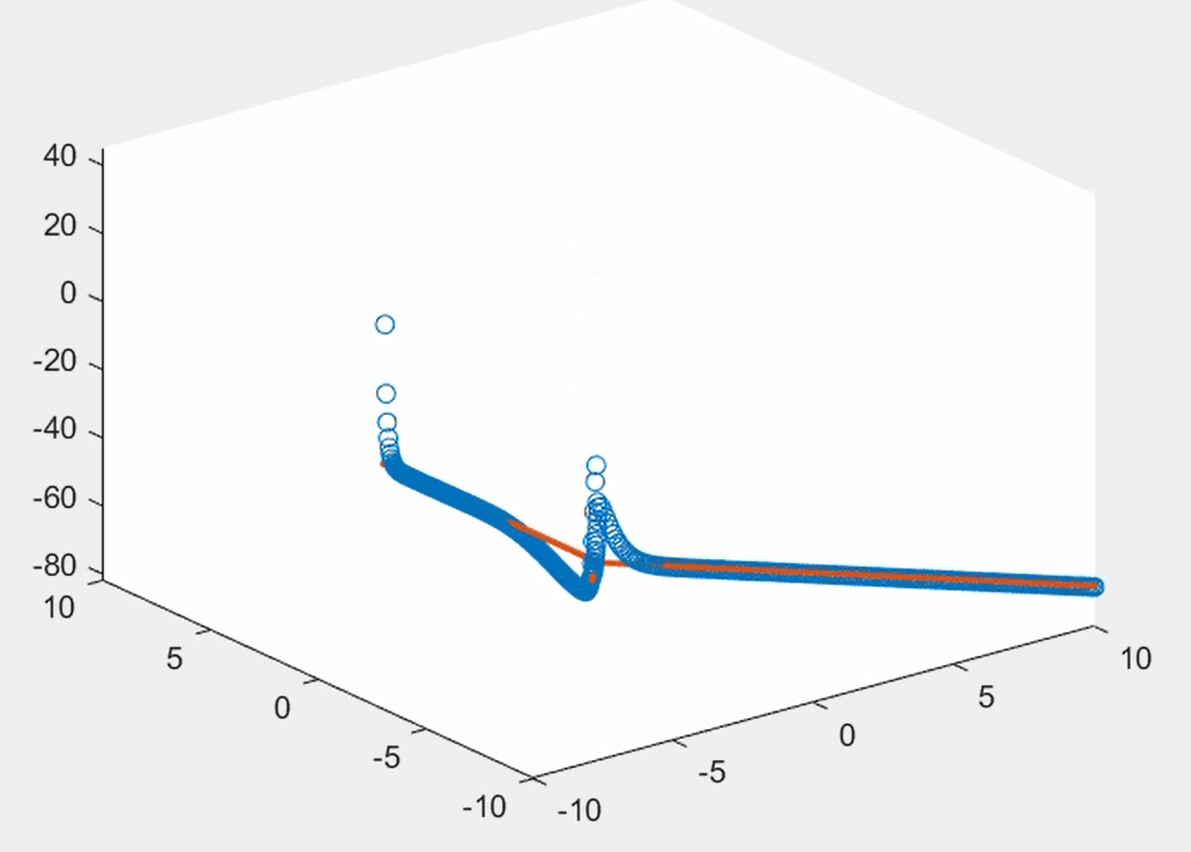}
\label{fig:cable3_5}
\end{subfigure}
\begin{subfigure}{0.15\textwidth}
\centering
\includegraphics[width=\textwidth,height=0.8\textwidth]{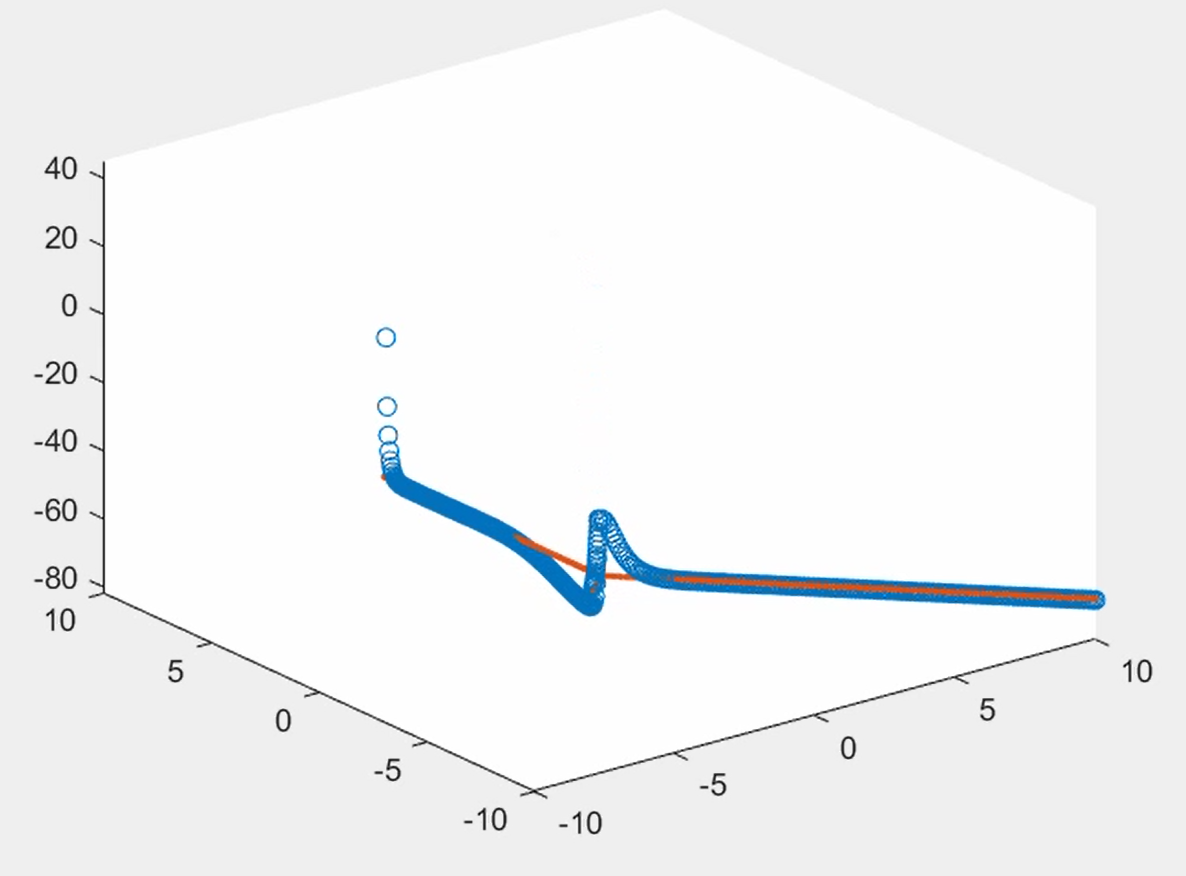}
\label{fig:cable3_6}
\end{subfigure}
\caption{3D action potential propagation for $r_1=0.0025$, $r_2=0.0075$, $r_3=0.0075$ showing snapshots at consecutive time steps from frame 1 to frame 6.}
\label{fig:cable3_propagation}
\end{figure}

Simulations of three-dimensional action potential propagation were performed for a branched axonal geometry with parent cable radius $r_1=0.0025$ and symmetric child branches $r_2=r_3=0.0075$ (figure \ref{fig:cable3_propagation}). Consecutive temporal snapshots (Frames 1 to 6) capture the spatiotemporal evolution of the propagating electrical signal. In Frame 1, the action potential initiates within the proximal segment of the parent axon and advances toward the bifurcation junction. By Frame 2, the depolarising wavefront arrives at the branch point, where current partitions between the two identical child branches due to their matched radii. Frames 3 and 4 show simultaneous, symmetric invasion of both child cables; the equal radii of $r_2$ and $r_3$ yield identical axial resistance and space constants, resulting in matching propagation speed along each branch. Frames 5 and 6 document continued unimpeded forward conduction within both distal branches, with no conduction failure or velocity asymmetry observed. The symmetric propagation behaviour arises from the matched cross-sectional area, axial resistance and electrotonic space constant $\lambda$ of the two daughter branches. In contrast to the asymmetric radius configurations examined in Experiments 1 and 2, equal branch radii eliminate preferential current loading onto one branch, removing the velocity reduction effect seen in heterogeneous bifurcations. Unlike the parent-radius threshold phenomenon observed in Experiment 3, the relatively narrow parent cable $r_1=0.0025$ does not exceed the critical radius for initiation failure, allowing the action potential to reliably reach the bifurcation and invade both child branches.


\begin{figure}
\centering
\begin{subfigure}{0.23\textwidth}
\centering
\includegraphics[width=\textwidth,height=0.8\textwidth]{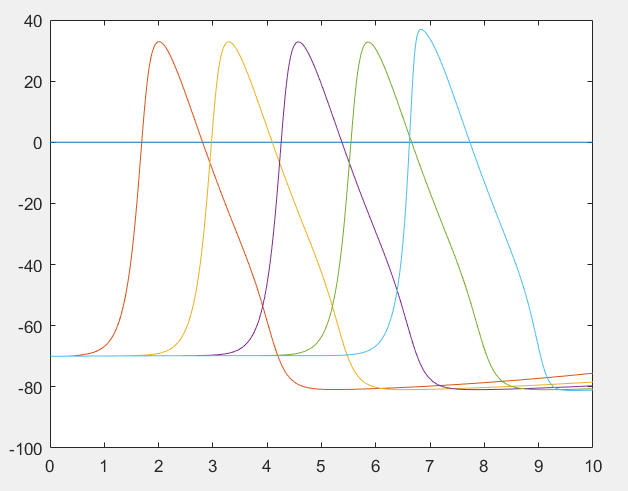}
\caption{$t=0$}
\label{fig:1d_r0004_0}
\end{subfigure}
\begin{subfigure}{0.23\textwidth}
\centering
\includegraphics[width=\textwidth,height=0.8\textwidth]{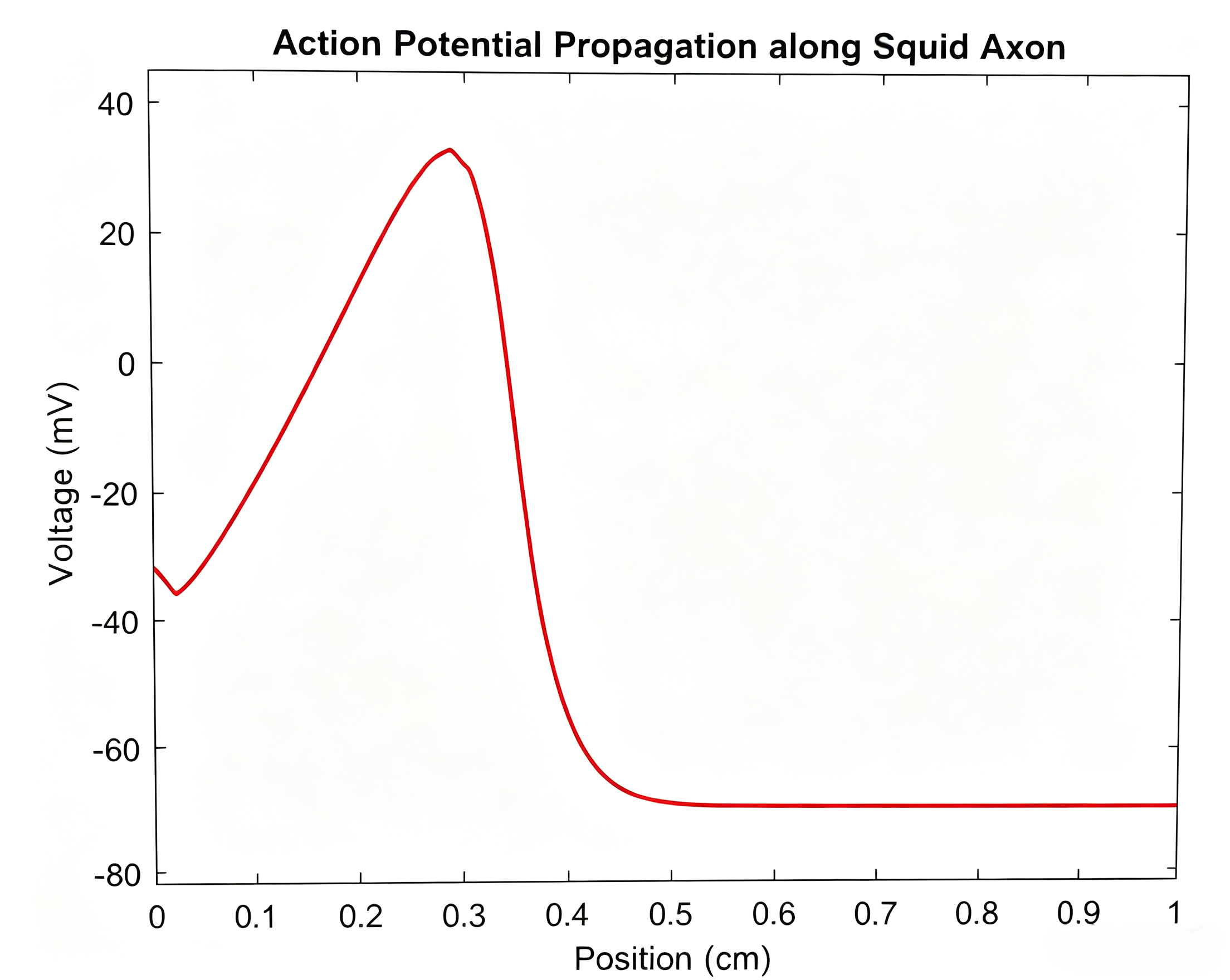}
\caption{$t=\Delta t$}
\label{fig:1d_r0004_1}
\end{subfigure}
\begin{subfigure}{0.23\textwidth}
\centering
\includegraphics[width=\textwidth,height=0.8\textwidth]{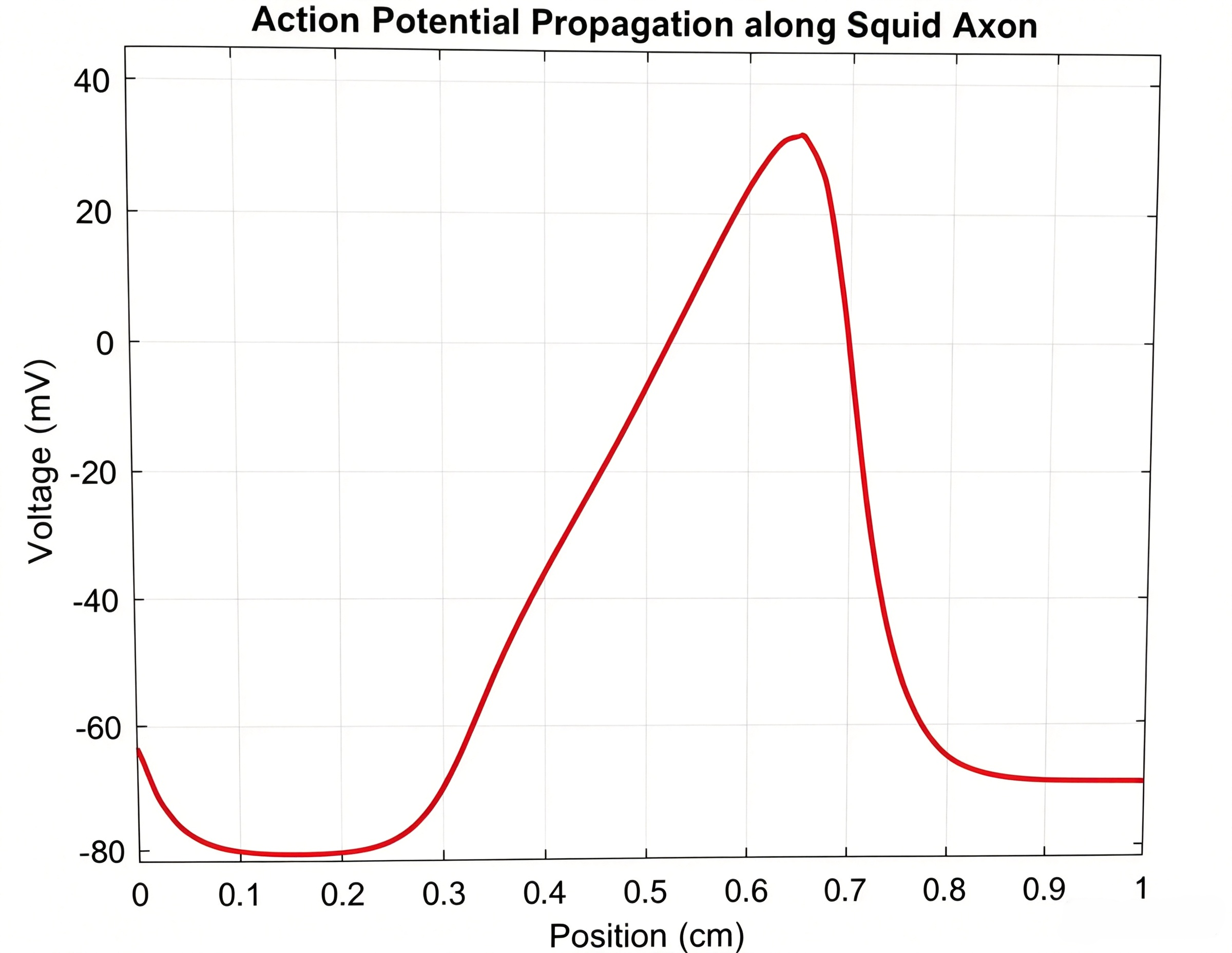}
\caption{$t=2\Delta t$}
\label{fig:1d_r0004_2}
\end{subfigure}
\begin{subfigure}{0.23\textwidth}
\centering
\includegraphics[width=\textwidth,height=0.8\textwidth]{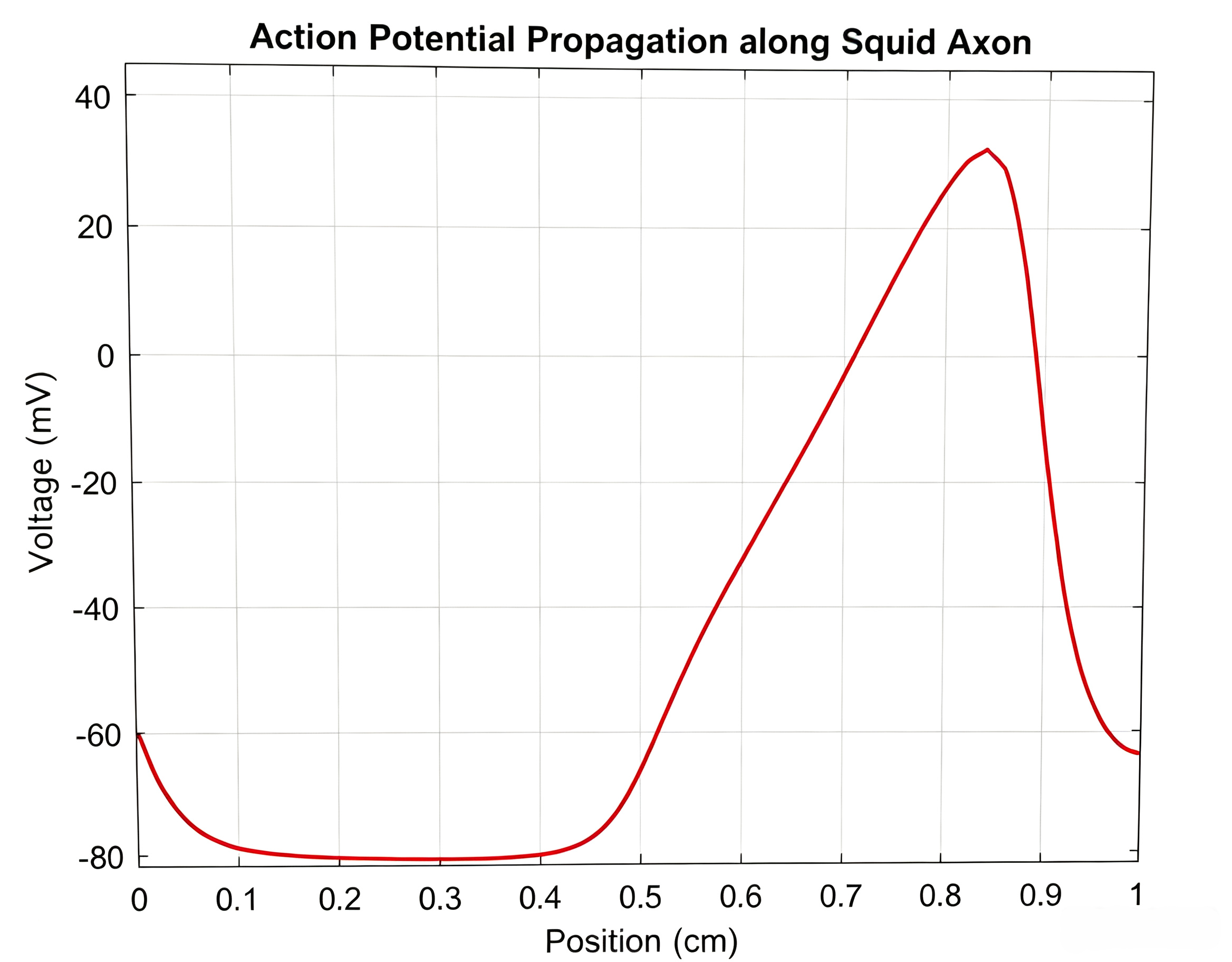}
\caption{$t=3\Delta t$}
\label{fig:1d_r0004_3}
\end{subfigure}
\caption{1D action potential propagation for $r=0.0004$. The evolution is shown from $t=0$ to $t=3\Delta t$.}
\label{fig:1d_r0004_propagation}
\end{figure}

Figure \ref{fig:1d_r0004_propagation} displays instantaneous spatial membrane potential profiles of a propagating action potential along a squid giant axon, obtained from Hodgkin-Huxley cable simulations. Each curve constitutes a temporal snapshot plotting transmembrane voltage against axial position, illustrating the stereotyped waveform formed by sodium-driven depolarization, peak overshoot, and potassium-mediated repolarization. Successive snapshots reveal the intact action potential waveform translates continuously along the axon toward distal regions, confirming unimpeded one-dimensional conduction. This unbranched baseline serves as a reference for evaluating propagation fidelity at axonal bifurcations with heterogeneous branch radii.
\section{Conclusion}
The paper investigated the impact of altering cable conductor geometry on action potential propagation through experiments involving gradual changes in cable radius and branching. Observations were made on how modifying the radius of various branches affected action potential propagation, with a parent and two child branches in the simulation. Fixing other initial conditions, increasing the diameter of a specific branch could either accelerate propagation in the modified child branch, slow it down in the unmodified child branch, or lead to propagation failure beyond the region of geometry change. At the branch point \( x_0 \) where geometry changes occur, continuity required matching currents \( I_i \) from the left and right sides, governed by the relation $I_i = -\frac{\pi d_1^2}{4 R_i}\frac{dV}{dx}_{x_{0-}} = -\frac{\pi d_2^2}{4 R_i}\frac{dV}{dx}_{x_{0+}}$ where \( d_1 \) and \( d_2 \) are diameters. A geometric ratio \( GR = \sum_i \frac{d^{3/2}}{d_a^{3/2}} \), comparing the diameter \( d_a \) of the source branch to \( d_i \) at all other branches on the opposite side of the junction, was used to classify equivalent cylinders.

This work extends conventional quasi-static neuronal cable theory by constructing a fully coupled multi-physics framework that unites Maxwell’s electrodynamic equations, finite-difference time-domain (FDTD) field solvers, and modified Hodgkin-Huxley/Fitzhugh-Nagumo membrane dynamics. The integrated model incorporates magnetic induction, Lorentz force ionic coupling, electromagnetic trans-membrane currents $I_{\text{EM}}$, magnetic perturbations to ion channel gating variables, and supplementary quantum corrections for nanoscale axonal and dendritic segments, addressing key simplifying assumptions omitted from standard core-conductor formulations.

A series of controlled numerical experiments were performed on branched axonal architectures featuring parent-child bifurcation geometries with variable branch radii. Simulations confirm that electromagnetic effects introduce measurable departures from purely geometric propagation predictions. First, inductive magnetic currents amplify impedance mismatch at junctions, reducing the critical branch diameter threshold that triggers complete action potential conduction failure relative to classical cable results. Second, externally applied transverse magnetic fields break symmetric signal invasion within geometrically identical child branches via Lorentz force interactions, producing differential conduction velocities absent in zero-field quasi-static simulations. Third, ideal current continuity at branch points is violated by transient displacement and magnetic fluxes, requiring the revised electromagnetic geometric ratio $GR_{\text{EM}}$ to accurately quantify junction impedance matching instead of the traditional $d^{3/2}$ scaling rule. Fourth, parent axon conduction velocity deviates from the canonical $\sqrt{d}$ space-constant scaling: electromagnetic inductive feedback accelerates signal transmission for intermediate radii, while large-diameter cables generate hyper-polarizing electromagnetic offsets that initiate premature propagation blockage.

The supporting numerical visualizations summarized in Figure \ref{fig:em_propagation_experiments} validate that quasi-static cable models systematically underestimate electrodynamic corrections to wavefront speed, action potential waveform morphology, bifurcation transmission reliability, and symmetric branch invasion dynamics. The proposed coupled Maxwell-cable framework resolves these limitations by self-consistently evolving electric and magnetic fields alongside active membrane voltage dynamics on a Yee FDTD grid, while modified junction boundary conditions preserve multi-physics current balance at geometric discontinuities.
Future work will extend this framework to heterogeneous neural networks with chemical and electrical synapses, incorporate stochastic electromagnetic thermal noise, and validate model outputs against high-resolution magnetophysiology experimental recordings. We demonstrate that full electrodynamic coupling is necessary to capture complete action potential propagation behavior in complex branched neuronal tissue, offering an advanced computational tool for biophysically realistic neural signal modeling.
\bibliographystyle{elsarticle-num}
\bibliography{ref}
\end{document}